\documentclass[a4paper,twoside]{book}
\newif\ifdraft \global\drafttrue

\usepackage[T1]{fontenc}
\usepackage{a4wide}
\usepackage{times}
\usepackage{graphicx}
\usepackage{theorem}
\usepackage[colorlinks=true,hyperindex=true]{hyperref}
\usepackage{framed}
\usepackage{tikz}
\usepackage{pgfplots}
\pgfplotsset{soldot/.style={color=blue,only marks,mark=*}}
\pgfplotsset{holdot/.style={color=blue,fill=white,only marks,mark=*}}
\usetikzlibrary{arrows.meta}
\tikzset{>={Latex[width=2.5mm,length=2.5mm]}}
\usetikzlibrary{patterns}

\usepackage{color}
\usepackage{fancyhdr}
\usepackage{amsmath}
\usepackage{bbm}
\usepackage{amsfonts}
\usepackage{enumerate}
\usepackage{cancel}
\ifdraft
\fi

\numberwithin{equation}{chapter}
\newtheorem{theorem}{Theorem}[chapter]
\newtheorem{proposition}[theorem]{Proposition}
\newtheorem{lemma}[theorem]{Lemma}
\newtheorem{definition}[theorem]{Definition}
\newtheorem{corollary}[theorem]{Corollary}
\theoremstyle{plain}
\theorembodyfont{\upshape}
\newtheorem{example}{Example}[chapter]

\newtheorem{remark}{Remark}[chapter]

\newcounter{smallarabics}
\newenvironment{arabicenumerate}
{\begin{list}{{\normalfont\textrm{(\arabic{smallarabics})}}}
{\usecounter{smallarabics}\setlength{\itemindent}{0cm}
\setlength{\leftmargin}{5ex}\setlength{\labelwidth}{4ex}
\setlength{\topsep}{0.75\parsep}\setlength{\partopsep}{0ex}
\setlength{\itemsep}{0ex}}}
{\end{list}}
\renewcommand\theexercise{\arabic{chapter}.\arabic{exercise}}
\newlength{\exoskip}
\newlength{\exopskip}
\newcounter{exopcnt}[exercise]
\newcommand{\exop}{
\stepcounter{exopcnt}
\ifnum\value{exopcnt}>0\vskip\exopskip\fi
\noindent\arabic{exopcnt}.\ }
\definecolor{shadecolor}{gray}{0.95}

\newenvironment{exo}
{\refstepcounter{exercise}%
\fboxsep=1.2\FrameSep
\MakeFramed{\advance\hsize-20pt\FrameRestore}%
\noindent\textbf{Exercise~\theexercise. }
\ignorespaces
}{\endMakeFramed}

\def\bel{\begin{lemma}}
\def\eel{\end{lemma}}
\def\bec{\begin{corollary}}
\def\eec{\end{corollary}}
\def\bet{\begin{theorem}}
\def\eet{\end{theorem}}
\def\bed{\begin{definition}}
\def\eed{\end{definition}}
\def\bep{\begin{proposition}}
\def\eep{\end{proposition}}
\def\ben{\begin{arabicenumerate}}  
\def\een{\end{arabicenumerate}}
\def\beq{\begin{equation}}
\def\eeq{\end{equation}}  
\def\demo{{\noindent\bf Proof.\ }}
\def\qed{\hfill$\square$}
\newcommand{\e}{\mathrm{e}}

\renewcommand{\i}{\mathrm{i}}

\renewcommand{\d}{\mathrm{d}}
\def\nn{{\mathbb N}}

\def\rr{{\mathbb R}}
\def\cc{{\mathbb C}}

\def\bPt{\bar P^{\kern1pt t}}
\def\bQt{\bar Q^{\kern1pt t}}

\def\mP{{ P}}
\def\mQ{{Q}}

\def\Im{\mathrm{Im}\,}

\def\bar{\overline}
\def\ubar{\underline}
\def\cal{\mathcal}

\def\mE{{ E}}
\def\supp{{\rm supp}}

\begin{document}
\title{\bf Lectures on Entropy. Part I.}
\author{\sc Vojkan Jak\v{s}i\'c\\ \\ \\
\\ \\ \\ 
Department of Mathematics and Statistics\\ 
McGill University\\
805 Sherbrooke Street West \\
Montreal,  QC,  H3A 2K6, Canada
 \\ \\
\copyright\; 2018 Vojkan Jak\v{s}i\'c\\
All Rights Reserved
}

\maketitle
\thispagestyle{empty}
\tableofcontents

\chapter{Introduction}
These lecture notes concern information-theoretic notions  of entropy. They are intended for, and 
have been successfully taught to, undergraduate students interested in research careers. 
Besides basic notions of analysis related to convergence that are  typically taught in  the first or second year 
of undergraduate studies, no other background is needed to read the notes.
The notes might  be  also of  interest to  any mathematically inclined reader  who wishes to learn basic facts about notions 
of entropy in 
an elementary setting.

As the title indicates,  this is the first in  a planned series of four  lecture notes. The Part II concerns notions of  entropy in  study 
of statistical mechanics, and  III/IV are the  quantum information theory/quantum statistical mechanics counterparts of I/II. 
All four  parts target similar audience and 
are  on a similar technical level. Eventually,  Parts I-IV  together are intended to be an introductory chapter to  a comprehensive 
volume dealing with the topic of entropy from a certain point of view on which I will elaborate 
below.

The research  program that leads to these lecture notes concerns the elusive notion of entropy in non-equilibrium 
statistical mechanics. It is for this pursuit that the notes are preparing a research-oriented reader, and it is the pursuit to 
which the later more advanced topics hope to contribute. Thus, it is important to emphasize that the choice of topics 
and their presentation have a specific motivation which may not be obvious until at least the Part II  of the lecture notes 
is completed. Needless to say, the lecture notes can be read independently of its motivation, as they provide 
a concise, elementary, and mathematically rigorous introduction to the topics they cover. 

The theme of this Part I is the Boltzmann--Gibbs--Shannon  (BGS) entropy of a finite probability distribution $(p_1, \cdots, p_n)$, and 
its various deformations such as the R\'enyi entropy,  the relative entropy, and the relative R\'enyi entropy. The BGS entropy and 
the relative  entropy have intuitive and beautiful axiomatic characterizations discussed in Section \ref{sec-e-natural} and Chapter \ref{sec-re-nat}.   The R\'enyi entropies also have axiomatic 
characterizations, but those are perhaps less natural, and we shall not discuss them in detail. Instead, 
we shall motivate the R\'enyi entropies by  the 
so-called Large Deviation Principle (LDP) in probability theory.  The link between the LDP and notions of 
entropy runs deep and will play a central role in this lecture notes.  For this reason Cram\'er's theorem is proven right away in the introductory 
Chapter \ref{sec-probability} (the more involved proof of  Sanov's theorem is given  in Section 
\ref{sec-sanov}). It is precisely 
this emphasis on the LDP that makes this lecture notes somewhat unusual in comparison with other introductory presentations 
of the information-theoretic entropy.

The  Fisher entropy and a related topic of  parameter estimation are also an important part of this lecture notes. The historical background and most 
of applications of these topics are in the field of statistics. There  is a hope that they  may play an important role in 
study of entropy in non-equilibrium statistical mechanics, and that is the reason for including them in the lecture notes. 
Again, Chapters  \ref{sec-fisher-entropy} and \ref{sec-estimation} can be read independently of this motivation by anyone interested in an elementary introduction 
to the Fisher entropy and parameter estimation. 

These notes are work in progress, and additional topics  may be added in the future.

The notes benefited from the comments of numerous McGill undergraduate students who attended the seminars and courses in which 
I have taught the presented material. I am grateful  for their help and for their enthusiasm which to a large 
extent  motivated my decision to prepare the notes for publication. In particular, I am grateful to Sherry Chu, Wissam Ghantous, 
and Jane Panangaden
 whose McGill's undergraduate summer research  projects were linked to the topics of the  lecture notes and whose 
research  reports  helped 
me in writing parts of the notes. 
I am also grateful to Laurent Bruneau, No\'e Cuneo, Tomas Langsetmo, Renaud Raqu\'epas and Armen Shirikyan  for comments and suggestions. I wish to thank 
Jacques Hurtubise and David Stephens who, as the chairmans of the McGill Department of Mathematics and 
Statistics, enabled me to 
teach the material of the notes in a course fomat. Finally, I am grateful to Marisa Rossi  for her 
exceptional hospitality and support during the period when Chapter \ref{sec-estimation} was written.

This research that has led to this lecture notes was partly funded by NSERC, {\it Agence Nationale de la Recherche\/} through the grant NONSTOPS (ANR-17-CE40-0006-01, ANR-17-CE40-0006-02, ANR-17-CE40-0006-03), the CNRS collaboration grant {\it Fluctuation theorems in stochastic systems\/}, and the {\it Initiative d'excellence Paris-Seine\/}.

\section{Notes and references.}

Shannon's seminal 1948 paper \cite{Sha}, reprinted in \cite{ShaWe}, 
 remains a must-read for anyone interested in notions of entropy.  Khintchine's  
 reworking of the mathematical foundations of Shannon's theory in early 1950's, summarized in 
 the monograph \cite{Khi},   provides  a perspective on the early 
mathematically  rigorous developments of the subject. For further historical perspective we refer the reader to   \cite{Ver} and the 
detailed list of references provided there. There are many books dealing with entropy and information theory. 
The textbook \cite{CovTh} is an excellent introduction to the subject, \cite{Bill, Gra, Shi} are recommended to mathematically 
more advanced reader. Another instructive reference is \cite{CsiK\"o}, where a substantial part of the material covered in this
lecture notes is left as an exercise for the reader!

Discussions of a link between  information and  statistical mechanics preceded Shannon's work. Although Weaver's remark\footnote{" Dr. Shannon's work roots back, as von Neumann has pointed out,
to Boltzmann's observation, in some of his work on statistical physics
(1894), that entropy is related to "missing information," inasmuch as it is
related to the number of alternatives which remain possible to a physical
system after all the macroscopically observable information concerning it
has been recorded."} on page 3 of   \cite{ShaWe}  appears to be historically inaccurate, the discussions of the role of  information in foundations 
of  statistical mechanics  goes back at least to the work of L. Szillard \cite{Szi}  in 1929, see also \url{https://plato.stanford.edu/entries/information-entropy/},
and remains to this day a hotly disputed subject; 
see \cite{GHLS} for a recent discussion. An early discussion can be found in \cite{Jay1, Jay2}. The textbook 
\cite{Mer} gives  an additional   perspective on this topic. 

In contrast to equilibrium statistical mechanics whose mathematically rigorous foundations, based on the 19th century works 
of Boltzmann and Gibbs,  were laid in 1960's and 70's, the physical and mathematical theory of 
non-equilibrium statistical mechanics remains  in its infancy.  The introduction of non-equilibrium steady states and the discovery 
of the fluctuation relations in context of chaotic dynamical systems in early 1990's (see \cite{JPR} for references)
revolutionized our understanding of some important corners of the field, and have generated an enormous amount 
of  theoretical, experimental, and numerical works with applications extending to chemistry and biology. The research program 
of Claude-Alain Pillet and myself mentioned in the introduction is rooted in these developments.\footnote{The  references to 
results of this program are not relevant for this Part I  of the lectures and they will be listed in the latter installements.} In  this program, 
the search for a notion of entropy for systems out of equilibrium plays a central role. The planned four parts  lecture notes are meant as an 
introduction to this search, with this Part I focusing on the   information-theoretic notions of entropy.

\chapter{Elements of probability}
\label{sec-probability}
\section{Prologue: integration on finite sets}
Let $\Omega$ be a finite set. Generic element of $\Omega$ is denoted by $\omega$. When needed, we will enumerate elements 
of $\Omega$ as $\Omega=\{\omega_1, \cdots, \omega_L\}$, where $|\Omega|=L$.

A measure on $\Omega$ is a map 
\[\mu: \Omega \rightarrow \rr_+=[0, \infty[.\]
The pair $(\Omega, \mu)$ is called measurable  space. The measure of $S\subset \Omega$ is 
\[
\mu(S)=\sum_{\omega\in S}\mu(\omega).
\]
By definition, $\mu(\emptyset)=0$.

Let $f:\Omega \rightarrow \cc$ be a function. The integral of $f$ over $S\subset \Omega$ is defined by 
\[
\int_S f\d\mu=\sum_{\omega \in S}f(\omega)\mu(\omega).
\]
Let $\Omega$ and ${\cal E}$ be two finite sets and $T:\Omega \rightarrow {\cal E}$ a map. Let $\mu$ be a measure 
on $\Omega$. For $\zeta \in {\cal E}$ set 
\[\mu_T(\zeta)=\mu(T^{-1}(\zeta))=\sum_{\omega: T(\omega)=\zeta}\mu(\omega).
\]
$\mu_T$ is a measure on ${\cal E}$ induced by $(\mu,T)$. If $f:{\cal E}\rightarrow \cc$, then 
\[\int_{\cal E}f \d\mu_T=\int_{\Omega}f \circ T\d\mu.
\]
If $f:\Omega\rightarrow \cc$, we denote by  $\mu_f$ the measure on the set of values ${\cal E}=\{f(\omega)\,|\,  \omega\in \Omega\}$ induced 
by $(\Omega, f)$. $\mu_f$ is called the distribution measure of the function $f$.

We denote by 
\[
\Omega^N=\{\omega=(\omega_1, \cdots, \omega_N)\,|\, \omega_k \in \Omega\},
\]
\[\mu_N(\omega=(\omega_1, \cdots, \omega_N))=\mu(\omega_1)\cdots\mu(\omega_N),
\]
the $N$-fold product set and measure of the pair $(\Omega, \mu)$.

Let $\Omega_{l/r}$ be two finite sets and $\mu$ a measure on $\Omega_l \times \Omega_r$. The marginals of $\mu$ are measures 
$\mu_{l/r}$ 
on $\Omega_{l/r}$ defined by  
\[
\mu_l(\omega)=\sum_{\omega^\prime\in \Omega_r}\mu(\omega, \omega^\prime),\qquad \omega\in \Omega_l,
\]
\[
\mu_r(\omega)=\sum_{\omega^\prime\in \Omega_l}\mu(\omega^\prime, \omega),\qquad \omega\in \Omega_r.
\]
If $\mu_{l/r}$ are measures on $\Omega_{l/r}$. we denote by $\mu_{l}\otimes \mu_r$ the product measure defined 
by 
\[\mu_l\otimes\mu_r(\omega, \omega^\prime)=\mu_l(\omega)\mu_r(\omega^\prime).\]

The support of the measure $\mu$ is the set
\[\supp\,\mu=\{\omega\,|\, \mu(\omega)\not=0\}.\]

Two measures $\mu_1$ and $\mu_2$ are mutually singular, denoted $\mu_1\perp \mu_2$, iff $\supp\, \mu_1\cap \supp \,\mu_2=\emptyset$.
A measure $\mu_1$ is absolutely continuous w.r.t. another measure $\mu_2$, denoted $\mu_1\ll\mu_2$, iff 
$\supp \mu_1\subset  \supp \mu_2$, that is, iff $\mu_2(\omega) =0\Rightarrow \mu_1(\omega)=0$. 
If  $\mu_1\ll \mu_2$, the Radon-Nikodym derivative of $\mu_1$ w.r.t. $\mu_2$ is defined by 
\[
\Delta_{\mu_1|\mu_2}(\omega) = \begin{cases}  \frac{\mu_1(\omega)}{\mu_2(\omega)} &\mbox{if } \omega \in \supp \,\mu_1 \\ 
0 & \mbox{if } \omega \not\in \supp\, \mu_1. \end{cases} 
\]
Note that 
\[
\int_\Omega f \Delta_{\mu_1|\mu_2}\d\mu_2=\int_\Omega f\d\mu_1.
\]
Two measures $\mu_1$ and $\mu_2$ are called equivalent iff $\supp\, \mu_1= \supp \,\mu_2$.

Let $\mu, \rho$ be two measures on $\Omega$. Then there exists a unique decomposition (called the  Lebesgue decomposition)
$\mu=\mu_1+\mu_2$, where $\mu_1\ll\rho$ and $\mu_2\perp \rho$. Obviously, 
\[
\mu_1(\omega) = \begin{cases} \mu(\omega) &\mbox{if } \omega \in \supp \,\rho \\ 
0 & \mbox{if } \omega \not\in \supp \,\rho,\end{cases} \qquad 
\mu_2(\omega) = \begin{cases} 0 &\mbox{if } \omega \in \supp\, \rho \\ 
\mu(\omega) & \mbox{if } \omega \not\in \supp\, \rho.\end{cases}
\]
A measure $\mu$ is called faithful if $\mu(\omega)>0$ for all $\omega \in \Omega$.

\bep \label{Cheb}Let $f: \Omega\rightarrow \rr_+$, $a>0$, and $S_a=\{\omega\,|\, f(\omega)\geq a\}$. Then
\[\mu(S_a)\leq\frac{1}{a}\int_\Omega f\d\mu.
\]
\eep
\demo The statement is obvious is $S_a=\emptyset$. If $S_a$  is non-empty,
\[
\mu(S_a)=\sum_{\omega \in S_a}\mu(\omega)\leq \frac{1}{a}\sum_{\omega \in S_a} f(\omega)\mu(\omega)\leq 
\frac{1}{a}\int_\Omega f\d\mu.
\]
\qed

We recall the Minkowski inequality 
\[\left(\int_{\Omega}|f+ g|^p\d \mu\right)^{1/p} \leq \left(\int_{\Omega}|f|^p \d\mu \right)^{1/p} + \left(\int_{\Omega}|g|^p \d\mu \right)^{1/p},
\]
where $p\geq 1$, and the  H\"older inequality 
\[
\int_{\Omega} fg \d\mu \leq \left(\int_{\Omega}|f|^p \d\mu \right)^{1/p}\left(\int_{\Omega}|g|^q \d\mu \right)^{1/q},
\]
where $p, q\geq 1$, $p^{-1}+ q^{-1}=1$. For $p=q=2$ the H\"older inequality reduces to the Cauchy-Schwarz inequality.

If $f: \Omega \rightarrow ]-\infty, \infty]$ or $[-\infty, \infty[$, we again set $\int_\Omega f\d\mu =\sum_\omega f(\omega)\mu(\omega)$ with 
the convention that $0\cdot (\pm \infty)=0$.

\section{Probability on finite sets}
We start with a change of vocabulary adapted to the probabilistic interpretation of measure theory.

A measure $\mP$ on a finite set $\Omega$ is called  a probability measure if $\mP(\Omega)=\sum_{\omega \in \Omega}\mP(\omega)=1$. 
The pair $(\Omega, \mP)$ is called probability space. 
A set $S\subset \Omega$ is called an event and $\mP(S)$ is the probability of the event $S$. Points $\omega \in \Omega$ 
are sometimes called elementary events.

A perhaps most basic example of a probabilistic  setting is a   fair coin experiment, where a coin is tossed $N$ times and the outcomes are recorded as ${\rm Head}=1$ and 
${\rm Tail}=-1$. The set  of outcomes is 
\[\Omega=\{\omega=(\omega_1, \cdots, \omega_N)\,|\, \omega_k=\pm 1\},\]
 and 
\[\mP(\omega=(\omega_1, \cdots, \omega_N))=\frac{1}{2^N}.\]
Let $S$ be the event that $k$ Heads and $N-k$ Tails are observed. The binomial formula gives 
\[ \mP(S)=\binom{N}{k}\frac{1}{2^N}.\]
As another example, let 
\[ S_j=\left\{\omega=(\omega_1,\cdots,\omega_N)\,\big|\, \sum_{k}\omega_k=j\right\},
\]
where $-N\leq j\leq N$. $P(S_j)=0$ if $N+j$ is odd. If $N+j$ is even, then 
\[\mP(S_j)=\binom{N}{\frac{N+j}{2}}\frac{1}{2^N}.
\]

A function $X:\Omega \rightarrow \rr$ is called random variable. 

The measure $\mP_X$  induced  by $(\mP, X)$ is called the probability distribution 
of $X$. The expectation of $X$ is 
\[\mE(X)=\int_\Omega X\d \mP.\]
The moments of $X$ are 
\[ M_k=\mE(X^k), \qquad k=1,2\cdots,
\]
and the moment generating function is 
\[ M(\alpha)=\mE(\e^{\alpha X})=\sum_{\omega \in \Omega}\e^{\alpha X(\omega)}\mP(\omega),
\]
where $\alpha \in \rr$. Obviously, 
\[
M_k=\frac{\d^k}{\d \alpha^k}M(\alpha)\big|_{\alpha=0}.
\]
The cumulant generating function of $X$  is 
\[ C(\alpha)=\log \mE(\e^{\alpha X})=\log \left(\sum_{\omega \in \Omega}\e^{\alpha X(\omega)}\mP(\omega)\right).\]
The cumulants of $X$ are 
\[ C_k=\frac{\d^k}{\d \alpha^k}C(\alpha)\big|_{\alpha=0}, \qquad k=1,2, \cdots.
\]
$C_1=M_1=\mE(X)$ and 
\[ C_2= \mE(X^2)-\mE(X)^2=\mE((X-\mE(X))^2).\]
$C_2$ is called the variance of $X$ and is denoted by ${\rm Var}(X)$. Note that ${\rm Var}(X)=0$ iff $X$ is  constant on 
$\supp\, \mP$. When we wish to indicate the dependence of the expectation and variance on the underlying measure $\mP$, we shall 
write $\mE_\mP(X)$, ${\rm Var}_\mP(X)$, etc.

\begin{exo}The sequences $\{M_k\}$ and $\{C_k\}$ determine each other, {\sl i.e.}, there are functions 
$F_k$ and $G_k$ such that 
\[C_k=F_k(M_1, \cdots, M_k), \qquad M_k=G_k(C_1, \cdots, C_k).
\]
Describe recursive relations that determine $F_k$ and $G_k$. 
\end{exo}
In probabilistic setup Proposition \ref{Cheb} takes the form 
\beq 
\mP(\{\omega \in \Omega\,|\, |X(\omega)|\geq a\})\leq \frac{1}{a}\mE(|X|), 
\label{markov}
\eeq
and is often called Markov or Chebyshev inequality.
We shall often use a shorthand and abbreviate the l.h.s in \eqref{markov} as $\mP\{|X(\omega)|\geq a\}$, etc.

 \section{Law of large numbers}
 Let $(\Omega, \mP)$ be a probability space and $X: \Omega \rightarrow \rr$ a random variable. 
 On the product probability space $(\Omega^N, \mP_N)$ we define 
 \[ {\cal S}_N(\omega=(\omega_1, \cdots, \omega_N))=\sum_{k=1}^N X(\omega_k).
 \]
 We shall refer to the following results as  the {\em Law of large numbers (LLN)}.
 \bep\label{LLN} For any $\epsilon >0$,
 \[
 \lim_{N\rightarrow \infty} \mP_N\left\{\left| \frac{{\cal S}_N(\omega)}{N}- \mE(X)\right|\geq \epsilon\right\}=0.
 \]
 \eep
 \begin{remark}
 An equivalent formulation of the LLN is that for any $\epsilon >0$, 
 \[
 \lim_{N\rightarrow \infty} \mP_N\left\{\left| \frac{{\cal S}_N(\omega)}{N}- \mE(X)\right|\leq \epsilon\right\}=1.
 \]
 \end{remark}
 \demo Denote by $\mE_N$ the expectation w.r.t. $P_N$. Define $X_k(\omega)=X(\omega_k)$ and note that 
 $\mE_N(X_k)=\mE(X)$, $E_N(X_k^2)=E(X^2)$, $\mE_N(X_k X_j)=\mE(X)^2$ for $k\not=j$. Then 
 \[
 \begin{split} 
 \mP_N\left\{\left| \frac{{\cal S}_N(\omega)}{N}- \mE(X)\right|\geq \epsilon\right\}&=
 \mP_N\left\{\left( \frac{{\cal S}_N(\omega)}{N}- \mE(X)\right)^2\geq \epsilon^2\right\}\\
 &\leq \frac{1}{\epsilon^2}
 \mE_N\left(\left(\frac{{\cal S}_N(\omega)}{N}- \mE(X)\right)^2\right)\\
 &=\frac{1}{N^2\epsilon^2}\mE_N\left(\sum_{k,j}(X_k-\mE (X_k))(X_j-\mE (X_j))\right)\\
 &=\frac{1}{N\epsilon^2}{\rm Var}(X),
 \end{split}
 \]
 and the statement follows. \qed
\section{Cumulant generating function}
\label{sec-cum}
Let $(\Omega, P)$ be a probability space and $X:\Omega\rightarrow \rr$ a random variable. In this section we shall study 
in some detail the properties of the cumulant generating function 
\[ C(\alpha)=\log \mE(\e^{\alpha X}).
\]
To avoid discussion of trivialities, until the end of this chapter  we shall assume that $X$ is not constant on $\supp\, \mP$, 
{\sl i.e.} that $X$ assumes at least two distinct values on $\supp \,\mP$. Obviously, the function $C(\alpha)$ is infinitely differentiable and 
\beq \begin{split}
\lim_{\alpha \rightarrow \infty}C^\prime(\alpha)&=\max_\omega X(\omega),\\
\lim_{\alpha \rightarrow -\infty}C^\prime(\alpha)&=\min_\omega X(\omega).
\end{split}
\label{cum-M-m}\eeq

\bep\label{C-convex} $C^{\prime\prime}(\alpha)>0$ for all $\alpha$. In particular, the function $C$ is strictly convex.
\eep
\begin{remark} By strictly convex we mean that $C^\prime$ is strictly increasing, {\sl i.e.}, that the graph of $C$ does 
not have a flat piece. 
\end{remark}
\demo Set  
\beq \label{def-Q-alpha}\mQ_\alpha(\omega)=\frac{\e^{\alpha X(\omega)}\mP(\omega)}{\sum_{\omega} \e^{\alpha X(\omega)}\mP(\omega)},
\eeq
and note that $\mQ_\alpha$ is a probability measure on $\Omega$ equivalent to $\mP$.
 
One easily verifies that
\[ C^{\prime}(\alpha)=\mE_{\mQ_\alpha}(X), \qquad C^{\prime\prime}(\alpha)={\rm Var}_{\mQ_\alpha}(X).
\]
The second identity yields the statement. \qed 

\bep\label{C-analytic} $C$ extends to an analytic function in the strip
\beq\label{strip}|\Im \alpha|<\frac{\pi}{2}\frac{1}{\max_\omega |X(\omega)|}.
\eeq
\eep
\demo  Obviously, the function $\alpha \mapsto \mE(\e^{\alpha X})$ is entire analytic. If $\alpha =a+\i b$, then 
\[\mE(\e^{\alpha X})=\sum_{\omega \in \Omega}\e^{a X(\omega)}\cos (bX(\omega))\mP(\omega) + \i \sum_{\omega \in \Omega}
\e^{a X(\omega)}\sin (bX(\omega))\mP(\omega). 
\]
If $|bX(\omega)|<\pi/2$ for all $\omega$, then the real part of $\mE(\e^{\alpha X})$ is strictly positive. It follows that the function 
\[{\rm Log}\,\mE(\e^{\alpha X}),\]
where ${\rm Log}$ is the principal branch of complex logarithm, is analytic in the strip  
(\ref{strip}) and the statement follows. \qed

\begin{remark}Let $\Omega=\{-1, 1\}$, $P(-1)=P(1)=1/2$, $X(1)=1$, $X(-1)=-1$. Then 
\[ C(\alpha)=\log \cosh \alpha.\]
Since $\cosh(\pi \i/2)=0$, we see that Proposition \ref{C-analytic} is an optimal result.
\label{ln-march}
\end{remark}

\section{Rate function}
\label{sec-rate}
We continue with the framework of the previous section. The {\em rate function} of the random variable $X$ is 
defined  by 
\[I(\theta)=\sup_{\alpha \in \rr}\,(\alpha \theta - C(\alpha)), \qquad \theta \in \rr.\]
In the language of convex analysis, $I$ is the Fenchel-Legendre transform of the cumulant generating function $C$. Obviously, 
$I(\theta)\geq 0$ for all $\theta$.
Set 
\[ m=\min_\omega X(\omega), \qquad M=\max_\omega X(\omega),\]
and recall the relations (\ref{cum-M-m}). By the intermediate value theorem, for any $\theta$ in $]m, M[$ there exists unique 
$\alpha(\theta)\in \rr$ such that 
\[ \theta=C^\prime(\alpha(\theta)).
\]
The function 
\[\alpha(\theta)= (C^\prime)^{-1}(\theta)\]
is infinitely differentiable on $]m, M[$,  strictly increasing on $]m, M[$, 
$\alpha(\theta)\downarrow -\infty$ iff $\theta \downarrow m$, and $\alpha(\theta)\uparrow\infty$ iff $\theta \uparrow M$. 

\begin{exo} Prove that the function $]m, M[\,\ni \theta \mapsto \alpha(\theta)$ is real-analytic.\newline
Hint: Apply the analytic implicit function  theorem.
\end{exo}

\bep\label{prop-rate} \begin{enumerate}[{\rm (1)}]
\item For $\theta \in ]m,M[$, 
\[ I(\theta)=\alpha(\theta)\theta - C(\alpha(\theta)).
\]
\item The function $I$ is infinitely differentiable on $]m, M[$.

\item $I^\prime(\theta)=\alpha(\theta)$. In particular, $I^\prime$ is strictly increasing on $]m, M[$ and 
\[\lim_{\theta \downarrow m}I^\prime(\theta)=-\infty,\qquad \lim_{\theta \uparrow M}I^\prime(\theta)=\infty.
\]
\item $I^{\prime\prime}(\theta)=1/C^{\prime\prime}(\alpha(\theta))$. 
\item $I(\theta)=0$ iff $\theta=\mE(X)$.

\end{enumerate}
\eep
\demo To prove (1), note that  for $\theta \in ]m, M[$ the function 
\[\frac{\d}{\d \alpha}(\alpha \theta -C(\alpha))=\theta - C^\prime(\alpha)\]
vanishes at $\alpha(\theta)$, is positive for $\alpha<\alpha(\theta)$, and is negative for $\alpha >\alpha(\theta)$. 
Hence, the function $\alpha \mapsto \alpha \theta - C(\alpha)$ has the global maximum at $\alpha =\alpha(\theta)$ and Part (1) follows.
Parts (2), (3) and (4) are obvious. To prove (5), note that if $I(\theta)=0$ for some $\theta \in ]m, M[$, then, since $I$ is non-negative, 
we  also have
$0=I^\prime(\theta)=\alpha(\theta)$, and the relation  $\theta = C^\prime(\alpha(\theta))=C^\prime(0)=\mE(X)$ follows. On the 
other hand, if $\theta = \mE(X)=C^\prime(0)$, then $\alpha(\theta)=0$, and $I(\theta)=-C(0)=0$. \qed

\begin{exo} Prove that the function $I$ is real-analytic in $]m, M[$. 
\end{exo}

Let 
\[ S_m=\{\omega\in \Omega\,|\, X(\omega)=m\}, \qquad S_M= \{\omega \in \Omega\,|\, X(\omega)=M\}.\]
\bep \begin{enumerate}[{\rm (1)}]
\item $I(\theta)=\infty$ for $\theta \not\in[m, M]$.
\item 
\[\begin{split}
I(m)&=\lim_{\theta \downarrow m}I(\theta)=-\log \mP(S_m),\\
I(M)&=\lim_{\theta \uparrow M}I(\theta)=-\log \mP(S_M).
\end{split}
\]

\end{enumerate}
\eep
\demo (1) Suppose that $\theta >M$. Then 
\[
\frac{\d}{\d \alpha}(\alpha \theta -C(\alpha))=\theta -C^\prime(\alpha)>\theta -M.
\]
Integrating this inequality over $[0, \alpha]$ we derive
\[\alpha\theta - C(\alpha)>(\theta-M)\alpha,
\]
and so 
\[ I(\theta)=\sup_{\alpha \in \rr}(\alpha \theta -C(\alpha))=\infty.
\]
The case $\theta <m$ is similar.

(2) We shall prove only the second formula, the proof of the first is similar. Since the function $\alpha M- C(\alpha)$ is increasing, 
\[ I(M)=\lim_{\alpha \rightarrow \infty}(\alpha M- C(\alpha)).\]
Since
\beq  C(\alpha)=\alpha M +\log \mP(S_M)+ \log(1 + A(\alpha)),
\label{rate-swiss-1}
\eeq
where 
\[
A(\alpha)=\frac{1}{\mP(S_M)}\sum_{\omega \not\in S_M}\e^{\alpha(X(\omega)-M)}\mP(\omega),
\label{rate-swiss-2}
\]
we derive that $I(M)=-\log \mP(S_M)$. 

Since $C^\prime(\alpha(\theta))=\theta$,  
Part (1) of Proposition \ref{prop-rate} gives that
\[
\lim_{\theta \uparrow M}I(\theta)=\lim_{\alpha \rightarrow \infty} (\alpha C^\prime (\alpha)-C(\alpha)).
\]
Write
\beq  C^\prime(\alpha)=M\frac{1 + B(\alpha)}{1+ A(\alpha)},
\label{rate-swiss-4}
\eeq
where 
\[ B(\alpha)=\frac{1}{M\mP(S_M)}\sum_{\omega \not\in S_M}X(\omega)\e^{\alpha(X(\omega)-M)}\mP(\omega).\]
The formulas (\ref{rate-swiss-1}) and (\ref{rate-swiss-4}) yield 
\[
\alpha C^\prime(\alpha)-C(\alpha)= \alpha M\frac{B(\alpha)-A(\alpha)}{1+A(\alpha)}-\log \mP(S_M)-\log (1+ A(\alpha)).
\]
Since $A(\alpha)$ and $B(\alpha)$  converge to $0$ as $\alpha \rightarrow \infty$, 
\[\lim_{\theta \uparrow M}I(\theta)=\lim_{\alpha \rightarrow \infty} (\alpha C^\prime (\alpha)-C(\alpha))=-\log \mP(S_M).\] 
\qed

\bep 
\beq \label{inv-rate}
 C(\alpha)=\sup_{\theta \in \rr}\,(\theta \alpha -I(\theta)).
 \eeq
\eep
\demo To avoid confusion, fix $\alpha=\alpha_0$. Below, $\alpha(\theta)= (C^\prime)^{-1}(\theta)$ is as in 
Proposition \ref{prop-rate}.

The supremum in (\ref{inv-rate}) is  achieved at $\theta_0$ satisfying 
\[ \alpha_0 =I^\prime(\theta_0).\]
Since $I^{\prime}(\theta_0)=\alpha(\theta_0)$, we have  $\alpha_0=\alpha(\theta_0)$, 
and 
\[I(\theta_0)=\theta_0\alpha(\theta_0)-C(\alpha(\theta_0))=\theta_0\alpha_0- C(\alpha_0).
\]
Hence 
\[\sup_{\theta \in \rr}\,(\theta \alpha_0 -I(\theta))=\alpha_0\theta_0- I(\theta_0)=C(\alpha_0).
\]
\qed

Returning to the example of Remark \ref{ln-march},  $m=-1$, $M=1$, 
$C(\alpha)=\log \cosh \alpha$, and $C^{\prime}(\alpha)=\tanh \alpha$. Hence, for $\theta \in ]-1, 1[$,
\[\alpha(\theta)=\tanh^{-1}(\theta)=\frac{1}{2}\log \frac{1+\theta}{1-\theta}.
\]
It follows that 
\[I(\theta)=\theta \alpha(\theta)-C(\alpha(\theta))= 
\frac{1}{2}(1+\theta)\log (1+\theta) +\frac{1}{2}(1-\theta)\log(1-\theta).
\]
\section{Cram\'er's theorem} 
\label{sub-sec-cramer}
This section is devoted to the proof of  Cram\'er's theorem:
\bet\label{Cramer} For any interval $[a, b]$,
\[\lim_{N\rightarrow \infty}\frac{1}{N}\log \mP_N\left\{ \frac{{\cal S}_N(\omega)}{N}\in [a, b]\right\}=
-\inf_{\theta \in [a, b]} I(\theta).
\]
\eet
\begin{remark} To prove this result without loss of generality we may assume that $[a, b]\subset [m, M]$. 
\end{remark}
\begin{remark} Note that 
\[  \inf_{\theta\in [a, b]} I(\theta)= \begin{cases}0&\mbox{if } {\mathbb E}(X) \in [a, b] \\ 
I(a) & \mbox{if } a >{\mathbb E}(X)\\
I(b) &\mbox{if }  b <{\mathbb E}(X),
\end{cases} \]
and that 
\[\lim_{N\rightarrow \infty}\frac{1}{N}\log \mP_N\left\{ \frac{{\cal S}_N(\omega)}{N}= M\right\}=\log P(S_M)=-I(M),\]
\[\lim_{N\rightarrow \infty}\frac{1}{N}\log \mP_N\left\{ \frac{{\cal S}_N(\omega)}{N}= m\right\}=\log P(S_m)=-I(m).\]
\end{remark}

We start the proof with 
\bep\label{bound-upper}
\begin{enumerate}[{\rm (1)}] 
\item 
For $\theta \geq {\mathbb E}(X)$,
\[\limsup_{N\rightarrow \infty}\frac{1}{N}\log \mP_N\left\{ \frac{{\cal S}_N(\omega)}{N}\geq \theta \right\}\leq -I(\theta).
\]
\item For $\theta \leq  {\mathbb E}(X)$, 
\[\limsup_{N\rightarrow \infty}\frac{1}{N}\log \mP_N\left\{ \frac{{\cal S}_N(\omega)}{N}\leq  \theta \right\}\leq -I(\theta).
\]
\end{enumerate} 
\eep
\begin{remark} Note that if $\theta <{\mathbb E}(X)$, then by the LLN
\[\lim_{N\rightarrow \infty}\frac{1}{N}\log P_N\left\{ \frac{{\cal S}_N(\omega)}{N}\geq \theta \right\}=0.
\]
Similarly, if $\theta >{\mathbb E}(X)$, 
\[\lim_{N\rightarrow \infty}\frac{1}{N}\log \mP_N\left\{ \frac{{\cal S}_N(\omega)}{N}\leq \theta \right\}=0.
\]
\end{remark}
\demo For $\alpha >0$,
\[
\begin{split}
\mP_N\left\{{\cal S}_N(\omega)\geq N\theta\right\}&= \mP_N\left\{ \e^{\alpha {\cal S}_{N}(\omega)}\geq \e^{\alpha N\theta}\right\}\\[2mm]
&\leq \e^{-\alpha N\theta}{\mathbb E}_N\left(\e^{\alpha{\cal S}_N(\omega)}\right)\\[2mm]
&= \e^{-\alpha N\theta}{\mathbb E}\left(\e^{\alpha X}\right)^N\\[2mm]
&=\e^{N(  C(\alpha)-\alpha \theta)}.
\end{split}
\]
It follows that 
\[
\limsup_{N\rightarrow \infty}\frac{1}{N}\log\mP_N\left\{ \frac{{\cal S}_N(\omega)}{N}\geq \theta \right\}\leq \inf_{\alpha >0}\,(C(\alpha)
-\alpha\theta)=-\sup_{\alpha>0}\,(\alpha \theta- C(\alpha)).
\]
If $\theta \geq {\mathbb E}(X)$, then $\alpha \theta -C(\alpha)\leq 0$ for $\alpha \leq 0$ and 
\[
\sup_{\alpha>0}\,(\alpha \theta- C(\alpha))=\sup_{\alpha\in \rr}\,(\alpha \theta- C(\alpha))=I(\theta).
\]
This yields Part (1). Part (2) follows by applying Part (1) to the random variable $-X$. \qed

\begin{exo} Using Proposition \ref{bound-upper} prove that for any $\epsilon >0$ there exist $\gamma_\epsilon >0$ and 
$N_\epsilon$ such that for $N\geq N_\epsilon$, 
\[
\mP_N\left\{\left| \frac{{\cal S}_N(\omega)}{N}- \mE(X)\right|\geq \epsilon\right\}\leq \e^{-\gamma_\epsilon N}.
\]
\label{ex-strong-LLN}
\end{exo}
\bep\label{bound-lower}
\begin{enumerate}[{\rm (1)}] 
\item 
For $\theta \geq {\mathbb E}(X)$,
\[\liminf_{N\rightarrow \infty}\frac{1}{N}\log \mP_N\left\{ \frac{{\cal S}_N(\omega)}{N}\geq \theta \right\}\geq -I(\theta).
\]
\item For $\theta \leq  {\mathbb E}(X)$, 
\[\liminf_{N\rightarrow \infty}\frac{1}{N}\log \mP_N\left\{ \frac{{\cal S}_N(\omega)}{N}\leq \theta \right\}\geq -I(\theta).
\]
\end{enumerate} 
\eep
\begin{remark} Note that Part (1) trivially holds if $\theta <{\mathbb E}(X)$. Similarly, Part (2) trivially 
holds if $\theta >{\mathbb E}(X)$.
\end{remark}

\demo We again need to prove only Part (1) (Part (2) follows by applying Part (1) to the random variable $-X$). If $\theta \geq M$, 
the statement is obvious and so without loss of generality we may assume that $\theta \in [{\mathbb E}(X), M[$. Fix 
such $\theta$ and choose $s$ and $\epsilon>0$ such that $\theta <s-\epsilon <s+\epsilon <M$. 

Let $\mQ_\alpha$ be the  probability measure introduced in the proof of Proposition \ref{C-convex}, and  let $\mQ_{\alpha, N}$ be the induced product probability  measure on $\Omega^N$. The measures $\mP_N$ and $\mQ_{\alpha, N}$ are equivalent, and for 
$\omega \in \supp \, \mP_N$
\[\Delta_{\mP_N|\mQ_{\alpha, N}}(\omega)=\e^{-\alpha {\cal S}_N(\omega) + N C(\alpha)}.
\]

We now consider the measure $\mQ_{\alpha, N}$ for $\alpha=\alpha(s)$.  Recall that 
\[C^\prime(\alpha(s))=s={\mathbb E}_{\mQ_{\alpha(s)}}(X).\]
Set 
\[ T_N=\left\{ \omega\in \Omega^N\,\big|\,\frac{{\cal S}_N(\omega)}{N}\in [s-\epsilon, s+\epsilon]\right\},\]
and note that the LLN implies 
\beq\label{mm-1}
\lim_{N\rightarrow \infty} \mQ_{\alpha(s), N}(T_N)=1.
\eeq
The estimates 
\[
\begin{split}
\mP_{N}\left\{\frac{{\cal S}_N(\omega)}{N}\geq \theta\right\}\geq \mP_{ N}(T_N)&=\int_{T_N}\Delta_{\mP_N|\mQ_{\alpha(s), N}}\d \mQ_{\alpha(s), N}\\[2mm]
&=\int_{T_N}\e^{-\alpha(s) {\cal S}_N+ N C(\alpha(s))}\d \mQ_{\alpha(s), N}\\[2mm]
&\geq \e^{N( C(\alpha(s))- s\alpha(s)- \epsilon|\alpha(s)|}\mQ_{\alpha(s), N}(T_N)
\end{split}
\]
and  (\ref{mm-1}) give
\[\liminf_{N\rightarrow \infty}\frac{1}{N}\log \mP_N\left\{ \frac{{\cal S}_N(\omega)}{N}\geq  \theta \right\}\geq 
C(\alpha(s))- s\alpha(s)-\epsilon|\alpha(s)|=-I(s)-\epsilon|\alpha(s)|.
\]
The statement now follows by taking first $\epsilon \downarrow 0$ and then $s \downarrow \theta$. \qed

Combining Propositions \ref{bound-upper} and \ref{bound-lower} we derive 
\begin{corollary}\label{no-way-paris} For $\theta \geq {\mathbb E}(X)$,
\[\lim_{N\rightarrow \infty}\frac{1}{N}\log \mP_N\left\{ \frac{{\cal S}_N(\omega)}{N}\geq \theta \right\}= -I(\theta).
\]
For $\theta \leq  {\mathbb E}(X)$, 
\[\lim_{N\rightarrow \infty}\frac{1}{N}\log \mP_N\left\{ \frac{{\cal S}_N(\omega)}{N}\leq  \theta \right\}= -I(\theta).
\]
\label{mm-2}
\end{corollary}
We are now ready to complete 

{\bf Proof of Theorem \ref{Cramer}.} If  ${\mathbb E}(X)\in ]a, b[$ the result follows from the LLN. Suppose that 
$M> a\geq {\mathbb E}(X)$. Then 
\[
\mP_N\left\{ \frac{{\cal S}_N(\omega)}{N}\in [a, b] \right\}=\mP_N\left\{ \frac{{\cal S}_N(\omega)}{N}\geq a\right\}- 
\mP_N\left\{ \frac{{\cal S}_N(\omega)}{N} >b\right\}.
\]
It follows from Corollary \ref{mm-2} that 
\beq\label{mm-3}
\lim_{N\rightarrow \infty}\frac{1}{N}\log \left[1 - \frac{\mP_N\left\{ \frac{{\cal S}_N(\omega)}{N} >b\right\}}{\mP_N\left\{ \frac{{\cal S}_N(\omega)}{N}\geq a\right\}}
\right]=0,
\eeq
and so 
\[
\lim_{N\rightarrow \infty}\frac{1}{N}\log \mP_N\left\{ \frac{{\cal S}_N(\omega)}{N}\in [a, b] \right\}=
\lim_{N\rightarrow \infty}\frac{1}{N}\log \mP_N\left\{ \frac{{\cal S}_N(\omega)}{N}\geq a \right\}=-I(a).
\]
The case $m < b\leq   {\mathbb E}(X)$ is similar. \qed

\begin{exo} Write down the proof of (\ref{mm-3}) and of the case $m<b\leq {\mathbb E}(X)$. 
\end{exo}

\begin{exo} Consider the example introduced in Remark \ref{ln-march} and prove  Cram\'er's theorem in 
this special case by using  Stirling's formula and a direct combinatorial argument. \newline
Hint: See  Theorem 1.3.1 in \cite{Ell}.
\label{ellis}
\end{exo}

\section{Notes and references}
Although it is assumed that the student reader had no previous exposure to  probability theory,  
a reading of additional material could be helpful at this point. Recommended 
textbooks are  \cite{Chu, RohSa, Ross}.

For  additional information and original references regarding  Cramer's  theorem we refer 
the reader to Chapter 2 of  \cite{DeZe}. 
Reader interested to learn more about theory of large deviations may consult  classical references 
\cite{dHoll, DeZe, Ell}, and the  lecture notes of S.R.S. Varadhan \url{https://math.nyu.edu/~varadhan/LDP.html}.

It is possible to give a  combinatorial proof of Theorem \ref{Cramer}, as indicated in the Exercise \ref{ellis}. 
The advantage of the argument  presented in this chapter   is that it naturally extends to a proof of  much more general results (such as 
the G\"artner-Ellis theorem) which will be discussed in the Part II of the lecture notes.

\chapter{Boltzmann--Gibbs--Shannon entropy}
\label{sec-ENT}
\section{Preliminaries}
Let $\Omega$ be a finite set, $|\Omega|=L$, and let ${\cal P}(\Omega)$ be the collection of all probability measures on $\Omega$. 
${\cal P}(\Omega)$ is naturally identified with   the set 
\beq
{\cal P}_L=\left\{(p_1, \cdots, p_L)\,|\, p_k\geq 0,\,\sum_{k=1}^L p_k=1\right\}
\label{iden-L}\eeq
(the identification map 
is $\mP\mapsto (\mP(\omega_1), \cdots \mP(\omega_L))$.  We shall often use this identification without further notice. A convenient metric 
on ${\cal P}(\Omega)$ is the variational distance 
\beq
d_V(\mP, \mQ)=\sum_{\omega\in \Omega}|\mP(\omega)-\mQ(\omega)|.
\label{var-dist}
\eeq
  We denote by ${\cal P}_{\rm f}(\Omega)$ the set of all faithful probability measures on ${\cal P}(\Omega)$ (recall 
  that $P\in {\cal P}_{\rm f}(\Omega)$ iff $P(\omega)>0$ for all $\omega \in \Omega$).
  ${\cal P}_{\rm f}(\Omega)$ coincides with the interior of ${\cal P}(\Omega)$ and is  identified with  
\[
{\cal P}_{L, {\rm f}}=\left\{(p_1, \cdots, p_L)\,|\, p_k> 0,\,\sum_{k=1}^L p_k=1\right\}.
\]
Note that ${\cal P}(\Omega)$ and ${\cal P}_{\rm f}(\Omega)$  are  convex sets.

The probability measure $\mP$ is called {\em pure} if $\mP(\omega)=1$ for some $\omega \in \Omega$. 
The {\em chaotic} probability measure is $\mP_{\rm ch}(\omega)=1/L$, $\omega \in \Omega$. 

We shall often make use of Jensen's inequality. This inequality states that if $f: [a, b]\rightarrow \rr$ is concave, then 
for $x_k\in [a, b]$, $k=1, \cdots, n$, and $(p_1, \cdots, p_n)\in {\cal P}_{n, {\rm f}}$  we have 
\beq
\sum_{k=1}^n p_k f(x_k)\leq f\left(\sum_{k=1}^n p_k x_k\right).
\label{jensen}
\eeq
Moreover, if $f$ is strictly concave the inequality is strict unless $x_1=\cdots =x_n$. A similar statement 
holds for convex functions. 

\begin{exo} Prove Jensen's inequality. 
\end{exo}
\section{Definition and basic properties}
\label{sec-def-ent}
The {\em entropy function} (sometimes called the {\em information function}) of $\mP\in {\cal P}(\Omega)$ is\footnote{
Regarding the choice of logarithm, in the introduction of \cite{Sha}  Shannon comments: "(1) It is practically more useful. Parameters of engineering importance such as time, bandwidth, number of relays, etc., tend to vary linearly with the logarithm of the number of possibilities. For example, adding one relay to a group doubles the number of possible states of the relays. It adds 1 to the base 2 logarithm of this number. Doubling the time roughly squares the number of possible messages, or doubles the logarithm, etc.
(2)  It is nearer to our intuitive feeling as to the proper measure. This is closely related to (1) since we intuitively measure entities by linear comparison with common standards. One feels, for example, that two punched cards should have twice the capacity of one for information storage, and two identical channels twice the capacity of one for transmitting information.
(3)  It is mathematically more suitable. Many of the limiting operations are simple in terms of the logarithm but would require clumsy restatement in terms of the number of possibilities."}  
\beq  S_\mP(\omega)=-c\log \mP(\omega),
\label{formula-log}
\eeq
where $c>0$ is a constant that does not depend on $P$ or $\Omega$, and $-\log 0=\infty$. The function $S_\mP$ takes values in $[0, \infty]$. The {\em Boltzmann--Gibbs--Shannon entropy} (in the sequel we will often call 
it just {\em entropy}) of $\mP$ is 
\beq
S({\mP})=\int_\Omega S_\mP\d \mP=-c\sum_{\omega\in \Omega}\mP(\omega)\log \mP(\omega). 
\label{formula-ent}
\eeq
The value of the constant $c$ is linked to the choice of units (or equivalently, the base of logarithm). The natural 
choice in the information theory is $c=1/\log 2$ (that is, the logarithm is taken in the base $2$). The value of $c$ plays 
no role in these lecture notes, and from now on we set $c=1$ and call 
\[  S(P)=-\sum_{\omega\in \Omega}\mP(\omega)\log \mP(\omega)
\]
the Boltzmann--Gibbs--Shannon entropy of $P$. We note, however,  that the constant $c$ will reappear  in the axiomatic 
characterizations of entropy given in Theorems \ref{khin} and \ref{ax-entropy}.

The basic properties of entropy are:
\bep
\label{e-basic}
 \begin{enumerate}[{\rm (1)}]
\item $S({P})\geq 0$ and $S({P})=0$ iff $P$ is pure. 
\item $S({P})\leq \log L$ and $S({P})=\log L$ iff $P=P_{\rm ch}$.
\item The map ${\cal P}(\Omega)\ni P\mapsto S({P})$ is continuous and concave, that is, if $p_k$'s are as in \eqref{jensen} and 
$P_k \in {\cal P}(\Omega)$, then 
\beq p_1 S(P_1) + \cdots +p_nS(P_n)\leq S(p_1P_1 + \cdots p_nP_n),
\label{conc-en}
\eeq
with equality iff $P_1=\cdots =P_n$. 
\item The concavity inequality \eqref{conc-en}  has the following "almost convexity" counterpart:
\[S(p_1P_1 + \cdots +p_nP_n) \leq p_1 S(P_1) + \cdots +p_nS(P_n) + S(p_1, \cdots, p_n),
\]
with equality iff $\supp\, P_k \cap \supp\, P_j=\emptyset$ for $k\not=j$.
\end{enumerate}
\eep
\demo Parts (1) and (3) follow from the obvious fact  that the function $[0,1]\ni x \mapsto -x\log x$ is continuous, strictly concave, 
non-negative, and vanishing iff $x=0$ or $x=1$. Part (2) follows from Jensen's inequality. Part (4) follows from the monotonicity 
of $\log x$: 
\[
\begin{split}S(p_1P_1 + \cdots +p_nP_n) &=\sum_{\omega \in \Omega}\sum_{k=1}^n -p_kP_k(\omega)\log \left(\sum_{j=1}^n p_jP_j(\omega)\right)\\[2mm]
&\leq \sum_{\omega \in \Omega}\sum_{k=1}^n- p_kP_k(\omega)\log \left(p_kP_k(\omega)\right)\\[2mm]
&=\sum_{k=1}^n p_k \left(\sum_{\omega \in \Omega}-P_k(\omega)\log P_k(\omega)\right) - \sum_{k=1}^n 
\left(\sum_{\omega \in \Omega}P_k(\omega)\right)p_k\log p_k\\[2mm]
&=\sum_{k=1}^n p_k S(P_k) + S(p_1, \cdots, p_n).
\end{split}
\]
The equality holds if for all $\omega$ and $k\not=j$, $p_kP_k(\omega)>0\,\, \Rightarrow\,\, p_jP_j(\omega)=0$, which is 
equivalent to $\supp\, P_k \cap \supp\, P_j=\emptyset$ for all $k\not=j$.
 \qed

Suppose that $\Omega=\Omega_l \times \Omega_r$ and let $P_{l/r}$ be the marginals of $P\in {\cal P}(\Omega)$. 
For a given $\omega\in \supp P_l$ the conditional probability measure $P_{r|l}^{\omega}$ on $\Omega_r$ is defined by 
\[
P_{r|l}^{\omega}(\omega^\prime)=
\frac{P(\omega, \omega^\prime)}{P_l(\omega)}.
\]
Note that 
\[\sum_{\omega\in \supp P_l}P_{l}(\omega)P_{r|l}^{\omega}=P_r.\]
\bep \label{e-subad}\begin{enumerate}[{\rm (1)}]
\item
\[ S({P})= S(P_l) +\sum_{\omega \in \Omega_l}P_{l}(\omega)S(P_{r|l}^{\omega}).
\]
\item The entropy is strictly sub-additive:
\[S({P})\leq S(P_l) + S(P_r), 
\]
with the equality iff $P=P_l\otimes P_r$. 
\end{enumerate}
\label{paris-cold}
\eep
\demo Part (1)  and the identity $S(P_l\otimes P_r)=S(P_l) + S(P_r)$ follow by direct computation. To prove (2), note that Part (3) of Proposition 
\ref{e-basic} gives 
\[
\sum_{\omega \in \supp P_l}P_{l}(\omega)S(P_{r|l}^{\omega})\leq S\left(\sum_{\omega\in \supp P_l}P_{l}(\omega)P_{r|l}^{\omega}\right)= S(P_r),
\]
and so it follows from Part (1) that $S(P)\leq S(P_l) + S(P_r)$ 
with the equality iff all the probability measures $P_{r|l}^{\omega}$, 
$\omega \in \supp P_l$, are equal.  Thus, if the equality holds, then for all $(\omega, \omega^\prime)\in \Omega_l\times 
\Omega_r$, $P(\omega, \omega^\prime)=C(\omega^\prime)P_l(\omega)$. Summing over $\omega$'s gives that 
$P=P_l\otimes P_r$. \qed

\begin{exo}The Hartley entropy of $P \in {\cal P}(\Omega)$ is defined by 
\[S_H({P})=\log |\{\omega\,|\, P(\omega)>0\}|.\]
\exop Prove that the Hartley entropy is also strictly sub-additive: $S_H({P})\leq S_H(P_l) + S_H(P_r)$,  with the equality iff $P=P_l\otimes P_r$.
\exop  Show that  the map $P\mapsto S_H(P)$ is not continuous if $L\geq 2$. 
\label{hartley}\end{exo}


\section{Covering exponents and source coding}
\label{sec-cover}
To gain further insight into the concept of entropy,  assume that $P$ is faithful and consider the product probability space $(\Omega^N, P_N)$. For given $\epsilon >0$ let 
\[ \begin{split}
T_{N, \epsilon}&=\left\{\omega=(\omega_1, \cdots, \omega_N)\in \Omega^N\,\big|\, 
\left|\frac{S_P(\omega_1)+ \cdots S_P(\omega_N)}{N}- S({P})\right|<\epsilon\right\}\\[2mm]
&=\left\{\omega\in \Omega^N\,\big|\, \left|-\frac{\log P_N(\omega)}{N}- S({P})\right|<\epsilon\right\}\\[2mm]
&=\left\{\omega \in \Omega^N\,\big|\, \e^{-N(S({P}) +\epsilon)} <P_N(\omega) <\e^{-N(S({P})-\epsilon)}\right\}.
\end{split}
\]
The LLN gives 
\[\lim_{N\rightarrow \infty}P_N(T_{N, \epsilon})=1.
\]
We also have the following obvious bounds on the cardinality of $T_{N, \epsilon}$:
\[P_N(T_{N, \epsilon})\e^{N(S({P})-\epsilon)}< |T_{N, \epsilon}|< \e^{N(S({P}) +\epsilon)}.\]
It follows that 
\[ S({P})- S({P}_{\rm ch})-\epsilon \leq \liminf_{N\rightarrow \infty}\frac{1}{N}
\log \frac{|T_{N, \epsilon}|}{|\Omega|^N}\leq \limsup_{N\rightarrow \infty}\frac{1}{N}
\log \frac{|T_{N, \epsilon}|}{|\Omega|^N}\leq S({P})- S({P}_{\rm ch})+\epsilon.
\]

This estimate  implies that if $P\not=P_{\rm ch}$, then, as $N\rightarrow \infty$, the measure $P_N$ is  "concentrated" and "equipartioned" on the set 
$T_{N, \epsilon}$ whose size is "exponentially small" with respect to the size of $\Omega^N$.

We continue with the analysis of the above concepts. Let $\gamma \in ]0,1[$ be fixed. The $(N,\gamma)$ covering exponent is 
defined by 
\beq\label{mm-4} c_N(\gamma)=\min\left\{ |A|\,|\, A\subset \Omega^N,\, P_N(A)\geq \gamma\right\}.
\eeq
One can find $c_N(\gamma)$ according to the following algorithm:

\begin{enumerate}[{\rm (a)}]
\item List the events $\omega=(\omega_1, \cdots, \omega_N)$ in order of decreasing probabilities.

\item Count the events until the first time the total probability is $\geq \gamma$.
\end{enumerate}

\bep\label{how-to} For all $\gamma \in ]0,1[$, 
\[ \lim_{N\rightarrow \infty}\frac{1}{N}\log c_N(\gamma)= S({P}).
\]
\eep
\demo Fix $\epsilon >0$ and recall the definition of $T_{N, \epsilon}$. For $N$ large enough, $P_N(T_{N, \epsilon})\geq \gamma$, and 
so for such $N$'s,
\[ c_N(\gamma)\leq |T_{N, \epsilon}|\leq \e^{N(S({P})+ \epsilon)}.
\]
It follows that  
\[
\limsup_{N\rightarrow \infty}\frac{1}{N}\log c_N(\gamma)\leq S({P}).
\]

To prove the lower bound, let $A_{N, \gamma}$ be a set for which the minimum in (\ref{mm-4}) is achieved. Let $\epsilon >0$. 
Note that 
\beq\label{mm-5}\liminf_{N\rightarrow \infty}P_N(T_{N, \epsilon}\cap A_{N, \gamma})\geq \gamma.
\eeq
Since for $P_N(\omega)\leq \e^{-N(S({P})-\epsilon)}$ for $\omega \in T_{N, \epsilon}$, 
\[
P_N(T_{N, \epsilon}\cap A_{N, \gamma})=\sum_{\omega \in T_{N, \epsilon}\cap A_{N,\gamma}}P_N(\omega)\leq 
\e^{-N(S({P})-\epsilon)}|T_{N, \epsilon}\cap A_{N, \gamma}|.
\]
Hence, 
\[
|A_{N, \gamma}|\geq \e^{N(S({P})-\epsilon)}P_N(T_{N, \epsilon}\cap A_{N, \gamma}),
\]
and it follows from (\ref{mm-5}) that 
\[
\liminf_{N\rightarrow \infty}\frac{1}{N}\log c_{N}( \gamma)\geq S({P})-\epsilon.
\]
Since $\epsilon >0$ is arbitrary, 
\[
\liminf _{N\rightarrow \infty}\frac{1}{N}\log c_N(\gamma)\geq  S({P}),
\]
and the proposition is proven. \qed

We finish this section with a discussion of Shannon's source coding theorem. 
Given a pair of positive integers $N, M$, the {\em encoder} is a map 
\[ F_{N}:\Omega^N \rightarrow \{0,1\}^M.
\]
The {\em decoder} is a map 
\[G_{N}: \{0,1\}^M\rightarrow \Omega^N.\]
The error probability of the coding  pair $(F_{N}, G_{N})$ is 
\[P_{N}\left\{ G_{N}\circ F_{N}(\omega)\not=\omega\right\}.
\]
If this probability is  less than some prescribed $1>\epsilon>0$, we shall say that the coding pair is $\epsilon$-good. Note that 
to any $\epsilon$-good coding pair one can associate the  set 
\[ A=\{\omega\,|\, G_{N}\circ F_{N}(\omega)=\omega\}\]
which satisfies 
\beq\label{mm-6}
P_N(A)\geq 1-\epsilon, \qquad |A|\leq 2^M.
\eeq
On the other hand, if $A\subset \Omega^N$ satisfies (\ref{mm-6}), we can associate to it an $\epsilon$-good pair 
$(F_{N}, G_{N})$ by setting $F_{N}$ to be one-one on $A$ (and arbitrary otherwise), and $G_{N}= F_{N}^{-1}$ on 
$F_{N}(A)$ (and arbitrary otherwise). 

In the source coding we wish to find $M$ that minimizes the compression coefficients $M/N$ subject to an allowed $\epsilon$-error probability. 
Clearly, the optimal $M$ is 
\[
M_N=\left[\log_2 \min\left\{ |A|\,|\, A\subset \Omega^N\, P_N(A)\geq 1-\epsilon\right\}\right],
\]
where $[\,\cdot\,]$ denotes the greatest integer part.  Shannon's source coding theorem now follows 
from Proposition \ref{how-to}: the  limiting optimal compression coefficient is 
\[
\lim_{N\rightarrow \infty}\frac{M_N}{N}=\frac{1}{\log 2}S({P}).
\]

\section{Why is the entropy natural?}
\label{sec-e-natural}
Set ${\cal P}=\cup_{\Omega} {\cal P}(\Omega)$. In this section we shall consider functions ${\mathfrak   S}: {\cal P}\rightarrow  \rr$ 
that satisfy properties that correspond intuitively to those of {\em entropy}  as a measure of {\em randomness} of probability measures. 
The goal is to show that those intuitive natural demands uniquely specify ${\mathfrak S}$ up to a choice of units, that is, 
that for some $c>0$ and all $P\in {\cal P}$, ${\mathfrak S}(P)= c S(P)$. 

We describe first three  basic  properties that any candidate for ${\mathfrak S}$ should satisfy. 
The first is the positivity and non-triviality requirement:  ${\mathfrak S}(P)\geq 0$ and this inequality is strict for at least one $P\in {\cal P}$.
The second  is that if $|\Omega_1|= |\Omega_2|$ and $\theta:\Omega_1\rightarrow \Omega_2$ is a bijection, 
then for any $P\in {\cal P}(\Omega_1)$, ${\mathfrak S}(P)={\mathfrak S}(P\circ \theta)$. In other words, 
the entropy of $P$ should not depend on the labeling  of the elementary events. 
This second requirement gives  that ${\mathfrak S}$ is completely specified by its restriction 
${\mathfrak  S}: \cup_{L\geq 1} {\cal P}_L\rightarrow [0, \infty[$ which  satisfies 
\beq  {\mathfrak S}(p_1, \cdots, p_L)= {\mathfrak S}(p_{\pi(1)}, \cdots, p_{\pi(L)})
\label{perm-inv}
\eeq
for any $L\geq 1$ and any  permutation $\pi$ of $\{1, \cdots, L\}$. In the proof of Theorem \ref{ax-entropy}
we shall also assume that  
\beq
\label{exten}
 {\mathfrak S}(p_1, \cdots, p_L, 0)= {\mathfrak S}(p_1, \cdots, p_L)
\eeq
for all $L\geq 1$ and $(p_1,\cdots, p_L)\in {\cal P}_L$.
In the literature, the common sense assumption \eqref{exten}   is sometimes called {\em expansibility}. 

Throughout this section we shall assume that the above three properties hold. We remark that the assumptions
of Theorem \ref{ax-entropy} actually imply the  positivity and non-triviality requirement.


\subsection{Split additivity characterization}

If $\Omega_1, \Omega_2$ are two disjoint sets, we denote by $\Omega_1\oplus \Omega_2$ their  union (the symbol 
$\oplus$ is used to emphasize the fact  that the sets are disjoint). 
If $\mu_1$ is a measure on $\Omega_1$ and $\mu_2$ is a measure on $\Omega_2$, then $\mu=\mu_1\oplus\mu_2$ is a measure on 
$\Omega_1\oplus\Omega_2$ defined by $\mu(\omega)=\mu_1(\omega)$ if $\omega\in \Omega_1$ and $\mu(\omega)=\mu_2(\omega)$ 
if $\omega \in \Omega_2$. Two measurable spaces $(\Omega_1, \mu_1)$, $(\Omega_2, \mu_2)$ are called 
disjoint if  the sets $\Omega_1$, $\Omega_2$, are disjoint.

The split additivity characterization has its roots in the identity 
\[
S(p_1P_1 + \cdots +p_nP_n) = p_1 S(P_1) + \cdots +p_nS(P_n) + S(p_1, \cdots, p_n)
\]
which holds if $\supp P_k \cap \supp P_j=\emptyset$ for $k\not=j$.
\bet
\label{khin}
Let ${\mathfrak S}: {\cal P}\rightarrow [0,\infty[$ be a function such that:
\begin{enumerate}[{\rm (a)}]

\item ${\mathfrak S}$ is continuous on ${\cal P}_2$.  

\item For any finite collection of disjoint probability spaces $(\Omega_j, P_j)$, $j=1, \cdots, n$, and any 
$(p_1, \cdots, p_n)\in {\cal P}_n$, 
\beq  {\mathfrak  S}\left(\bigoplus_{k=1}^n p_kP_k\right)=\sum_{k=1}^n p_k {\mathfrak S}(P_k) + {\mathfrak  S}(p_1, \cdots, p_n).
\label{rain-sun}
\eeq
\end{enumerate}
Then there exists $c>0$ such that for all $P\in {\cal P}$, 
\beq  {\mathfrak S}({P})=c S({P}).
\label{long-dm}
\eeq
\eet
\begin{remark}  If the  positivity and non-triviality assumptions are dropped, then  the proof gives that  \eqref{long-dm} holds for some $c\in \rr$. 
\end{remark}
\begin{remark} The split-additivity property \eqref{rain-sun}  is sometimes called the chain rule 
for entropy. It  can be verbalized as follows: if the initial choices 
$(1, \cdots, n)$, realized with  probabilities $(p_1, \cdots, p_n)$, are split into sub-choices described by probability spaces 
$(\Omega_k, P_k)$, $k=1, \cdots, n$, then the new entropy is  the sum of the initial entropy and the entropies of  
sub-choices weighted by their  probabilities. 
\label{split-ver}
\end{remark}
\demo 
In what follows, $\bar P_n\in {\cal P}_n$ denotes the chaotic probability measure
\[
\bar P_n=\left(\frac{1}{n}, \cdots, \frac{1}{n}\right),
\]
and 
\[
f(n)={\mathfrak S}(\bar P_n)={\mathfrak S}\left(\frac{1}{n}, \cdots, \frac{1}{n}\right).
\]

We split the argument into six steps.

{\bf Step 1.} ${\mathfrak  S}(1)= {\mathfrak S}(0,1)=0$. 

Suppose that $|\Omega|=2$ and let $P=(q_1, q_2)\in  {\cal P}_2$. Writing $\Omega=\Omega_1\oplus \Omega_2$ where $|\Omega_1|=|\Omega_2|=1$  and taking $P_1=(1)$, $P_2=(1)$, $p_1=q_1$, 
$p_2=q_2$, we get 
${\mathfrak  S}(q_1, q_2)={\mathfrak S}(1) + {\mathfrak  S}(q_1,q_2)$, and so ${\mathfrak S}(1)=0$. Similarly, the 
relations
\[
\begin{split}
{\mathfrak S }(0, q_1, q_2)&=q_1{\mathfrak  S}(0,1) +q_2{\mathfrak  S}(1) + {\mathfrak  S}(q_1, q_2),\\[2mm]
{\mathfrak  S}(0,q_1, q_2)&=0\cdot {\mathfrak  S}(1) + 1\cdot {\mathfrak S}(q_1, q_2)+ {\mathfrak S}(0,1),
\end{split}
\]
yield that ${\mathfrak  S}(0,1)=q_1{\mathfrak  S}(0,1)$ for all $q_1$, and so ${\mathfrak  S}(0,1)=0$.

{\bf Step 2.} $f(nm)=f(n)+ f(m)$.

Take $\Omega=\Omega_1\oplus\cdots \oplus \Omega_{m}$ with $|\Omega_k|=n$ for all $1\leq k\leq m$, and set 
$P_k=\bar P_n$, $p_k=1/m$.  It then follows from  \eqref{rain-sun} that $
f(nm)=m \cdot \frac{1}{m}f(n)+ f(m)= f(n) + f(m)$. 

{\bf Step 3.} $\lim_{n\rightarrow \infty}(f(n)-f(n-1))=0$.

In the proof of this step we shall make use of the following elementary result regarding convergence 
of the Ces\`aro means: if $(a_n)_{n\geq 1}$ is  a converging sequence 
of real numbers and $\lim_{n\rightarrow \infty}a_n=a$, then 
\[
\lim_{n\rightarrow \infty}\frac{1}{n}\sum_{k=1}^n a_k=a.
\]
As an exercise, prove this result. 

Set $d_n=f(n)-f(n-1)$, $\delta_n={\mathfrak S}(\frac{1}{n}, 1-\frac{1}{n})$. Since $f(1)={\mathfrak S}(1)=0$, 
\[f(n)=d_{n} +\cdots +d_{2}.\]
The relation \eqref{rain-sun} gives 
\[f(n)= \left(1-\frac{1}{n}\right)f(n-1) +\delta_n,\]
and so 
\[
n\delta_n= nd_n + f(n-1).
\]
It follows that 
\[
\sum_{k=2}^n k \delta_k= nf(n)=n(d_n + f(n-1))=n(n\delta_n-(n-1)d_n),
\]
which yields 
\[
d_n =\delta_n -\frac{1}{n(n-1)}\sum_{k=2}^{n-1}k\delta_k.
\]
By Step 1, $\lim_{n\rightarrow \infty} \delta_n=0$. Obviously, 
\[0\leq \frac{1}{n(n-1)}\sum_{k=2}^{n-1}k\delta_k\leq \frac{1}{n}\sum_{k=2}^{n-1}\delta_k,
\]
and  we derive 
\[\lim_{n\rightarrow \infty} \frac{1}{n(n-1)}\sum_{k=2}^{n-1}k\delta_k=0.
\]
It follow that    $\lim_{n\rightarrow \infty} d_n=0$.

{\bf Step 4.} There is a constant $c$ such that $f(n)=c \log n$ for all $n$. 

By Step 2, for any $k\geq 1$, 
\[
\frac{f(n^k)}{\log n^k}=\frac{n}{\log n}.
\]
Hence, to prove the statement it suffices to show that the limit 
\[c=\lim_{n\rightarrow \infty}\frac{f(n)}{\log n}
\]
exists. To prove that, we will show that $g(n)$ defined by 
\beq g(n)= f(n) - \frac{f(2)}{\log 2} \log n
\label{flashy}\eeq
satisfies 
\[
\lim_{n\rightarrow \infty}\frac{g(n)}{\log n}=0.
\]
The choice of integer $2$ in \eqref{flashy} is irrelevant, and the the argument works with $2$ replaced by any 
integer $m\geq 2$. 

Obviously, $g(nm)=g(n) + g(m)$ and $g(1)=g(2)=0$.  
Set $\xi_m= g(m)-g(m-1)$ if $n$ is odd,  $\xi_m=0$ if $m$ is even. By Step 3, $\lim_{m\rightarrow 
\infty} \xi_m=0$. 
Let $n> 1$ be given. Write $n=2n_1 + r_1$, where $r_1=0$ or $r_1=1$. Then 
\[
g(n) =\zeta_n + g(2n_1)=\zeta_n + g(n_1),
\]
where we used that $g(2)=0$. If $n_1>1$, write again $n_1=2n_1 + r_2$, where $r_2=0$ or $r_2=1$, so that 
\[g(n_1) =\zeta_{n_1}  + g(n_2).
\] This procedure terminates after $k_0$ steps, that is, when  we reach $n_{k_0}=1$. Obviously, 
\[k_0 \leq \frac{\log n}{\log 2}, \qquad g(n)=\sum _{k=0}^{k_0-1} \zeta_{n_k},\]
where we set $n_0=n$. Let $\epsilon >0$ and  $m_\epsilon$ be such that for $m\geq m_\epsilon$ we have $|\xi_m|<\epsilon/\log 2$. 
Then 
\[
\frac{|g(n)|}{\log n}\leq \frac{1}{\log n}\left(\sum_{m\leq m_\epsilon}|\xi_m|\right)+ \epsilon \frac{k_0\log 2 }{\log n} \leq \frac{1}{\log n}\left(\sum_{m\leq m_\epsilon}|\xi_m|\right)+ \epsilon.
\]
It follows that 
\[
\limsup_{n\rightarrow \infty}\frac{|g(n)|}{\log n}\leq \epsilon.
\]
Since $\epsilon >0$ is arbitary, the proof is complete. 

{\bf Step 5.} If  $c$ is as in Step 4, then 
\[{\mathfrak  S}(q_1, q_2)= cS(q_1, q_2).
\]

 Let $\Omega=\Omega_1\oplus\Omega_2$ with $|\Omega_1|= m$, $|\Omega_2|=m-n$. Applying 
\eqref{rain-sun} to $P_1=\bar P_n$, $P_2=\bar P_{n-m}$, $p_1=\frac{n}{m}$, $p_2=\frac{m-n}{m}$, we derive 
\[
f(m)=\frac{n}{m}f(n) + \frac{m-n}{m}f(m-n) + {\mathfrak  S}\left(\frac{n}{m}, \frac{m-n}{m}\right).
\]
Step 4 gives that 
\[ 
{\mathfrak  S}\left(\frac{n}{m}, \frac{m-n}{m}\right)= c S\left(\frac{n}{m}, \frac{m-n}{m}\right).
\]
Since this relation holds for any $m<n$, the continuity of ${\mathfrak  S}$ and $S$ on ${\cal P}_2$  yields the statement.

{\bf Step 6.} We now complete the proof by induction on $|\Omega|$. Suppose that  ${\mathfrak S}(P)=cS(P)$ holds 
for all $P\in {\cal P}(\Omega)$ with $|\Omega|=n-1$, where $c$ is as in Step 4. Let $P=(p_1, \cdots, p_n)$ be a probability measure 
on $\Omega=\Omega_{n-1}\oplus \Omega_1$, where $|\Omega_{n-1}|=n-1$, $|\Omega_1|=1$. Without loss of generality 
we may assume that $q_n <1$. Applying  \eqref{rain-sun} with 
\[ P_1=\left(\frac{q_1}{1-q_n}, \cdots, \frac{q_{n-1}}{1-q_n}\right),\]
$P_2=(1)$, $p_1=1-q_n$, $p_2=q_n$, we derive   
\[
{\mathfrak S}(P)= cS(P_1) + cS(p_1, p_2)= cS (P).
\]
This completes the proof. The non-triviality assumption  yields that $c>0$. \qed

\subsection{Sub-additivity characterization}

The sub-additivity of entropy described in Proposition \ref{e-subad} is certainly a very intuitive property. 
If the entropy quantifies randomness of a probability measure $P$, or equivalently, the amount 
of information gained by an outcome of a probabilistic experiment described by $P$, than 
 the product of marginals $P_l\otimes P_r$ is certainly more random then $P\in {\cal P}(\Omega_l\times\Omega_r)$. 
 The Boltzmann--Gibbs--Shannon entropy $S$ and the Hartley entropy $S_H$ introduced in Exercise \ref{hartley} 
 are strictly sub-additive, and so is any linear combination 
 \beq {\mathfrak S}=cS + CS_H,
 \label{sub-beau}
 \eeq
 where $c\geq 0$, $C\geq 0$, and at least one of these constants is strictly positive. It is a  remarkable 
 fact that the strict sub-additivity requirement together with the obvious assumption \eqref{exten} selects \eqref{sub-beau} 
 as the only possible choices for entropy. We also note   the strict sub-additivity assumption selects the sign of the constants in \eqref{sub-beau}, 
and  that here we can  omit the assumption (a)  of Theorem \ref{khin}.
\bet\label{ax-entropy} Let ${\mathfrak  S} : {\cal P}\rightarrow [0, \infty[ $ be a strictly sub-additive map, namely  if $\Omega=\Omega_l \times \Omega_r$ and $P\in {\cal P}(\Omega)$, 
then 
\[{\mathfrak S}({P})\leq {\mathfrak S}(P_l) + {\mathfrak S}(P_r)\]
with equality iff $P=P_{l}\otimes P_r$.
Then there are   constants $c\geq0 , C\geq 0$, $c+C>0$,  such that for all $P\in {\cal P}$,
\beq {\mathfrak  S}({P}) = cS({P}) + CS_H(P).
\label{freezing-r}
\eeq
If in addition  ${\mathfrak S}$ is continuous on ${\cal P}_2$, then 
 $C=0$ and ${\mathfrak S}=c S$ for some $c>0$.
\eet

\demo We denote by ${\mathfrak  S}_n$ the restriction of ${\mathfrak S}$ to ${\cal P}_n$. 
Note that the sub-additivity  implies that 
\beq \begin{split}
{\mathfrak  S}_{2n}(p_{11}, p_{12}, \cdots, p_{n1}, p_{n2})&\leq {\mathfrak  S}_2(p_{11}+\cdots+ p_{n1}, p_{12} +\cdots +p_{n2}) \\[2mm]
&\qquad + 
{\mathfrak S}_n(p_{11}+ p_{12}, \cdots, p_{n1}+p_{n2}).
\end{split}
\label{sub-simp}
\eeq
For $x\in [0,1]$ we   set $\bar x=1-x$. The function 
\beq F(x)={\mathfrak S}_2(\bar x, x)
\label{inf-function}
\eeq
will play an important role in the proof. It follows  from  \eqref{perm-inv} that  $F(x)=F(\bar x)$. 
By taking $P_l=P_r=(1,0)$, we see that 
\[
2F(0)= {\mathfrak S}(P_l)+ {\mathfrak S}(P_r)= {\mathfrak S}(P_l\otimes P_r)= {\mathfrak S}(1,0,0,0)= {\mathfrak  S}(1,0)=F(0),
\]
and so $F(0)=0$. 

We split the proof  into eight steps. 

{\bf Step 1.} For all $q, r \in [0,1]$ and $(p,p_3,\cdots, p_n)\in {\cal P}_{n-1}$, $n\geq 3$, one has 
\beq \label{paris-night}
\begin{split}
{\mathfrak  S}_2(\bar q, q)- {\mathfrak S}_2(\bar p\,\bar q+p\bar r, \bar p q + pr)&\leq {\mathfrak S}_n(p\bar q, pq, p_3, \cdots, p_n) - {\mathfrak S}_n(p\bar r, pr, p_3, \cdots, p_n)\\[2mm]
&\leq {\mathfrak S}_2(\bar p \,\bar r +p\bar q, \bar p r + rq)- {\mathfrak S}_2(\bar r, r).
\end{split}
\eeq

By interchanging $q$ and $r$, it suffices to prove the first inequality in \eqref{paris-night}. We have 
\[
\begin{split}
{\mathfrak S}_2(\bar q, q)+ {\mathfrak S}_{n}(p\bar r, pr, p_3, \cdots, p_n) &= {\mathfrak S}_{2n}(\bar q p \bar r, qp\bar r, \bar q pr, qpr, \bar qp_3, qp_3, \cdots, 
\bar q p_n, qp_n)\\[2mm]
&={\mathfrak S}_{2n}(\bar q p \bar r, \bar q pr, qp\bar r, qpr, \bar qp_3, qp_3, \cdots, \bar q p_n, qp_n)\\[2mm]
&\leq {\mathfrak S}_2(\bar q p\bar r +qp\bar r + \bar q(p_3+\cdots+p_n), \bar q p r + q p r + q(p_3+\cdots+p_n))\\[2mm]
&\qquad +{\mathfrak S}_n(\bar q p\bar r +\bar q pr, qp\bar r + qpr, \bar q p_3 + qp_3, \cdots, \bar q p_n + qp_n)\\[2mm]
&={\mathfrak S}_2(\bar p\,\bar q+p\bar r, \bar p q + pr) + S_n(p\bar r, pr, p_3, \cdots, p_n).
\end{split}
\]
The first equality follows from \eqref{perm-inv} and  the first inequality from \eqref{sub-simp}. The final equality is elementary 
(we used that $p+p_3+ \cdots p_n=1$). 

{\bf Step 2.} The function $F$, defined by \eqref{inf-function}, is increasing on $[0,1/2]$, decreasing on $[1/2, 1]$, and 
is continuous and concave on $]0,1[$. Morever, for $q\in ]0,1[$ the left and right derivatives 
\[
D^+F(q)=\lim_{h \downarrow 0} \frac{F(q+h)- F(q)}{h}, \qquad D^-F(q)=\lim_{h \uparrow 0} \frac{F(q+h)- F(q)}{h}
\]
exist, are finite, and $D^+F(q)\geq D^{-}F(q)$.

We first establish the monotonicity statement.  Note that the inequality of Step 1 
\beq 
{\mathfrak S}_2(\bar q, q)- {\mathfrak S}_2(\bar p\,\bar q+p\bar r, \bar p q + pr)\leq {\mathfrak S}_2(\bar p \,\bar r +p\bar q, \bar p r + rq)- {\mathfrak S}_2(\bar r, r)
\label{paris-mor}
\eeq
with $r=\bar q$ gives 
\[\begin{split}
2 {\mathfrak S}_2(\bar q, q)&\leq {\mathfrak S}_2((1-p)(1-q) + pq, (1-p)q + p(1-q))\\[2mm]
&\qquad + {\mathfrak S}_2((1-p)q + p(1-q), (1-p)(1-q) + pq),
\end{split}
\]
or equivalently, that 
\beq\label{paris-noise} F(q)\leq F((1-p)q +p(1-q)).\eeq
Fix $q\in [0, 1/2]$ and note that $[0,1]\ni p \mapsto (1-p)q  +p(1-q)$ is the parametrization of the interval $[q,1-q]$. Since 
$F(q)= F(1-q)$, we derive that  $F(q)\leq F(x)$ for $x\in [q, 1/2]$, and that $F(x)\geq F(1-q)$ for $x\in [1/2, q]$. Thus, $F$ is 
increasing on $[0,1/2]$ and decreasing on $[1/2,1]$. In particular,  for all $x\in [0,1]$, 
\beq\label{paris-bounds}
F(1/2)\geq F(x)\geq 0,
\eeq
where we used that $F(0)=F(1)=0$.

We now turn to the continuity and concavity, starting with continuity first. 
The inequality \eqref{paris-mor} with $p=1/2$ gives that for any $q, r\in [0,1]$,
\beq
\frac{1}{2}F(q) + \frac{1}{2}F(r)\leq F\left(\frac{1}{2}q +\frac{1}{2}r\right).
\label{conv-first}
\eeq
Fix now $q\in ]0, 1[$,  set $\lambda_n=2^{-n}$ and, starting with large enough $n$ so that $q\pm \lambda_n\in [0,1]$,
define
\[
\Delta_n^+(q)=\frac{F(q + \lambda_n)-F(q)}{\lambda_n}, \qquad \Delta_n^-(q)=\frac{F(q-\lambda_n)- F(-q)}{-\lambda_n}.
\]
It follows from  \eqref{conv-first} that the sequence $\Delta_n^+(q)$ is increasing, that the 
sequence $\Delta_n^{-}(q)$ is decreasing , and that $\Delta_n^+(q)\leq  \Delta_n^{-}(q)$ (write down 
the details!). Hence, the limits 
\[ 
\lim_{n\rightarrow \infty}\Delta_n^+(q), \qquad \lim_{n\rightarrow \infty}\Delta_n^-(q)
\]
exists,  are finite, and 
\beq
\lim_{n \rightarrow \infty}F(q\pm \lambda_n)=F(q).
\label{no-way}
\eeq
The established monotonicity properties of $F$ yield that the limits $\lim_{h\downarrow 0} F(q+h)$ and $
\lim_{h \uparrow 0}F(q+h)$ exist. Combining this observation with \eqref{no-way}, we derive that 
\[\lim_{h\rightarrow 0}F(q+h)=F(q),\]
and so  $F$ is  continuous 
on $]0,1[$. We now prove the concavity.  
Replacing $r$ with $(q +r)/2$ in \eqref{conv-first}, we get that 
\beq 
\lambda F(q) + (1-\lambda) F(r)\leq F\left( \lambda q +(1-\lambda) r\right)
\label{conc-gen}
\eeq
holds for $\lambda=3/4$, while replacing $q$ with $(q+r)/2$ shows that \eqref{conc-gen} holds for $\lambda=1/4$. 
Continuing in this way  shows that \eqref{conc-gen} holds for all dyadic fractions $\lambda=k/2^n$, $1\leq k \leq 2^n$, 
$n=1,2,\cdots$. 
Since dyadic fractions are dense in $[0,1]$, the continuity of $F$ 
yields that \eqref{conc-gen} holds for $\lambda \in [0,1]$ and $q, r\in ]0,1[$. Finally, to prove the statement 
about the derivatives, fix $q\in ]0,1[$ and for $h>0$ small enough consider the functions 
\[
\Delta^+(h)=\frac{F(q + h)-F(q)}{h}, \qquad \Delta^-(h)=\frac{F(q-h)- F(q)}{-h}.
\]
The concavity of $F$ gives that the function $h\mapsto \Delta^+(h)$ is increasing, that $h\mapsto \Delta^-(h)$ is increasing, 
and that $\Delta^+(h)\leq \Delta^-(h)$. This establishes the last claim of the Step 2 concerning left and right derivatives of 
$F$ on $]0,1[$. 

{\bf Step 3.} There exist functions ${\cal R}_n: {\cal P}_n\rightarrow \rr$, $n\geq 2$, such that 
\beq\label{cergy-cold}
{\mathfrak S}_n(p\bar q, pq, p_3, \cdots, p_n)= pF(q)+ {\cal R}_{n-1}(p, p_3, \cdots, p_n)
\eeq
for all $q\in ]0,1[$, $(p, p_3, \cdots, p_n)\in {\cal P}_{n-1}$ and $n\geq 2$. 

To prove this, note that the Step 1 and the relation $F(x)=F(\bar x)$ give
\beq \label{ihp}
\begin{split}
\frac{F(\bar p q +p q)- F(\bar p q +pr)}{q-r}
&\leq \frac{S_n(p\bar q, pq, p_3, \cdots, p_n)- S_n(p\bar r, pr, p_3, \cdots, p_n)}{q-r}\\[2mm]
&\leq \frac{F(pq +\bar p r)- F(pr +\bar p r)}{q-r}
\end{split}
\eeq
for $0<r <q <1$ and $(p, p_3, \cdots, p_n)\in {\cal P}_n$. Fix $(p, p_3, \cdots, p_n)\in {\cal P}_n$ and set 
\[
L(q)={\mathfrak S}_n(p\bar q, pq, p_3, \cdots, p_n).
\]
Taking $q\downarrow r$ in \eqref{ihp} we get 
\[ pD^-F(r)= D^{-}L(r),
\]
while taking $r\uparrow q$ gives 
\[p D^+ F(q)= D^+ L(q).\]
Since $D^\pm F(q)$ is finite by  Step 2, we derive that the function $L(q)- pF(q)$ is differentiable on 
$]0, 1[$ with vanishing derivative. Hence, for $q\in ]0,1[$, 
\[ L(q) =p F(q)+ {\cal R}_{n-1}(p, p_3, \cdots, p_n),
\]
where the constant  ${\cal R}_{n-1}$ depends on  the  values 
$(p, p_3, \cdots, p_n)$ we have fixed in the above argument. 

{\bf Step 4.} There exist constants $c\geq 0$ and $C$ such that for  all $q\in ]0,1[$, 
\beq\label{paris-late}  F(q) =cS(1-q, q) + C.\eeq

We start the proof by taking $(p_1, p_2, p_3)\in {\cal P}_{3, {\rm f}}$. Setting 
\[p=p_1+p_2, \qquad q=\frac{p_2}{p_1+ p_2},\]
we write 
\[{\mathfrak S}_3(p_1, p_2, p_3)= {\mathfrak S}_3(p\bar q, pq,p_3).\]
It then follows from Step 3 that 
\beq\label{if-ihp}
{\mathfrak S}_3(p_1, p_2, p_3)=(p_1+p_2){\mathfrak S}_2\left( \frac{p_1}{p_1+p_2}, \frac{p_2}{p_1+p_2}\right) + 
{\cal R}_2(p_1+p_2, p_3).
\eeq
By \eqref{perm-inv} we also have 
\beq\label{if-ihp-1}
{\mathfrak S}_3(p_1, p_2, p_3)={\mathfrak S}_3(p_1, p_3, p_2)=(p_1+p_3){\mathfrak S}_2\left( \frac{p_1}{p_1+p_3}, \frac{p_3}{p_1+p_3}\right) + 
{\cal R}_2(p_1+p_3, p_3).
\eeq
Setting $G(x)= {\cal R}_2(\bar x, x)$, $x=p_3$, $y=p_2$,  we rewrite \eqref{if-ihp}=\eqref{if-ihp-1} as 
\beq\label{jet-lag}
(1-x)F\left(\frac{y}{1-x}\right) + G(x) = (1-y)F\left(\frac{x}{1-y}\right) + G(y),
\eeq
where $x,y\in ]0,1[$ and $x+y <1$. The rest of the proof concerns analysis of the functional equation \eqref{jet-lag}.

Since $F$ is continuous on $]0,1[$, fixing one variable one easily deduces from \eqref{jet-lag} that $G$ is also continuous on $]0,1[$. 
Let $0 <a <b<1$ and fix $y\in ]0, 1-b[$.
It follows that (verify this!)
\[
\frac{x}{1-y}\in \left] a, \frac{b}{1-y}\right]\subset\,\,\, ]0,1[, \qquad \frac{y}{1-x}\in  \left]y, \frac{y}{1-b}\right]\subset \,\,\,]0,1[.
\]
Integrating \eqref{jet-lag} with respect to $x$ over $[a, b]$ we derive
\beq
\begin{split}
(b-a)G(y)&=\int_a^b G(y)\d x \\[2mm]
&=\int_a^b G(x)\d x + \int_a^b (1-x)F\left(\frac{y}{1-x}\right)\d x - (1-y)\int_a^b F\left(\frac{x}{1-y}\right) \d x\\[2mm]
&= \int_a^b G(x)\d x + y^2\int_{y/(1-a)}^{y/(1-b)} s^{-3}F(s)\d s - (1-y)^2\int_{a/(1-y)}^{b/(1-y)} F(t) \d t, 
\end{split}
\label{integrals}
\eeq
where we have used the change of variable 
\beq\label{change-v} 
s=\frac{y}{1-x}, \qquad t=\frac{x}{1-y}.
\eeq
It follows that $G$ is differentiable on $]0, b[$. Since $0 <b<1$ is arbitrary, $G$ is  differentiable  on $]0,1[$. 

The change of variable \eqref{change-v} maps bijectively $\{(x,y)\,|\, x, y>0\}$ to $\{(s, t)\,|\, s, t\in ]0,1[\}$ (verify this!), and in this  new 
variables the functional equation \eqref{jet-lag} reads
\beq\label{diff-F}
F(t)=\frac{1-t}{1-s}F(s) + \frac{1-st}{1-s}\left[ G\left(\frac{t-st}{1-st}\right)- G\left(\frac{s-st}{1-st}\right)\right].
\eeq
Fixing $s$, we see that the differentiablity of $G$ implies the differentiability of $F$ on $]0,1[$. Returning to 
\eqref{integrals}, we get that $G$ is twice differentiable on $]0,1[$, and then \eqref{diff-F} gives that $F$ is also 
twice differentiable on $]0,1[$. Continuing in this way we derive that both $F$ and $G$ are infinitely differentiable on $]0,1[$. Differentiating  
 \eqref{jet-lag} first with respect to $x$ and then with respect to $y$ gives 
 \beq\label{diff-2}
\frac{y}{(1-x)^2} F^{\prime\prime}\left(\frac{y}{1-x}\right)=\frac{x}{(1-y)^2}F^{\prime\prime}\left(\frac{x}{1-y}\right).
\eeq
The substitution \eqref{change-v} gives that for $s, t\in ]0,1[$, 
\[
s(1-s)F^{\prime\prime}(s)= t(1-t)F^{\prime\prime}(t).
\]
It follows that for some $c\in \rr$,
\[
t(1-t)F^{\prime\prime}(t)=-c.
\]
Integration gives
\[ 
F(t) = c S(1-t, t) + Bt + C.
\]
Since $F(t)=F(\bar t)$, we have $B=0$, and since $F$ is increasing on $[0,1/2]$, we have $c\geq 0$. This completes the 
proof of the Step 4. Note that as a by-product of the proof we have derived that for some constant $D$,
\beq\label{verify} G(x)= F(x) + D, \qquad x\in ]0,1[.
\eeq
To prove \eqref{verify}, note that \eqref{paris-late} gives that $F$ satisfies the functional equation
\[
(1-x)F\left(\frac{y}{1-x}\right) + F(x) = (1-y)F\left(\frac{x}{1-y}\right) + F(y).
\]
Combining this equation with \eqref{jet-lag} we derive that for $x,y>0$, $0<x+y <1$,
\[G(x)-F(x)=G(y)-F(y).\]
Hence, $G(x)- F(x)= D_y$ for $x\in ]0, 1-y[$. If $y_1<y_2$, we must have $D_{y_1}= D_{y_2}$, and 
so $D=D_y$ does not depend on $y$, which gives \eqref{verify}. 

{\bf Step 5.} For any $n\geq 2$ there exists constant $C(n)$ such that for $(p_1, \cdots, p_n)\in {\cal P}_{n, {\rm f}}$, 
\beq\label{cergy-1}
{\mathfrak S}_n(p_1,\cdots, p_n)= c S(p_1, \cdots, p_n) + C(n), 
\eeq
where $c\geq 0$ is the constant from the Step 4.

In the Step 4 we established \eqref{cergy-1} for $n=2$ (we set $C(2)=C$), and so we assume that $n\geq 3$. 
Set  $p=p_1+p_2$, $q=p_2/(p_1+p_2)$. It then follows from Steps 3 and 4 that 
\beq \label{cergy-cold}
\begin{split}
{\mathfrak S}_n(p_1,\cdots, p_n)&= (p_1 +p_2){\mathfrak S}_2\left(
\frac{p_1}{p_1+p_2}, \frac{p_2}{p_1+p_2}\right) + {\cal R}_{n-1}(p_1+p_2, p_3, \cdots, p_n)\\[2mm]
&=(p_1+p_2)c S\left(
\frac{p_1}{p_1+p_2}, \frac{p_2}{p_1+p_2}\right) +  \widehat {\cal R}_{n-1}(p_1+p_2, p_3, \cdots, p_n),
\end{split}
\eeq
where $\widehat {\cal R}_{n-1}(p, p_3, \cdots, p_n)= pC_2 + {\cal R}_{n-1}(p, p_3, \cdots, p_n)$. 
Note that since ${\cal R}_{n-1}$ is invariant under the permutations of the variables 
$(p_3, \cdots, p_n)$ (recall \eqref{cergy-cold}), so is $\widehat {\cal R}_{n-1}$. The invariance of ${\mathfrak S}_n$ under the 
permutation of the variables gives 
\[{\mathfrak S}_n(p_1,\cdots, p_n)
=(p_1+p_3)c S\left(
\frac{p_1}{p_1+p_3}, \frac{p_3}{p_1+p_3}\right) +  \widehat {\cal R}_{n-1}(p_1+p_3, p_2, p_4\cdots, p_n),
\]
and so 
\beq\label{cergy-hola}
\begin{split}
(p_1+p_2)c S&\left(
\frac{p_1}{p_1+p_2}, \frac{p_2}{p_1+p_2}\right) -(p_1+p_3)c S\left(
\frac{p_1}{p_1+p_3}, \frac{p_3}{p_1+p_3}\right)\\[2mm]
&=\widehat {\cal R}_{n-1}(p_1+p_2, p_3, \cdots, p_n)- \widehat {\cal R}_{n-1}(p_1+p_3, p_2, p_4\cdots, p_n).
\end{split}
\eeq
Until the the end of the proof when we wish to indicate the number of variables in the 
Boltzmann--Gibbs--Shannon entropy we will write  $S_n(p_1, \cdots, p_n)$. One easily verifies that 
\[
\begin{split}
S_n(p_1, \cdots, p_n)&=(p_1 +p_2)S_2\left(
\frac{p_1}{p_1+p_2}, \frac{p_2}{p_1+p_2}\right) + S_{n-1}(p_1+p_2, p_3, \cdots, p_n)\\[2mm]
&=(p_1 +p_3)S_2\left(
\frac{p_1}{p_1+p_3}, \frac{p_3}{p_1+p_3}\right) + S_{n-1}(p_1+p_3, p_2, p_4, \cdots, p_n),
\end{split}
\]
and so 
\beq\label{cergy-hola1}
\begin{split}
(p_1+p_2)S_2&\left(
\frac{p_1}{p_1+p_2}, \frac{p_2}{p_1+p_2}\right) -(p_1+p_3) S_2\left(
\frac{p_1}{p_1+p_3}, \frac{p_3}{p_1+p_3}\right)\\[2mm]
&= S_{n-1}(p_1+p_2, p_3, \cdots, p_n)- S_{n-1}(p_1+p_3, p_2, p_4\cdots, p_n).
\end{split}
\eeq
Since in the formulas \eqref{cergy-hola} and \eqref{cergy-hola1} $S= S_2$, we derive that 
the function 
\[
T_{n-1}(p, q, p_4, \cdots, p_n)=\widehat {\cal R}_{n-1}(p,q, p_4, \cdots, p_n)- cS_{n-1}(p, q, p_4, \cdots, p_n)
\]
satisfies 
\beq\label{paris-thu}
T_{n-1}(p_1+p_2, p_3, p_4, \cdots, p_n)=T_{n-1}(p_1+p_3, p_2, p_4, \cdots, p_n)
\eeq
for all $(p_1, \cdots p_n)\in {\cal P}_{n, {\rm f}}$. 
Moreover, by construction, $T_{n-1}(p,q, p_4, \cdots, p_n)$ is invariant under the permutation of the variables $(q,p_4, \cdots, p_n)$. 
Set $s=p_1+p_2+p_3$. Then  \eqref{paris-thu} reads 
as 
\[ T_{n-1}(s-p_3, p_3, p_4, \cdots, p_n)=T_{n-1}(s-p_2, p_2, p-p_4, \cdots, p_n).\]
Hence, the map 
\[ ]0, s[\ni p \mapsto T_{n-1}(s-p, p, p_4, \cdots, p_n)\]
is contant. By the permutation invariance, the maps
\[
 ]0, s[\ni p \mapsto T_{n-1}(s-p, p_3, \cdots, p_{m-1}, p, p_{m+1}, \cdots)
\]
are also constant. Setting $s=p_1+p_2 + p_3 +p_4$, we deduce that the map 
\[
(p_3, p_4)\mapsto T_{n-1}(s-p_3-p_4, p_3, p_4, \cdots, p_n)
\]
with domain $p_3>0, p_4>0$, $p_3 +p_4 <s$, is constant. Continuing inductively, we conclude that the map 
\[
(p_3, \cdots, p_n)\mapsto T_{n-1}(1 - (p_3+\cdots+p_n), p_3, p_4, \cdots, p_n)
\]
with domain $p_k>0$, $\sum_{k=3}^n p_k <1$ is constant. 
Hence, the map 
\[
{\cal P}_{n, {\rm f}}\ni (p_1, \cdots, p_n)\mapsto  T_{n-1}(p_1+p_2, p_3, \cdots, p_n)
\]
is constant, and we denote the value it assumes by $C(n)$.  Returning  now to \eqref{cergy-cold}, 
we conclude the proof of \eqref{cergy-1}:
\beq 
\begin{split}
{\mathfrak S}_n(p_1,\cdots, p_n)&=(p_1+p_2)c S_2\left(
\frac{p_1}{p_1+p_2}, \frac{p_2}{p_1+p_2}\right) +  \widehat {\cal R}_{n-1}(p_1+p_2, p_3, \cdots, p_n)\\[2mm]
&=(p_1+p_2)c S_2\left(\frac{p_1}{p_1+p_2}, \frac{p_2}{p_1+p_2}\right) + cS_{n-1}(p_1+p_2, p_3, \cdots, p_n) + C(n)\\[2mm]
&= c S_n(p_1, \cdots, p_n) + C(n).
\end{split}
\eeq

{\bf Step 6.} $C(n+m)= C(n) C(m)$ for  $n,m\geq 2$, and 
\beq\label{paris-wed}
\liminf_{n\rightarrow \infty}(C(n+1)-C(n))=0.
\eeq
If $P_l\in {\cal P}_n$ and $P_r\in {\cal P}_m$, then 
the identity ${\mathfrak S}_{nm}(P_l \times P_r)= {\mathfrak S}_n(P_l)+ {\mathfrak S}(P_r)$ and \eqref{cergy-1} give that $C(n+m)=C(n) + C(m)$. 
To prove \eqref{paris-wed}, suppose that $n\geq 3$ and take in \eqref{paris-night} $q=1/2$, $r=0$, $p=p_3=\cdots=p_n=1/(n-1)$. Then, combining 
\eqref{paris-night} with Step 5, we derive
\[
\begin{split}
F\left(\frac{1}{2}\right)&- F\left( \frac{n-2}{2(n-1)}\right)\\[2mm]
&\leq {\mathfrak S}_n\left(\frac{1}{2(n-1)}, \frac{1}{2(n-1)}, \frac{1}{n-1}, \cdots, \frac{1}{n-1}\right)- {\mathfrak S}_n
\left(\frac{1}{n-1}, 0, \frac{1}{n-1}, \cdots \frac{1}{n-1}\right)\\[2mm]
&=cS_n\left(\frac{1}{2(n-1)}, \frac{1}{2(n-1)}, \frac{1}{n-1}, \cdots, \frac{1}{n-1}\right)- cS_{n-1}
\left(\frac{1}{n-1}, \frac{1}{n-1}, \cdots \frac{1}{n-1}\right)\\[2mm]
&\qquad  + C(n)- C(n-1)\\[2mm]
&= \frac{\log 2}{n-1} + C(n)- C(n-1).
\end{split}
\]
The first inequality in  \eqref{paris-bounds} gives 
\[
0\leq \frac{\log 2}{n-1} + C(n)- C(n-1),
\]
and the statement follows. 

{\bf Step 7.} There is a constant $C\geq 0$ such that for all $n\geq 2$, $C(n)=C \log n$.

Fix $\epsilon >0$ and $n>1$. Let $k\in \nn$ be such that for all integers $p \geq n^k$, $C(p+1)-C(p)\geq -\epsilon$. It follows 
that for $p\geq p^k$ and $j\in \nn$,  
\[ C(p+ j)- C(p)= \sum_{i=1}^{j} (C(p+i)-C(p+i-1))\geq - j\epsilon.\]
Fix now $p\geq n^k$ and let $m\in \nn$ be such that $n^m\leq p <n^{m+1}$. Obviously, $m\geq k$. 
Write 
\[ p= a_mn^m + a_{m-1}n^{m-1} +\cdots + a_1p + a_0,
\]
where $a_k$'s are integers such that $1\leq a_m<n$  and $0\leq a_k<n$ for $k<m$. 
It follows that 
\[
C(p)> C(a_m n^m +\cdots + a_1 n)- n\epsilon = C(n) + C(a_mn^{m-1} +\cdots + a_2n + a_1)-n\epsilon.\]
Continuing inductively, we derive that 
\[
C(p)> (m-k+1)C(n) + C(a_mn^{k-1} +a_{m-1}n^{k-2}+  \cdots+ a_{m-k+1})- (m-k+1)\epsilon.
\]
If $M=\max_{2\leq j \leq n^{k+1}}|C(j)|$, then the last inequality gives 
\[ C(p)>(m-k+1)C(n) - M - (m-k+1)\epsilon.\]
By the choice of $m$, $\log p \leq (m+1)\log n$, and so 
\[
\liminf_{p\rightarrow \infty}\frac{C(p)}{\log p}\geq \frac{C(n)}{\log n}.
\]
Since 
\[
\liminf_{n\rightarrow \infty}\frac{C(p)}{\log p}\leq \liminf_{j \rightarrow \infty}\frac{C(n^j)}{\log n^j}= \frac{C(n)}{n},
\]
we derive that for all $n\geq 2$, 
\[ C(n)= C\log n,
\]
where 
\[
C=\liminf_{p\rightarrow \infty}\frac{C(p)}{p}.
\]
It remains to show that $C\geq 0$. Since 
\[
F(x) =c S_2(1-x, x) + C\log 2,
\]
we have  $\lim_{x\downarrow 0}F(x)=C\log 2$, and \eqref{paris-bounds} yields that $C\geq 0$.

{\bf Step 8.} We now conclude the proof. Let $P=(p_1, \cdots, p_n)\in {\cal P}_n$. Write 
\[P=(p_{j_1}, \cdots, p_{j_k}, 0, \cdots, 0),\]
where $p_{j_m}>0$ for $m=1, \cdots, k$. Then 
\[
{\mathfrak S}_n(P)={\mathfrak S}_k(p_{j_1}, \cdots, p_{j_k})=cS_k(p_{j_1}, \cdots, p_{j_k}) + C\log k=
cS_n(P) + CS_H(P).
\]
Since ${\mathfrak S}_n$ is strictly sub-additive, we must have $c+ C>0$. The final statement is a consequence of 
the fact that $S_H$ is not continuous on ${\cal P}_n$ for $n\geq 2$. 
\qed

\section{R\'enyi entropy}
Let $\Omega$ be a finite set and $P\in {\cal P}(\Omega)$. For $\alpha \in ]0, 1[$  we set 
\[S_\alpha({P})=\frac{1}{1-\alpha}\log \left(\sum_{\omega \in \Omega}P(\omega)^\alpha\right).
\]
$S_\alpha({P})$ is called the R\'enyi entropy of $P$. 
\bep\label{Renyi}\begin{enumerate}[{\rm (1)}]
\item $\lim_{\alpha \uparrow 1}S_\alpha({P})=S({P})$.

\item $\lim_{\alpha \downarrow 0}S_\alpha({P})=S_H({P})$.


\item $S_\alpha({P})\geq 0$  and $S_\alpha({P})=0$ iff $P$ is pure. 

\item $S_\alpha({P})\leq \log |\Omega|$ with equality iff $P=P_{\rm ch}$. 

\item The map  $]0, 1[ \ni \alpha \mapsto S_\alpha({P})$ is decreasing and is strictly decreasing  unless $P=P_{\rm ch}$. 

\item The map ${\cal P}(\Omega)\ni P \mapsto S_\alpha({P})$ is continuous and concave.

\item If $P=P_l\otimes P_r$ is a product measure on $\Omega=\Omega_l \times \Omega_r$, then 
$S_\alpha({P})=S_{\alpha}(P_l) + S_{\alpha}(P_r)$.

\item  The map $\alpha \mapsto S_\alpha({P})$ extends to a real analytic function on $\rr$ by the formulas
$S_1(P)=S(P)$ and 
\[S_\alpha({P})=\frac{1}{1-\alpha}\log \left(\sum_{\omega \in \supp P}P(\omega)^\alpha\right), \qquad \alpha\not=1.\]

\end{enumerate}
\eep

\begin{exo} Prove Proposition \ref{Renyi}. \end{exo}

\begin{exo} Describe properties of $S_\alpha(P)$ for $\alpha\not\in\, ]0,1[$. 
\end{exo}

\begin{exo} Let $\Omega=\{-1,1\}\times \{-1,1\}$, $0<p,q<1$, $p+q=1$, $p\not=q$, and 
\[P_\epsilon(-1,-1)=pq+\epsilon, \qquad P_\epsilon(-1,1)=p(1-q)-\epsilon, \]
\[P_\epsilon(1,-1)=(1-p)q-\epsilon, \qquad P_\epsilon(1,1)=(1-p)(1-q)+\epsilon.
\]
Show that for $\alpha \not=1$  and small non-zero $\epsilon$, 
\[S_\alpha({P}_\epsilon)>S_{\alpha}(P_{\epsilon, l}) + S_\alpha(P_{\epsilon, r}).\]
Hence, R\'enyi entropy is not sub-additive (compare with Theorem \ref{ax-entropy}).
\end{exo}
\section{Why is the R\'enyi entropy natural?}
\label{sec-ren-nat}
In introducing $S_\alpha({P})$ R\'enyi  was motivated by a concept of generalized means. Let $w_k>0$, $\sum_{k=1}^n w_k=1$ be 
weights and $G: ]0, \infty[\, \rightarrow\, ]0, \infty[$ a continuous strictly increasing function. We shall call such  $G$ a {\em mean function}. The $G$ -mean of strictly  positive real numbers 
$x_1, \cdots, x_n$ is 
\[ S_G(x_1, \cdots, x_n)=G^{-1}\left(\sum_{k=1}^n w_k G(x_k)\right).
\]
Set ${\cal P}_{\rm f}=\cup_{n\geq 1} {\cal P}_{n, {\rm f}}$.
 
One then has:
\bet \label{renyi-axiom} Let ${\mathfrak  S}: {\cal P}_{\rm f}\rightarrow [0, \infty[$ be a function with the following properties. 
\begin{enumerate}[{\rm (a)}]
\item If $P=P_l \otimes P_r$, then 
${\mathfrak  S}({P})={\mathfrak  S}(P_l) + {\mathfrak  S}(P_r)$.
\item There exists a mean function $G$ such that for all $n\geq 1$ and $P=(p_1, \cdots, p_n)\in 
{\cal P}_{n, {\rm f}}$, 
\[ {\mathfrak S}(p_1, \cdots, p_n)=G^{-1}\left({\mathbb E}_P(G(S_P))\right)=G^{-1}\left(\sum_{k=1}^{n} p_k G(-\log p_k)\right).
\]
\item ${\mathfrak S}(p, 1-p)\rightarrow 0$ as $p\rightarrow 0$.
\end{enumerate}
Then there exists $\alpha >0$ and a constant $c\geq 0$   such that for all $P\in P_{\rm f}$, 
\[
{\mathfrak  S}({P})= cS_\alpha({P}).
\]
\eet
\begin{remark} The assumption ({c}) excludes the possibility $\alpha \leq 0$. 
\end{remark}
\begin{remark}
 If in addition one requires that the map ${\cal P}_{n, {\rm f}} \ni P \rightarrow {\mathfrak S}({P})$ is concave for all $n\geq 1$, then 
${\mathfrak  S}({P})=cS_{\alpha}({P})$ for some  $\alpha \in ]0,1]$. 
\end{remark}

Although historically important, we find that Theorem \ref{renyi-axiom} (and any other axiomatic characterization 
of the R\'enyi entropy) is  less satisfactory then the powerful characterizations of the Boltzmann--Gibbs--Shannon 
entropy given in Section \ref{sec-e-natural}. Taking Boltzmann--Gibbs--Shannon entropy for granted, an alternative 
understanding  of the R\'enyi entropy arises through  Cram\'er's theorem for the entropy function $S_P$. For the purpose of 
this interpretation, without loss of generality we may assume that  $P\in {\cal P}(\Omega)$ is faithful.  Set 
\beq {\widehat S}_{\alpha}(P)=\log \left(\sum_{\omega\in \Omega}[P(\omega)]^{1-\alpha}\right), \qquad \alpha \in \rr.
\label{renyi-na}
\eeq
Obviously, for $\alpha \in \rr$, 
\beq \label{renyi-norm}
 {\widehat S}_{\alpha}(P)=\alpha S_{1-\alpha}(P).\eeq
The naturalness of the choice \eqref{renyi-na} stems from the fact 
that the  function $\alpha \mapsto {\widehat S}_\alpha(P)$ is the cumulant generating function of $S_P(\omega)=-\log P(\omega)$ with respect to $P$, 
\beq {\widehat S}_{\alpha}(P)= \log {\mathbb E}_P(\e^{\alpha S_P}).
\label{renyi-cl}
\eeq
Passing to the products $(\Omega^N, P_N)$, the LLN gives  that for any $\epsilon >0$,  
\beq\label{lln-entropy}
\lim_{N\rightarrow \infty}P_N\left\{\omega=(\omega_1, \cdots, \omega_N)\in \Omega^N\,\big|\, 
\left|\frac{S_P(\omega_1)+ \cdots S_P(\omega_N)}{N}- S({P})\right|\geq \epsilon\right\}=0.
\eeq
It follows from  Cram\'er's theorem that the  rate function 
\beq
I(\theta)=\sup_{\alpha \in \rr}(\alpha \theta - {\widehat S}_\alpha(P)), \qquad \theta \in \rr,
\label{cramer-renyi}
\eeq
controls  the fluctuations that accompany the limit \eqref{lln-entropy}: 
\beq
\label{ldp-ent}
\lim_{N\rightarrow \infty}\frac{1}{N}\log P_N\left\{\omega=(\omega_1, \cdots, \omega_N)\in \Omega^N\,\big|\, 
\frac{S_P(\omega_1)+ \cdots S_P(\omega_N)}{N}\in [a,b]\right\}=-\inf_{\theta \in [a,b]}I(\theta).
\eeq
We shall adopt a point of view that the relations \eqref{renyi-norm}, \eqref{cramer-renyi}, and \eqref{ldp-ent} constitute 
the foundational basis for introduction of the R\'enyi entropy. In accordance with this interpretation, the  traditional 
definition of the R\'enyi entropy is somewhat redundant, and one may as well work with $\widehat S_\alpha(P)$  from the beginning 
and call  it the {\em R\'enyi entropy} of $P$ (or $\alpha$-entropy of $P$ when there is a danger of confusion). 

The basic properties of the map $\alpha \mapsto {\widehat S}_\alpha(P)$ follow from \eqref{renyi-cl} and results 
described in Section \ref{sec-cum}. Note that $S_0(P)=0$ and $S_1(P)=\log|\Omega|$. The map 
${\cal P}_{\rm f}(\Omega) \ni P \mapsto {\widehat S}_\alpha(P)$ is convex for $\alpha \not\in [0,1]$ and 
concave for $\alpha \in ]0, 1[$.

\section{Notes and references}
\label{sec-notre-entropy}
The celebrated expression  \eqref{formula-ent} for entropy of a probability measure goes back to 1870's and  works of 
Boltzmann and Gibbs  on the foundations of statistical mechanics. This will be discussed in more detail 
in Part II of the lecture notes. Shannon has rediscovered this expression  in his work on foundations of mathematical information  theory 
\cite{Sha}. The results of Section \ref{sec-def-ent} and \ref{sec-cover} go back to 
this seminal work. Regarding Exercise \ref{hartley}, Hartley entropy was introduced in \cite{Har}. Hartley's work 
has partly motivated Shannon's \cite{Sha}.

Shannon was also first to give an axiomatization of entropy. The axioms in \cite{Sha}
are the continuity of ${\mathfrak S}$ on ${\cal P}_n$ for all $n$, the split-additivity \eqref{rain-sun}, and the monotonicity 
${\mathfrak S}(\bar P_{n+1})< {\mathfrak S}(\bar P_n)$, where $\bar P_k \in {\cal P}_k$ is the chaotic probability measures. 
Shannon then proved  that the only functions ${\mathfrak S}$ satisfying these  properties  are  $cS$, $c>0$.  Theorem 
\ref{khin} is in spirit of Shannon's axiomatization, with the monotonicity axiom ${\mathfrak S}(\bar P_{n+1})< {\mathfrak S}(\bar P_n)$ dropped and the continuity requirement relaxed; see Chapter 2 in \cite{AczDa}  for additional information 
and Theorem 2.2.3 in  \cite{Thi}  whose proof we roughly 
followed. We leave it as an exercise for the reader to simplify the proof of Theorem \ref{khin} under  additional Shannon's axioms. 

Shannon comments in \cite{Sha}
on the importance of his axiomatization  as 
\begin{quote}{\em This theorem, and the assumptions required for its proof, are in no way necessary for the present theory. It is given chiefly to lend a certain plausibility to some of our later definitions. The real justification of these definitions, however, will reside in their implications.
}
\end{quote}
The others beg to differ on its importance, and axiomatizations of entropies became an independent research direction, starting with early works 
of Khintchine \cite{Khi} and Faddeev \cite{Fadd}. Much of these efforts are summarized in the monograph \cite{AczDa}, see also 
\cite{Csi}. 

The magnificent 
Theorem \ref{ax-entropy} is due to Acz\'el, Forte, and Ng \cite{AcFoNg}. I was not able to simplify their arguments and the proof 
of Theorem \ref{ax-entropy} follows closely the original paper. The Step 7 is due to \cite{K\'at}. The proof of Theorem \ref{ax-entropy}
can be also found in \cite{AczDa}, Section 4.4.  An interesting exercise that may elucidate a line 
of thought that has   led  to  the proof of Theorem \ref{ax-entropy} is to  simplify various steps of the   the proof 
  by making additional regularity assumptions.

R\'enyi entropy has been introduced in \cite{R\'en}. Theorem \ref{renyi-axiom} was proven in  \cite{Dar}; see Chapter 5 in \cite{AczDa} for additional information.

\chapter{Relative entropy}

\section{Definition and basic properties}
\label{sec-rel-entropy}
Let $\Omega$ be a finite set and $P, Q\in {\cal P}(\Omega)$. If $P\ll Q$, the relative entropy function  of the pair 
$(P, Q)$ is defined 
for $\omega \in \supp P$ by 
\[ cS_{P|Q}(\omega)= cS_Q(\omega)- cS_P(\omega)=c\log P(\omega)-c \log Q(\omega)=c\log \Delta_{P|Q}(\omega),
\]
where $c>0$ is a constant that does not depend on $\Omega, P,Q$.
The  relative entropy of $P$ with respect to $Q$ is 
\beq S(P|Q)=  c\int_{{\rm supp} P} S_{P|Q}\d P= c \sum_{\omega \in \supp P}P(\omega)\log \frac{P(\omega)}{Q(\omega)}.
\label{def-rel-ent}
\eeq
If $P$ is not absolutely continuous with respect to $Q$ ({\sl i.e.}, $Q(\omega)=0$ and $P(\omega)>0$ for some $\omega$), 
we set 
\[S(P|Q)=\infty.\] 
The value of the constant $c$ will play no role in the sequel, and we set $c=1$. As in the case of entropy, the constant $c$ will reappear in the axiomatic 
characterizations of relative entropy (see Theorems \ref{khin-rel} and \ref{ax-rel-1}). 

Note that 
\[ S(P|P_{\rm ch})=- S({P}) +\log |\Omega|.
\]

\bep\label{pos} $S(P|Q)\geq 0$ and $S(P|Q)=0$ iff $P=Q$.
\eep
\demo  We need to consider only the case $P\ll Q$. By Jensen's inequality, 
\[\sum_{\omega \in \supp P}P(\omega)\log \frac{Q(\omega)}{P(\omega)}\leq 
\log \left(\sum_{\omega \in \supp P} Q(\omega)\right),
\]
and so 
\[
\sum_{\omega \in \supp P}P(\omega)\log \frac{Q(\omega)}{P(\omega)}\leq 0
\]
with equality iff $P=Q$. \qed 

The next result refines the previous proposition. Recall that the variational distance $d_V(P, Q)$ is defined by \eqref{var-dist}.

\bet\label{pos-fine}
\beq S(P|Q)\geq \frac{1}{2}d_V(P, Q)^2.
\label{an-s}
\eeq
The equality holds iff $P=Q$.
\eet
\demo 
We start with the elementary inequality 
\beq\label{elem}
(1+x)\log (1+x)- x \geq \frac{1}{2}\frac{x^2}{1+ \frac{x}{3}}, \qquad x\geq -1.
\eeq
This inequality obviously holds for $x=-1$, so we may assume that $x>-1$. Denote the l.h.s by $F(x)$ and the r.h.s. by $G(x)$. 
One verifies that $F(0)=F^\prime(0)= G(0)=G^\prime(0)=0$, and that 
\[ F^{\prime\prime}(x)=\frac{1}{1+x}, \qquad G^{\prime\prime}(x)=\left(1 +\frac{x}{3}\right)^{-3}.
\]
Obviously, $F^{\prime \prime}(x)> G^{\prime \prime}(x)$ for $x>-1, x\not=0$. Integrating this inequality we derive that $F^{\prime}(x)>G^\prime(x)$ for 
$x>0$ and $F^\prime(x)< G^\prime(x)$ for $x\in ]-1, 0[$. Integrating these inequalities we get $F(x)\geq G(x)$ and that 
equality holds iff $x=0$. 

We now turn to the proof of the theorem. We need only to consider the case $P\ll Q$. Set 
\[ X(\omega)=\frac{P(\omega)}{Q(\omega)}-1,\]
with the convention that $0/0=0$. Note that $\int_\Omega X \d Q=0$ and that 
\[ S(P|Q)=\int_\Omega \left( (X+1)\log (X+1) - X\right)\d Q.\]
The inequality  (\ref{elem}) implies 
\beq\label{elem1}S(P|Q)\geq \frac{1}{2}\int_\Omega \frac{X^2}{1+\frac{X}{3}}\d Q,
\eeq
with the equality  iff $P=Q$. Note that 
\[
\int_\Omega\left(1 +\frac{X}{3}\right)\d Q=1,
\]
and that Cauchy-Schwarz inequality gives 
\beq\label{elem2}
\int_\Omega \frac{X^2}{1+\frac{X}{3}}\d Q=\left(\int_\Omega\left(1 +\frac{X}{3}\right)\d Q\right)\left(\int_\Omega \frac{X^2}{1+\frac{X}{3}}\d Q\right)
\geq \left(\int_\Omega|X|\d Q\right)^2=d_V(P, Q)^2.
\eeq
Combining (\ref{elem1}) and (\ref{elem2}) we derive the statement. \qed
\begin{exo} Prove that the estimate \eqref{an-s} is the best possible  in the sense that 
\[\inf_{P\not=Q}\frac{S(P|Q)}{d_V(P, Q)^2}=\frac{1}{2}.
\]
\end{exo}

Set
\beq
\label{a-om}
{\cal A}(\Omega)=\{(P, Q)\,|\, P, Q\in {\cal P}(\Omega),\, P\ll Q\}.
\eeq
One easily verifies that ${\cal A}(\Omega)$ is a  convex subset of ${\cal P}(\Omega)\times {\cal P}(\Omega)$. Obviously, 
\[{\cal A}(\Omega)=\{(P,Q)\,|\, S(P|Q)<\infty\}.\] Note also that ${\cal P}(\Omega)\times {\cal P}_{\rm f}(\Omega)$ 
is a dense subset of ${\cal A}(\Omega)$.
\bep \label{rel-ent-paris} The map 
\[{\cal A}(\Omega) \ni (P, Q)\mapsto S(P|Q)\]
is continuous, and the map 
\beq  {\cal P}(\Omega)\times {\cal P}(\Omega)\ni (P, Q)\mapsto S(P|Q)
\label{lower}
\eeq
is lower semicontinuous.
\eep

\begin{exo} \label{f-cms-1}
Prove the above proposition. Show that if  $|\Omega|>1$ and $Q$ is a boundary point of 
${\cal P}(\Omega)$, then there is a sequence $P_n\rightarrow Q$ such that $\lim_{n\rightarrow \infty}S(P_n|Q)=\infty$. Hence, the map \eqref{lower} is 
not continuous except in the trivial case $|\Omega|=1$. 
\end{exo}
\bep
\label{convex}
The relative entropy is jointly convex: for $\lambda \in ]0,1[$ and 
$P_1, P_2, Q_1, Q_2 \in {\cal P}(\Omega)$, 
\beq S(\lambda P_1 +(1-\lambda)P_2|\lambda Q_1 +(1-\lambda) Q_2)\leq \lambda S(P_1|Q_1)+(1-\lambda)S(P_2|Q_2).
\label{convex-feb}
\eeq
Moreover, if the r.h.s. in \eqref{convex-feb} is finite, the equality holds iff for $\omega\in \supp\, Q_1\cap\, 
\supp\, Q_2$ we have $P_1(\omega)/Q_1(\omega)=P_2(\omega)/Q_2(\omega)$. 
\eep
\begin{remark} In particular, if $Q_1\perp Q_2$ and the r.h.s. in \eqref{convex-feb} is finite,  then $P_1\perp P_2$ and 
the equality holds in \eqref{convex-feb}. On the other hand, if $Q_1=Q_2=Q$  and $Q$ is faithful, 
\[
S(\lambda P_1 +(1-\lambda)P_2|Q)\leq \lambda S(P_1|Q)+(1-\lambda)S(P_2|Q).
\]
with the equality iff $P_1=P_2$. An analogous statement holds if $P_1=P_2=P$ and $P$ is faithful.
\end{remark}
\demo We recall the following basic fact:
if  $g: ]0, \infty[ \rightarrow  \rr $ is concave, then the function 
\beq  G(x, y)=x g\left(\frac{y}{x}\right)
\label{never-feb}
\eeq
is jointly concave on $]0, \infty[ \times ]0, \infty[$. Indeed, for $\lambda \in ]0,1[$, 
\beq\label{convex-surprise}
\begin{split}G(\lambda x_1 &+(1-\lambda)x_2, \lambda y_1 +(1-\lambda)y_2) \\[2mm]
&=(\lambda x_1 + (1-\lambda)x_2) g \left( 
\frac{\lambda x_1}{\lambda x_1 + (1-\lambda) x_2}\frac{y_1}{x_1} + \frac{(1-\lambda) x_2}{\lambda x_1 + (1-\lambda) x_2}\frac{y_2}{x_2} 
\right)\\[2mm]
&\geq \lambda G(x_1, y_1) + (1-\lambda)G(x_2, y_2),
\end{split}
\eeq
and  if $g$ is strictly concave, the inequality is strict unless $\frac{y_1}{x_1} =\frac{y_2}{x_2}$. 

We now turn to the proof. Without loss of generality we may assume that $P_1\ll Q_1$ and $P_2\ll Q_2$. 
One easily shows that then also $\lambda P_1 +(1-\lambda)P_2\ll \lambda Q_1 +(1-\lambda) Q_2$.
For any $\omega \in \Omega$ we have that 
\beq 
\begin{split}
\left(\lambda_1P_1(\omega) + (1-\lambda)P_2(\omega)\right)
&\log \frac{\lambda_1P_1(\omega) + (1-\lambda)P_2(\omega)}{\lambda_1Q_1(\omega) + (1-\lambda)Q_2(\omega)}\\[2mm]
&\leq \lambda P_1(\omega)\log \frac{P_1(\omega)}{Q_1(\omega)} + (1-\lambda)P_2(\omega)\log \frac{P_2(\omega)}{Q_2(\omega)}.
\label{foo-feb}
\end{split}
\eeq
To establish this relation, note that if   $P_1(\omega)=P_2(\omega)=0$,  then \eqref{foo-feb} holds with the equality. If $P_1(\omega)=0$ and $P_2(\omega)>0$, 
the inequality \eqref{foo-feb} is strict unless $Q_1(\omega)=0$, and similarly in the case $P_1(\omega)>0$, $P_2(\omega)=0$.
If $P_1(\omega)>0$ and $P_2(\omega)>0$, 
then taking $g(t)=\log t$ in \eqref{never-feb} and using the joint concavity of $G$ gives that \eqref{foo-feb} holds and that 
the inequality is strict unless $P_1(\omega)/Q_1(\omega)= P_2(\omega)/Q_2(\omega)$. Summing 
\eqref{foo-feb} over $\omega$ we derive the statement. The discussion of the cases where the equality holds in 
\eqref{convex-feb} is simple and is left to the reader.
\qed

The relative entropy is super-additive in the following sense:
\bep For any $P$ and $Q=Q_l\otimes Q_r$ in ${\cal P}(\Omega_l\times \Omega_r)$, 
\beq  S(P_l| Q_l) + S(P_r|Q_r)\leq S(P|Q).
\label{sup-add}
\eeq
Moreover, if the r.h.s. in  \eqref{sup-add} is finite, the equality holds iff $P=P_l\otimes P_r$.
\eep
\demo We may assume that $P\ll Q$, in which case one easily verifies that $P_l\ll Q_l$ and $P_r\ll Q_r$. 
One computes 
\[
S(P|Q)-S(P_l| Q_l) - S(P_r|Q_r)= S(P_l) + S(P_r)- S(P),
\]
and the result follows from Proposition \ref{e-subad}. \qed

In general, for $P, Q\in {\cal P}(\Omega_l\times \Omega_r)$ it is {\em not} true that 
$S(P|Q)\geq S(P_l|Q_l) + S(P_r|Q_r)$ even if $P=P_l\otimes P_r$. 
\begin{exo} Find an example of faithful $P=P_l\otimes P_r, Q\in {\cal P}(\Omega_l\times \Omega_r)$  where 
$|\Omega_l|=|\Omega_r|=2$ such that 
\[ S(P|Q)< S(P_l|Q_l) + S(P_r|Q_r).\]
\end{exo}

Let $\Omega=(\omega_1, \cdots, \omega_L)$, $\widehat \Omega=\{\hat \omega_1, \cdots, \hat \omega_{\hat L}\}$ be two finite 
sets. A matrix of real numbers $[\Phi(\omega, \hat \omega)]_{(\omega, \hat \omega)\in \Omega\times \widehat \Omega}$ 
is called {\em stochastic}  if $\Phi(\omega, \hat \omega)\geq 0$ for all pairs $(\omega, \hat \omega)$ and  
\[\sum_{\hat \omega\in \widehat \Omega}\Phi(\omega, \hat \omega)=1\]
for all $\omega\in \Omega$.   A stochastic matrix induces a map $\Phi: {\cal P}( \Omega)\rightarrow  {\cal P}(\widehat \Omega)$ by 
\[
\Phi({P})(\hat \omega)=\sum_{\omega\in  \Omega}P( \omega)\Phi(\omega, \hat \omega).
\]
We shall refer to $\Phi$ as the {\em stochastic map} induced by the stochastic matrix $[\Phi(\omega, \hat\omega)]$. 
One can interpret the elements of $\Omega$ and $\widehat \Omega$ as states of two  stochastic systems and $P(\omega)$ as probability that the state $\omega$  is realized. 
$\Phi(\omega, \hat \omega)$ is interpreted as the  {\em transition probability}, i.e. the probability that in a unit of time the system will 
make a transition from the state $\omega$ to the state $\hat \omega$. With this interpretation, the probability that the state $\hat \omega$ is realized 
after the transition has taken place is $\Phi({P})(\hat \omega)$. 

Note that if $[\Phi(\omega, {\hat \omega})]_{(\omega, \hat \omega)\in \Omega\times \widehat \Omega}$ and 
$[\widehat \Phi(\hat \omega, \hat {\hat \omega})]_{(\hat \omega, \hat{\hat \omega})\in \widehat \Omega\times\widehat{ \widehat \Omega}}$ are stochastic 
matrices, then their product is also stochastic matrix and that the induced stochastic map is $\widehat \Phi \circ \Phi$. 
Another elementary property of stochastic maps is:
\bep $d_V(\Phi(P), \Phi(Q))\leq d_V(P, Q)$.
\label{var-sto}
\eep
\begin{exo} Prove Proposition \ref{var-sto}. When the equality holds?
\end{exo}

The following result is deeper.
\bep 
\beq\label{data-pro} S(\Phi({P})|\Phi(Q))\leq S(P|Q).
\eeq
\label{data-sherry}
\eep
\begin{remark} In information theory, the inequality (\ref{data-pro}) is sometimes called  the {\em data processing 
inequality}. We shall refer to it as  the {\em stochastic monotonicity}. If the relative entropy is interpreted as a measure of {\em distinguishability} of two probability measures, then 
the inequality asserts that probability measures are less distinguishable after an application of a stochastic map. 
\end{remark}

\demo We start with the so called {\em log-sum} inequality: If $a_j, b_j$, $j=1, \cdots, M,$ are non-negative numbers, then 
\beq\label{log-sum}
\sum_{j=1}^M a_j \log \frac{a_j}{b_j}\geq \sum_{j=1}^M a_j \log\frac{\sum_{k=1}^M a_k}{\sum_{k=1}^M b_k}, 
\eeq
with the usual convention that $0 \log 0/x=0$. If $b_j=0$ and $a_j>0$ for some $j$, then l.h.s is $\infty$ and there is nothing 
to prove.  If $a_j=0$ for all $j$ 
again there is nothing to prove. Hence, without loss 
of generality we may assume that $\sum_j a_j >0$, $\sum b_j >0$, and $b_j=0 \Rightarrow a_j=0$. 
 Set $p=(p_1, \cdots, p_M)$, $p_k =a_k/\sum_j a_j$, 
$q=(q_1, \cdots, q_M)$, $q_k=b_k/\sum_j b_j$. Then the inequality (\ref{log-sum}) is equivalent to 
\[
S(p|q)\geq 0.
\]
This observation and Proposition \ref{pos} prove (\ref{log-sum}).

We now turn to the proof. Clearly, we need only to consider 
the case $P\ll Q$. Then 
\[
\begin{split}
S(\Phi({P})|\Phi(Q))&=\sum_{\hat \omega\in \hat \Omega}\Phi({P})(\hat \omega)\log \frac{\Phi({P})(\hat \omega)}{\Phi(Q)(\hat \omega)}\\[2mm]
&=\sum_{\hat \omega\in \hat \Omega}\sum_{\omega \in \Omega}P(\omega)\Phi(\omega, \hat \omega)\log 
\frac{\sum_{\omega^{\prime}\in \Omega} P(\omega^{\prime})\Phi(\omega^{\prime}, \hat \omega)}
{\sum_{\omega^{\prime}\in \Omega} Q(\omega^{\prime})\Phi(\omega^{\prime}, \hat \omega)}\\[2mm]
&\leq\sum_{\hat \omega \in\hat  \Omega} \sum_{\omega \in \Omega}P(\omega)\Phi(\omega, \hat \omega)\log \frac{P(\omega)}{Q(\omega)}
\\[2mm]
&=S(P|Q),
\end{split}
\]
where the third step follows from the log-sum inequality. \qed
 \qed
\begin{exo} \label{stoc-ent}
A stochastic matrix $[\Phi(\omega, \hat \omega)]$  is 
called doubly stochastic if 
\[\sum_{\omega\in \Omega}\Phi(\omega, \hat\omega)=\frac{|\Omega|}{| \hat \Omega|}\]
 for all $\hat \omega\in \hat \Omega$. Prove that 
$S(P)\leq S(\Phi(P))$ for all $P\in {\cal P}(\Omega)$ iff $[\Phi(\omega,\hat  \omega)]$ is doubly stochastic. \newline
Hint: Use that $\Phi(P_{\rm ch})= \hat P_{\rm ch}$ iff $[\Phi(\omega, \hat \omega)]$ is doubly stochastic.
\label{sto-e-f}
\end{exo}

\begin{exo} Suppose that $\Omega=\hat \Omega$. Let $\gamma=\min_{(\omega_1, \omega_2)} \Phi(\omega_1, \omega_2)$ and suppose 
that $\gamma >0$. 
\exop Show that $S(\Phi({P})|\Phi(Q))= S(P|Q)$ iff $P=Q$. 
\exop Show that 
\[ d_V(\Phi({P}), \Phi(Q))\leq(1-\gamma)d_V(P, Q).
\]
\exop Using Part 2 show that there exists unique probability measure $\bar Q$ such that $\Phi(\bar Q)=\bar Q$. 
Show that $\bar Q$ is faithful and that for any $P\in {\cal P}(\Omega)$, 
\[ d_V(\Phi^n({P}), \bar Q)\leq (1-\gamma)^nd_V(P, \bar Q),
\]
where $\Phi^2=\Phi \ \circ \Phi$, etc. 
\newline
Hint:  Follow the proof of the Banach fixed point theorem.
\end{exo}
\begin{exo} The stochastic monotonicity yields the following elegant proof of 
Theorem \ref{pos-fine}. 
\exop Let $P, Q\in {\cal P}(\Omega)$ be given, where $|\Omega|\geq 2$. Let  $T=\{\omega: P(\omega)\geq Q(\omega)\}$ and 
\[p=(p_1, p_2)=(P(T), P(T^c)), \qquad q=(q_1, q_2)=(Q(T), Q(T^c)),\]
 be  probability measures on $\widehat \Omega=\{1,2\}$. Find  a stochastic map $\Phi: {\cal P}(\Omega)\rightarrow 
{\cal P}(\widehat \Omega)$ such that $\Phi(P)=p$, $\Phi(Q)=q$.
\exop Since $S(P|Q)\geq S(p|q)$ and $d_V(P,Q)=d_V(p, q)$, observe that to prove Theorem \ref{pos-fine} 
 it suffices to show that for all 
$p, q\in {\cal P}(\widehat \Omega)$, 
\beq 
S(p|q)\geq \frac{1}{2}d_V(p,q)^2.
\label{an-observe}
\eeq
\exop Show that  \eqref{an-observe} is equivalent to the inequality
\beq 
x\log \frac{x}{y} + (1-x)\log \frac{1-x}{1-y}\geq 2(x-y)^2,
\label{an-recast}
\eeq
where $0\leq y\leq x\leq 1$. Complete the proof by establishing \eqref{an-recast}.\newline
Hint: Fix $x>0$ and consider the function 
\[F(y)=x\log \frac{x}{y} + (1-x)\log \frac{1-x}{1-y}- 2(x-y)^2\]
on $]0, x]$. Since  $F(x)=0$, it suffices to show that 
$F^\prime(y)\leq 0$ for $y \in ]0, x[$. Direct computation gives $F^\prime(y)\leq 0  \Leftrightarrow y(1-y)\leq \frac{1}{4}$ 
and the statement follows. 

\label{csi-ex}
\end{exo}
The log-sum inequality used in the proof Proposition \ref{data-sherry} leads to the following refinement 
of Proposition \ref{convex}.
\bep \label{convex-rel} Let $P_1, \cdots, P_n, Q_1, \cdots, Q_n\in {\cal P}(\Omega)$ and $p=(p_1, \cdots, p_n), q=(q_1, \cdots, q_n)
\in {\cal P}_n$. Then
\beq 
\label{convex-general}
S(p_1P_1+ \cdots+ p_nP_n|q_1Q_1+\cdots+ q_nQ_n)\leq p_1S(P_1|Q_1) + \cdots+ p_nS(P_n|Q_n) + S(p|q).
\eeq
If the r.h.s. in \eqref{convex-general} is finite, then the equality holds iff for all $j,k$ such that $q_j>0, q_k>0$, 
\[
\frac{p_jP_j(\omega)}{q_j Q_j(\omega)}=\frac{p_kP_k(\omega)}{q_k Q_k(\omega)}
\]
holds for all $\omega \in \supp\, Q_k\cap \supp \,Q_j$.
\eep
\begin{exo} Deduce Proposition \ref{convex-rel} from the log-sum inequality. 
\end{exo}
\section{Variational principles}
\label{sec-var-pri}
The relative entropy is characterized by the following variational principle.
\bep \label{var-prin-10}
\beq\label{var-prin}
S(P|Q)= \sup _{X:\Omega \rightarrow \rr} \left(\int_\Omega X\d P - \log \int_{\supp P} \e^X \d Q\right).
\eeq
If $S(P|Q)<\infty$, then the  supremum is achieved, and each maximizer is equal to  $S_{P|Q} + {\rm const}$ on $\supp P$ and is
arbitrary otherwise.
\eep 
\demo Suppose that $Q(\omega_0)=0$ and $P(\omega_0)>0$ for some $\omega_0 \in \Omega$. Set 
$X_n(\omega)=n$ if $\omega=\omega_0$ and zero otherwise. Then 
\[ \int_\Omega X_n \d P = nP(\omega_0), \qquad \int_{\supp P}\e^{X_n}\d Q= Q(\supp P).\]
Hence, if $P$ is not absolutely continuous w.r.t. $Q$ the relation  (\ref{var-prin}) holds since both sides are equal to $\infty$. 

Suppose now that $P\ll Q$. For given $X:\Omega \rightarrow \rr$ set 
\[ Q_X(\omega)= \frac{\e^{X(\omega)}Q(\omega)}{\sum_{\omega^\prime \in \supp P}\e^{X(\omega^\prime)}Q(\omega^\prime)}
\]
if $\omega \in \supp P$ and zero otherwise. $Q_X\in {\cal P}(\Omega)$ and 
\[
S(P|Q_X)=S(P|Q)- \left(\int_\Omega X\d P - \log \int_{\supp P} \e^X \d Q\right).
\]
Hence, 
\[
S(P|Q)\geq \int_\Omega X\d P - \log \int_{\supp P} \e^X \d Q
\]
with equality iff $P=Q_X$. Obviously, $P=Q_X$ iff $X=S_{P|Q} + {\rm const}$ on $\supp P$ and is arbitrary 
otherwise. \qed
\begin{exo} Show that 
\beq S(P|Q)= \sup _{X:\Omega \rightarrow \rr} \left(\int_\Omega X\d P - \log \int_\Omega \e^X \d Q\right).
\label{no-ho}
\eeq
When is the supremum achieved? Use  \eqref{no-ho} to prove that the map $(P, Q)\mapsto S(P|Q)$ is jointly 
convex.
\label{exe-var}
\end{exo}

\bep \label{var-prin-20} The following dual variational principle holds: for $X:\Omega \rightarrow \rr$ and 
$Q\in {\cal P}(\Omega)$, 
\[\log \int_\Omega \e^X \d Q=\max_{P \in {\cal P}(\Omega)}\left(\int_\Omega X\d P- S(P|Q)\right).\]
The maximizer is unique and is given by 
\[P_{X, Q}(\omega)=\frac{\e^{X(\omega)}Q(\omega)}{\sum_{\omega^\prime\in \Omega}\e^{X(\omega^\prime)}Q(\omega^\prime)}.
\]
\eep
\demo For any $P\ll Q$,
\[
\log \int_\Omega \e^X \d Q-\int_\Omega X\d P+ S(P|Q)= S(P|P_{X, Q}),
\]
and the result follows from Proposition \ref{pos}. \qed

Setting $Q=P_{\rm ch}$ in Propositions \ref{var-prin-10} and \ref{var-prin-20}, we derive the variational 
principle for entropy and the respective  dual variational principle.
\bep
\begin{enumerate}[{\rm (1)}]
\item
\[ S({P}) = \inf_{X: \Omega\rightarrow \rr}\left( \log \left(\sum_{\omega \in \Omega}\e^{X(\omega)}\right) - \int_\Omega X\d P\right).\]
The infimum   is achieved if $P$ is faithful  and $X=-S_{P} + {\rm const}$.
\item For any $X: \Omega \rightarrow \rr$, 
\[ \log \left(\sum_{\omega \in \Omega}\e^{X(\omega)}\right)= \max_{P\in {\cal P}(\Omega)}\left( 
\int_\Omega X\d P + S({P})\right).
\]
The maximizer is unique and is given by 
\[P(\omega)=\frac{\e^{X(\omega)}}{\sum_{\omega^\prime\in \Omega}\e^{X(\omega^\prime)}}.\]
\end{enumerate}
\eep


\section{Stein's Lemma}
Let $P, Q\in {\cal P}(\Omega)$ and let $P_N, Q_N$ be the induced product 
probability measures on $\Omega^N$. For  $\gamma \in ]0,1[$ the Stein exponents are defined by 
\beq\label{stein-exp} s_N(\gamma)=\min\left\{ Q_N(T)\,|\, T\subset \Omega^N, \, P_N(T)\geq \gamma\right\}.
\eeq
The following result is often called  {\em Stein's Lemma}.
\bet \label{stein-lemma}
\[\lim_{N\rightarrow \infty}\frac{1}{N}\log s_N(\gamma)=- S(P|Q).\]
\eet
\begin{remark} If $Q=P_{\rm ch}$, then  Stein's Lemma  reduces to Proposition \ref{how-to}. In fact, the proofs of the
two results are very similar. 
\end{remark}

\demo We deal first with the case $S(P|Q)<\infty$. Set $S_{P|Q}(\omega)=0$ for $\omega \not\in \supp P$ and 
\[{\cal S}_N(\omega=(\omega_1, \cdots, \omega_N))=\sum_{j=1}^N S_{P|Q}(\omega_j).\] 
For given $\epsilon >0$ let 
\[ R_{N, \epsilon}= \left\{\omega \in \Omega^N\,\big|\, \frac{S_N(\omega)}{N}\geq S(P|Q)-\epsilon\right\}.
\]
By the LLN, 
\[
\lim_{N\rightarrow \infty}P_N(R_{N, \epsilon})=1,
\]
and so for $N$ large enough, $P_N(R_{N, \epsilon})\geq \gamma$. 
We also have 
\[
Q_N(R_{N, \epsilon})=Q_N\left\{\e^{ {\cal S}_N(\omega)}\geq \e^{N S(P|Q)- N\epsilon}\right\}\leq 
\e^{N\epsilon - N S(P|Q)} {\mathbb E}_{Q_N}(\e^{{\cal S}_N}).
\]
Since 
\[ {\mathbb E}_{Q_N}(\e^{{\cal S}_N})= \left(\int_\Omega \Delta_{P|Q}\d Q\right)^N=1,
\]
we derive
\[
\limsup_{N\rightarrow \infty}\frac{1}{N}\log s_N(\gamma)\leq -S(P|Q) +\epsilon.
\]
Since $\epsilon>0$ is arbitrary, 
\[
\limsup_{N\rightarrow \infty}\frac{1}{N}\log s_N(\gamma)\leq -S(P|Q).
\]
To prove the lower bound, let $U_{N,\gamma}$ be the set for which the  minimum in (\ref{stein-exp}) is achieved. 
Let $\epsilon >0$ be given and let 
 \[ D_{N, \epsilon}= \left\{\omega \in \Omega^N\,\big|\, \frac{S_N(\omega)}{N}\leq  S(P|Q)+\epsilon\right\}.
\]
Again, by  the LLN, 
\[
\lim_{N\rightarrow \infty}P_N(D_{N, \epsilon})=1,
\]
and so for $N$ large enough, $P_N(D_{N, \epsilon})\geq \gamma$. We then have 
\[
\begin{split}
P_{N}(U_{N,\gamma}\cap D_{N, \epsilon})&=\int_{U_{N,\gamma}\cap D_{N, \epsilon}}\Delta_{P_N|Q_N}\d Q_N
=\int_{U_{N,\gamma}\cap D_{N, \epsilon}}\e^{{\cal S}_N}\d Q_N\\[2mm]
&\leq \e^{NS(P|Q) + N\epsilon} Q_N(U_{N,\gamma}\cap D_{N, \epsilon})\\[2mm]
&\leq \e^{NS(P|Q) + N\epsilon} Q_N(U_{N,\gamma}).
\end{split}
\]
Since 
\[\liminf_{N\rightarrow \infty}P_N(U_{N,\gamma}\cap D_{N, \epsilon})\geq \gamma,
\]
we have 
\[
\liminf_{N\rightarrow \infty}\frac{1}{N}s_N(\gamma)\geq -S(P|Q)-\epsilon.
\]
Since $\epsilon >0$ is arbitrary, 
\[
\liminf_{N\rightarrow \infty}\frac{1}{N}s_N(\gamma)\geq -S(P|Q).
\]
This proves  Stein's Lemma in the case $S(P|Q)<\infty$. 

We now deal with the case $S(P|Q)=\infty$. For $0 <\delta <1$ set 
$Q_\delta = (1-\delta)Q + \delta P$. Obviously, $S(P|Q_\delta)<\infty$.  Let $s_{N, \delta}(\gamma)$ be the Stein 
exponent of the pair $(P, Q_\delta)$. Then 
\[
s_{N, \delta}(\gamma)\geq (1-\delta)^N s_{N}(\gamma),
\]
and
\[-S(P|Q_\delta)=\lim_{N \rightarrow \infty}\frac{1}{N}\log s_{N, \delta}(\gamma)\geq \log (1-\delta) +
\liminf_{N\rightarrow \infty}\frac{1}{N}\log s_{N}(\gamma).
\]
The lower semicontinuity  of  relative entropy gives $\lim_{\delta\rightarrow 0} S(P|Q_\delta)=\infty$, and so 
\[
\lim_{N\rightarrow \infty}\frac{1}{N}\log s_{N}(\gamma)=\infty=-S(P|Q).
\]
\qed
\begin{exo} \label{stein-diff}Prove the following variant of Stein's Lemma. Let 
\[
\begin{split}
\ubar  s&=\inf_{(T_N)}\left\{\liminf_{N\rightarrow \infty}\frac{1}{N}Q_N(T_N)\,|\, \lim_{N\rightarrow \infty}P_N(T_N^c)=0\right\},\\[3mm]
\bar s &=\inf_{(T_N)}\left\{\limsup_{N\rightarrow \infty}\frac{1}{N}Q_N(T_N)\,|\, \lim_{N\rightarrow \infty}P_N(T_N^c)=0\right\},
\end{split}
\]
where the infimum is taken over all sequences $(T_N)_{N\geq 1}$ of sets such that $T_N\subset \Omega^N$ for all $N\geq 1$.
Then 
\[\ubar s=\bar s=-S(P|Q).
\]
\label{Stein-0}
\end{exo}

\section{Fluctuation relation}
\label{sec-fluct-rel}
Let $\Omega$ be a finite set and $P\in {\cal P}_{\rm f}(\Omega)$. Let 
$\Theta:\Omega \rightarrow \Omega$ be a bijection such that 
\beq\Theta^2(\omega)=\Theta\circ \Theta(\omega)=\omega
\label{cambridge-tired}
\eeq
for all $\omega$. We set $P_\Theta(\omega)=P(\Theta(\omega))$. Obviously, $P_\Theta \in {\cal P}_{\rm f}(\Omega)$.
The relative entropy function
\[S_{P|P_\Theta}(\omega)=\log \frac{P(\omega)}{P_\Theta(\omega)}\]
satisfies 
\beq
\label{cambridge-sunny}
S_{P|P_\Theta}(\Theta(\omega))=-S_{P|P_\Theta}(\omega),
\eeq
and so the set of values of $S_{P|P_\Theta}$is  symmetric with respect to the origin. On the other hand, 
\[
S(P|P_{\Theta})= {\mathbb E}_P(S(P|P_\Theta))\geq 0
\]
with equality iff $P=P_\Theta$. Thus, the probability measure $P$ "favours" positive values of $S_{P|P_\Theta}$. 
Proposition \ref{fluct-rel} below is a refinement of this  observation.

Let $Q$ be the probability distribution of the random variable $S(P|P_\Theta)$ w.r.t. $P$. We recall that $Q$ is defined by 
\[Q(s)=P\left\{ \omega\,|\, S_{P|P_\Theta}(\omega)=s\right\}.\]
Obviously, $Q(s)\not=0$ iff $Q(-s)\not=0$.

The following result is known as the  {\em fluctuation relation}.
\bep  
\label{fluct-rel}
For all $s$, 
\[Q(-s)= \e^{-s}Q(s).\]
\eep

\demo For any $\alpha$, 
\[ 
\begin{split}
{\mathbb E}_P\left(\e^{-\alpha S_{P|P_\Theta}}\right)&=\sum_{\omega \in \Omega}[P_\Theta(\omega)]^{\alpha}[P(\omega)]^{1-\alpha}\\[2mm]
&=\sum_{\omega \in \Omega}[P_\Theta(\Theta(\omega))]^{\alpha}[P(\Theta(\omega))]^{1-\alpha}\\[2mm]
&=\sum_{\omega \in \Omega}[P(\omega)]^{\alpha}[P_\Theta(\omega)]^{1-\alpha}\\[2mm]
&={\mathbb E}_P\left(\e^{-(1-\alpha) S_{P|P_\Theta}}\right).
\end{split}
\]
Hence, if ${\cal S}=\{ s\, |\, Q(s)\not=0\}$, 
\[
\sum_{s\in {\cal S}}\e^{-\alpha s}Q(s)=\sum_{s\in {\cal S}}\e^{-(1-\alpha)s} Q(s)= \sum_{s\in {\cal S}} \e^{(1-\alpha)s} Q(-s),
\]
and so 
\beq
\label{cambridge-sat}
\sum_{s\in {\cal S}}\e^{-\alpha s}(Q(s)- \e^s Q(-s))=0.
\eeq
Since (\ref{cambridge-sat}) holds for all real $\alpha$, we must have that $Q(s)- \e^s Q(-s)=0$ for all $s\in {\cal S}$, and 
the statement follows. \qed

\begin{remark}
The  assumption that $P$ is faithful can be omitted if one assumes in addition that $\Theta$ preserves $\supp P$. If this is the case, one can replace $\Omega$ with $\supp P$, and the above proof applies. 
\end{remark}

\begin{exo} Prove that the fluctuation relation implies (\ref{cambridge-sunny}).
\end{exo}

\begin{exo}
 This exercise  is devoted to a generalization of the fluctuation relation which has also 
found fundamental application in physics. Consider a family $\{P_X\}_{X\in \rr^n}$ of probability measures on $\Omega$ indexed by 
vectors $X=(X_1, \cdots, X_n)\in \rr^n$.  Set 
\[
{\cal E}_X(\omega)=\log\frac{ P_X(\omega)}{P_X(\Theta_X(\omega))},
\]
where $\Theta_X$  satisfies (\ref{cambridge-tired}). 
Suppose that  ${\cal E}_0=0$ and consider a decomposition 
\beq
{\cal E}_X=\sum_{k=1}^n X_k {\mathfrak F}_{X, k},
\label{cambridge-linear}
\eeq
where the random variables ${\mathfrak  F}_{X, k}$ satisfy 
\beq
{\mathfrak F}_{X,k}\circ \Theta_X=-{\mathfrak  F}_{X, k}.
\label{cambridge-linear-1}
\eeq
We denote by ${\cal Q}_X$ the probability distribution of the vector random variable $({\mathfrak  F}_{X,1}, \cdots, {\mathfrak  F}_{X, n})$ with respect to 
$P_X$: for $s=(s_1, \cdots, s_n)\in \rr^n$, 
\[Q_X(s)=P_X\left\{ \omega\in \Omega\,|\, {\cal F}_{X, 1}=s_1, \cdots, {\cal F}_{X, n}=s_n\right\}.
\]
We also denote ${\cal S}=\{ s\in \rr^n \, |\, Q_X(s)\not=0\}$ and, for $Y=(Y_1, \cdots, Y_n)\in \rr^n$,  set 
\[ G(X, Y)=\sum_{s\in {\cal S}}\e^{-\sum_k s_k Y_k}Q_X(s).
\]
\exop Prove that a decomposition (\ref{cambridge-linear}) satisfying (\ref{cambridge-linear-1}) always exists and that, except 
in trivial cases, is never unique.
\exop Prove that $Q_X(s)\not=0$ iff $Q_X(-s)\not=0$. 
\exop Prove that 
\[G(X, Y)=G(X, X-Y).\]
\exop Prove that 
\[ Q_X(-s)=\e^{-\sum_k s_k X_k}Q_X(s).\]
\end{exo}
{
\section{Jensen-Shannon entropy and metric}
The Jensen-Shannon entropy of two probability measures $P, Q\in {\cal P}(\Omega)$ is 
\[
\begin{split}
S_{ JS}(P|Q)&=S(M(P, Q))- \frac{1}{2}S({P})- \frac{1}{2}S(Q)\\[2mm]
&=
\frac{1}{2}\left( S\left(P|M(P, Q)\right)+  S\left(Q|M(P,Q)\right)\right),
\end{split}
\]
where 
\[M(P,Q)=\frac{P+Q}{2}.
\]
The Jensen-Shannon entropy can be viewed as a measure of concavity of the entropy. 
Obviously, $S_{ JS}(P|Q)\geq 0$ with equality iff $P=Q$. In addition:
\bep \begin{enumerate}[{\rm (1)}]
\item 
\[ S_{JS}(P|Q)\leq \log 2,
\]
with equality  iff $P\perp Q$.
\item 
\[ \frac{1}{8}d_V(P, Q)^2 \leq S_{\rm JS}(P|Q)\leq d_V(P, Q)\log \sqrt 2.
\]
The first inequality is saturated iff $P=Q$ and the second iff $P=Q$ or $P\perp Q$.
\end{enumerate}
\eep
\demo  Part (1) follows from 
\[
\begin{split}
S_{JS}(P|Q)&=\frac{1}{2}\sum_{\omega \in \Omega}\left(P(\omega)\log\left(
\frac{2P(\omega)}{P(\omega) + Q(\omega)}\right) + Q(\omega)\log\left(
\frac{2Q(\omega)}{P(\omega) + Q(\omega)}\right)\right)\\[2mm]
&\leq \frac{1}{2}\sum_{\omega \in \Omega}(P(\omega) + Q(\omega))\log 2\\[2mm]
&=\log 2.
\end{split}
\]

To prove (2), we start with the lower bound:
\[
\begin{split}
S_{JS}(P|Q) &=\frac{1}{2}S(P|M(P,Q))+ \frac{1}{2} S(Q|M(P|Q))\\[2mm]
&\geq \frac{1}{4}d_V(P,M(P,Q))^2 + \frac{1}{4}d_V(Q, M(P,Q))^2\\[2mm]
&=\frac{1}{8}\left(\sum_{\omega \in \Omega}|P(\omega)-Q(\omega)|\right)^2= \frac{1}{8}d_V(P|Q)^2,
\end{split}
\]
where the inequality follows from Theorem \ref{pos-fine}. 

To prove the upper bound, set 
$S_+=\{\omega\,|\, P(\omega)\geq Q(\omega)\}$, $S_-=\{\omega\,|\, P(\omega)<Q(\omega)\}$.  Then 
\[
\begin{split}
S_{JS}(P|Q)&=
\frac{1}{2}\sum_{\omega \in S_+}\left(P(\omega)\log\left(
\frac{2P(\omega)}{P(\omega) + Q(\omega)}\right) - Q(\omega)\log\left(
\frac{P(\omega) + Q(\omega)}{2 Q(\omega)}\right)\right)\\[2mm]
&\qquad + \frac{1}{2}\sum_{\omega \in S_-}\left(Q(\omega)\log\left(
\frac{2Q(\omega)}{P(\omega) + Q(\omega)}\right) - P(\omega)\log\left(
\frac{P(\omega) + Q(\omega)}{2 P(\omega)}\right)\right) \\[2mm]
&\leq \frac{1}{2}\sum_{\omega \in S_+}(P(\omega)-Q(\omega))\log\left(\frac{2P(\omega)}{P(\omega) + Q(\omega)}\right)\\[2mm]
&\qquad + \frac{1}{2}\sum_{\omega \in S_-}(Q(\omega)-P(\omega))\log\left(\frac{2Q(\omega)}{P(\omega) + Q(\omega)}\right)\\[2mm]
&\leq\frac{1}{2} \sum_{\omega\in S_-}(P(\omega)-Q(\omega))\log 2 + \frac{1}{2}\sum_{\omega\in S_-}(Q(\omega)-P(\omega))\log 2\\[2mm]
&= d_V(P,Q)\log \sqrt 2.
\end{split}
\]
In the first inequality we have used that for $P(\omega)\not=0$ and $Q(\omega)\not=0$, 
\[
\frac{P(\omega)+ Q(\omega)}{2P(\omega)}\geq \frac{2Q(\omega)}{P(\omega) + Q(\omega)},
\]
and the same inequality with $P$ and $Q$ interchanged. 

The cases where equality holds  in Parts (1) and  (2)  are easily identified from the above 
argument and we leave the formal proof as  an exercise for the reader. \qed

Set
\[d_{JS}(P, Q)= \sqrt{S_{ JS}(P, Q)}.
\]

\bet $d_{JS}$ is a metric on ${\cal P}(\Omega)$.
\label{sj-metric}
\eet
\begin{remark}If $|\Omega|\geq 2$, then $S_{JS}$ is not a metric on ${\cal P}(\Omega)$. To see that, pick $\omega_1, \omega_2\in \Omega$ and 
define $P, Q, R\in {\cal P}(\Omega)$ by  $P(\omega_1)=1$, $Q(\omega_2)=1$, $R(\omega_1)=R(\omega_2)=\frac{1}{2}$. Then 
\[S_{JS}(P|Q)=\log 2 >\frac{3}{2}\log \frac{4}{3}=S_{JS}(P|R) + S_{JS}(R|Q).\]
\end{remark}
\begin{remark}
In the sequel we shall refer to $d_{SJ}$ as the {\em Jensen-Shannon} metric.
\end{remark}

\demo Note that only the triangle inequality needs to be proved. Set $\rr_+=]0, \infty[$. 

For $p, q\in \rr_+$ let 
\[ L(p,q)=p\log\left(\frac{2p}{p+q}\right) + q\log \left(\frac{2q}{p+q}\right). \]
Since the function $F(x)=x\log x$ is strictly convex, writing 
\[
L(p, q)=(p+q)\left[ \frac{1}{2}F\left(\frac{2p}{p+q}\right) +  \frac{1}{2}F\left(\frac{2q}{p+q}\right)\right]
\]
and applying the Jensen inequality to the expression in the brackets, we derive that $L(p,q)\geq 0$ with 
 equality iff $p=q$. Our goal is to prove that for all $p,q,r\in \rr_+$, 
\beq\label{ma-sj}
L(p,q)\leq \sqrt{L(p,r)} +\sqrt{L(r,q)}.
\eeq
This yields the triangle inequality for $d_{JS}$ as follows. If $P, Q, R\in {\cal P}_{\rm f}(\Omega)$, 
\eqref{ma-sj} and Minkowski's inequality give
\[
\begin{split}
d_{JS}(P, Q)&=\left(\sum_{\omega \in \Omega}\sqrt{L(P(\omega), Q(\omega))}^2\right)^{\frac{1}{2}}\\[2mm]
&\leq \left(\sum_{\omega \in \Omega}\left(\sqrt{L(P(\omega), R(\omega))} + \sqrt{L(R(\omega), Q(\omega)})\right)^2\right)^{\frac{1}{2}}\\[2mm]
&\leq \left(\sum_{\omega \in \Omega}\sqrt{L(P(\omega), R(\omega))}^2\right)^{\frac{1}{2}} + \left(\sum_{\omega \in \Omega}\sqrt{L(R(\omega), Q(\omega))}^2\right)^{\frac{1}{2}}\\[2mm]
&=d_{JS}(P,R) + d_{JS}(R, Q).
\end{split}
\]
This yields the triangle inequality on ${\cal P}_{\rm f}(\Omega)$. Since the map $(P, Q)\mapsto d_{JS}(P, Q)$ is continuous, 
the triangle inequality extends to ${\cal P}(\Omega)$. 

The proof of \eqref{ma-sj} is an elaborate calculus exercise. The relation is obvious if $p=q$. Since $L(p,q)=L(q,p)$, it 
suffices to consider the case $p<q$. We fix such $p$ and $q$ and set 
\[
f(r)= \sqrt{L(p,r)} +\sqrt{L(r,q)}.
\]
Then 
\[
f^\prime(r)=\frac{1}{2\sqrt{L(p,r)}}\log \left(\frac{2r}{p+r}\right) + \frac{1}{2\sqrt{L(r,q)}}\log \left(\frac{2r}{r+q}\right). 
\]
Define $g: \rr_+\setminus\{1\}\rightarrow \rr$ by 
\[ g(x)=\frac{1}{\sqrt{L(x,1)}}\log \left(\frac{2}{x+1}\right),
\]
One easily verifies that 
\beq
f^\prime(r)=\frac{1}{2\sqrt{r}} \left(g\left(\frac{p}{r}\right)+ g\left(\frac{q}{r}\right)\right).
\label{fi-da}
\eeq
We shall need the following basic properties of $g$, clearly displayed in the above graph:
\begin{figure}
\centering
\begin{tikzpicture}
\begin{axis}[
 axis x line=middle, axis y line=middle,
 ymin=-1.3, ymax=1.5, ytick={-1,...,1}, ylabel=$g(x)$,
 xmin=0, xmax=10, xtick={1,3,...,9}, xlabel=$x$,
 samples=150, 
]

\addplot[domain=0:0.95,blue,thick]
{ln(2/((x)+1))*1/(sqrt((x)*ln(2*(x)/((x)+1))+ln(2/((x)+1))))};
\addplot[domain=1.05:10,blue,thick]
{ln(2/((x)+1))*1/(sqrt((x)*ln(2*(x)/((x)+1))+ln(2/((x)+1))))};
\draw[dotted] (axis cs:1,-1) -- (axis cs:1,1);
\addplot[holdot] coordinates{(1,1)(1,-1)};

\end{axis}
\end{tikzpicture}
\end{figure}

\begin{enumerate}[{\rm (a)}]
\item $g>0$ on $]0,1[$, $g<0$ on $]1,\infty[$.

\item $\lim_{x\uparrow 1}g(x)=1$, $\lim_{x\downarrow 1}g(x)=-1$. This follows from $\lim_{x\rightarrow 1}[g(x)]^2=1$, which can 
be established 
by applying l'Hopital's rule twice.

\item $g^\prime(x)>0$ for $x\in \rr_+\setminus\{1\}$. To prove this one computes 
\[
g^\prime(x)=-\frac{h(x)}{(x+1)L(x,1)^{3/2}},
\]
where 
\[h(x)= 2x\log\left(\frac{2x}{x+1}\right) + 2\log\left(\frac{2}{x+1}\right) + (x+1)\log\left(
\frac{2x}{x+1}\right)\log\left(\frac{2}{x+1}\right).
\]
One further computes 
\[h^\prime(x)=\log \left(\frac{2x}{x+1}\right)\log\left(\frac{2}{x+1}\right) + 
\log\left(\frac{2x}{x+1}\right) + \frac{1}{x}\log\left(\frac{2}{x+1}\right),
\]
\[
h^{\prime\prime}(x)=-\frac{1}{x+1}\log \left(\frac{2x}{x+1}\right)-\frac{1}{x^2(x+1)}\log\left(\frac{2}{x+1}\right).
\]
Note that $h(1)=h^\prime(1)=h^{\prime\prime}(1)=0$. The inequality $\log t\geq (t-1)/t$, which holds for 
all $t>0$, gives 
\[
h^{\prime\prime}(x)\leq -\frac{1}{x+1}\left(1-\frac{x+1}{2x}\right)-\frac{1}{x^2(x+1)}\left( 
1-\frac{x+1}{2}\right)=-\frac{(x-1)^2}{2x^2(x+1)}.
\]
Hence $h^{\prime\prime}(x)<0$ for $x\in \rr_+\setminus\{1\}$, and the statement follows. 
\item Note that (a),  (b)  and (c) give that $0<g(x)<1$ on $]0,1[$ and $-1<g(x)<0$ on $]1, \infty[$.
\end{enumerate}

If follows from (a) that $f^\prime(r)<0$ for $r\in ]0, p[$, $f^\prime(r)>0$ for $r>q$, and so  $f(r)$ is decreasing on $]0, p[$ and 
increasing on $]q, \infty[$. Hence, for $r<p$ and $r>q$, $f(r)>f(p)$, which qives \eqref{ma-sj} for those $r$'s. To deal with the case 
$p<r<q$, set $m(r)=g(p/r) + g(q/r)$. It follows from (b) that $m^\prime(r)<0$ for $p<r<q$, while (b) and (d) give 
$m(p+)=1 + g(q/p)>0$, $m(q-)=-1+ g(p/q)<0$. Hence $f^\prime(r)$ has precisely one zero $r_m$ in the interval $]p,q[$. Since
$f^\prime(p+)>0$, $f^\prime(q-)>0$, $f(r)$ is increasing  in $[p, r_m]$ and decreasing on $[r_m, q]$. On the 
first interval, $f(r)\geq f(p)$, and on the second interval $f(r)\geq f(q)$, which gives that \eqref{ma-sj} also holds 
for $p<r<q$. \qed

The graph of $r\mapsto f(r)$ is plotted below for $p=\frac{1}{10}$ and $q=\frac{2}{3}$. In this case $r_m\approx 0.28$. 
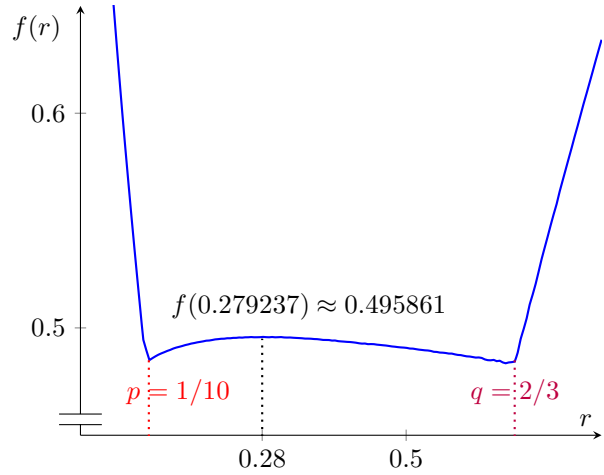
\begin{figure}[h]
\centering
\begin{tikzpicture}
\begin{axis}[
 axis x line=middle, axis y line=middle,
 ymin=0.45, ymax=0.65, ytick={0.5,0.6}, ylabel=$f(r)$,
 y label style={at={(-0.15,1)}},
 xmin=0, xmax=0.8, xtick={0.279,0.5,1}, xlabel=$r$,
 axis y discontinuity=parallel,
 samples=175 
]

\addplot[domain=0:0.8,blue,thick,smooth,tension=.1]
{sqrt(1/10*ln(2/10*1/(1/10+(x)))+x*ln(2*(x)/(1/10+(x))))+sqrt(x*ln(2*(x)*1/(2/3+(x)))+2/3*ln(4/3/(2/3+(x))))};
\draw[dotted,thick] (axis cs:.279,0) -- (axis cs:.279,.495);
\draw[dotted,red,thick] (axis cs:.105,0) -- (axis cs:.105,.485);
\draw[dotted,purple,thick] (axis cs:2/3,0) -- (axis cs:2/3,.485);
\node[] at (axis cs: .35,.51) {$f(0.279237) \approx 0.495861$};
\node[red] at (axis cs: .15,.47) {$p = 1/10$};
\node[purple] at (axis cs: 2/3,.47) {$q = 2/3$};
\end{axis}
\end{tikzpicture}
\end{figure}

\section{R\'enyi's relative entropy}
Let $\Omega$ be a finite set and $P, Q\in {\cal P}(\Omega)$. For $\alpha \in ]0,1[$ we set  
\[ S_\alpha(P|Q)=\frac{1}{\alpha-1}\log \left(\sum_{\omega \in \Omega}P(\omega)^\alpha Q(\omega)^{1-\alpha}\right).
\]
$S_\alpha(P|Q)$ is called R\'enyi's relative entropy of $P$ with respect to $Q$. 
Note that 
\[ S_\alpha (P|P_{\rm ch})= S_\alpha (P)+ \log |\Omega|.
\]
\bep\label{renyi-relative} \begin{enumerate}[{\rm (1)}]
\item $S_\alpha(P|Q)\geq 0$.
\item  $S_\alpha(P|Q)=\infty$ iff $P\perp Q$ and $S_\alpha(P|Q)=0$ iff $P=Q$. 
\item 
\[ S_\alpha(P|Q)=\frac{\alpha}{1-\alpha}S_{1-\alpha}(Q|P).\]
\item  
\[
\lim_{\alpha \uparrow 1} S_\alpha(P|Q)=S(P|Q).
\]
\item Suppose that $P\not\perp Q$.  Then  the    function $]0,1[\ni \alpha \mapsto S_\alpha(P|Q)$ is strictly increasing
\item The map $(P, Q)\mapsto S_\alpha(P|Q)\in [0, \infty]$ is continuous and jointly convex.

\item Let  $\Phi: {\cal P}(\Omega)\rightarrow {\cal P}(\hat \Omega)$ be a stochastic map. Then for all 
$P, Q\in {\cal P}(\Omega)$, 
\[S_\alpha(\Phi({P})|\Phi(Q))\leq S_\alpha(P|Q).
\]
\item If $S(P|Q)<\infty$, then  $\alpha \mapsto S_\alpha(P|Q)$ extends to a real-analytic function on 
$\rr$. 

\end{enumerate}
\eep
\demo Obviously, $S_\alpha(P|Q)=\infty$ iff $P\perp Q$. In what follows, if $P\not\perp Q$,  we set
 \[T=\supp \,P\cap \supp\,Q.\]
 An application of Jensen's inequality gives 
\[
\begin{split}
\sum_{\omega\in \Omega}P(\omega)^\alpha Q(\omega)^{1-\alpha}
&=Q(T)\sum_{\omega \in T} \left(\frac{P(\omega)}{Q(\omega)}\right)^\alpha \frac{Q(\omega)}{Q(T)}\\[2mm]
&\leq Q(T)\left(\sum_{\omega\in T}\frac{P(\omega)}{Q(\omega)}\frac{Q(\omega)}{Q(T)}\right)^\alpha\\[2mm]
&=
Q(T)^{1-\alpha} P(T)^\alpha.
\end{split}
\]
Hence,  $\sum_{\omega\in \Omega}P(\omega)^\alpha Q(\omega)^{1-\alpha}\leq 1$ with the equality iff $P=Q$, and Parts (1), (2) 
follow.

 Part (3) is obvious. To prove (4), note that 
\[
\lim_{\alpha\uparrow 1}\sum_{\omega}P(\omega)^\alpha Q(\omega)^{1-\alpha}= P(T),
\]
and that $P(T)=1$ iff $P\ll Q$. Hence, if $P$ is not absolutely continuous with respect to $Q$, then 
$\lim_{\alpha \uparrow 1}S_\alpha(P|Q)=\infty=S(P|Q)$. If $P\ll Q$, an application of L'Hopital rule gives 
$\lim_{\alpha \uparrow 1}S_\alpha (P|Q)= S(P|Q)$.

To prove (5), set 
\[ F(\alpha)=\log \left(\sum_{\omega \in \Omega}P(\omega)^{\alpha}Q(\omega)^{1-\alpha}\right),
\]
and note that $\rr\ni \alpha \mapsto F(\alpha)$ is a real-analytic strictly convex function satisfying $F(0)\leq 0$, $F(1)\leq 0$. We have 
\[
\frac{\d S_\alpha(P|Q)}{\d \alpha} = \frac{F^\prime(\alpha)(\alpha-1)- (F(\alpha)- F(1))}{(\alpha-1)^2} - \frac{F(1)}{(\alpha-1)^2}.
\]
By the mean-value theorem, $F(\alpha)- F(1)= (\alpha-1)F^\prime(\zeta_\alpha)$ for some $\zeta_\alpha \in ]\alpha, 1[$. Since 
$F^\prime$ is strictly increasing, $F^\prime(\alpha)< F^\prime(\zeta_\alpha)$ and 
\[
\frac{\d S_\alpha(P|Q)}{\d \alpha}>0
\]
for $\alpha \in ]0,1[$. 

The continuity part of (6) are obvious. The proof of the joint convexity is the same 
as the proof of Proposition \ref{convex} (one now takes $g(t)=t^\alpha$) and is left as an exercise for the reader.

We now turn to  Part (7). First, we have 
\[
\left[\Phi({P})(\hat \omega)\right]^\alpha\left[ \Phi(Q)(\hat \omega)\right]^{1-\alpha}\geq 
\sum_{\omega}P(\omega)^\alpha Q(\omega)^{1-\alpha}\Phi(\omega,\hat \omega).
\]
This  inequality is obvious if the r.h.s. is equal to zero. Otherwise, let 
\[R=\{\omega\,|\, P(\omega) Q(\omega)\Phi(\omega, \hat \omega)>0\}.\]
Then 
\[
\begin{split}
\left[\Phi({P})(\hat \omega)\right]^\alpha\left[ \Phi(Q)(\hat \omega)\right]^{1-\alpha}&\geq 
\left(\sum_{\omega\in R}P(\omega)\Phi(\omega, \hat \omega)\right)^\alpha
\left(\sum_{\omega \in R}Q(\omega)\Phi(\omega, \hat \omega)\right)^{1-\alpha}\\[2mm]
&=\left(\frac{\sum_{\omega\in R}P(\omega)\Phi(\omega, \hat \omega)}{
\sum_{\omega \in R}Q(\omega) \Phi(\omega, \hat \omega)}\right)^\alpha
\sum_{\omega \in R}Q(\omega) \Phi(\omega, \hat \omega)\\[2mm]
&\geq \sum_{\omega}P(\omega)^{\alpha}Q(\omega)^{1-\alpha}\Phi(\omega, \hat \omega),
\end{split}
\]
where in the last step we have used the joint concavity of the function $(x, y)\mapsto x (y/x)^\alpha$ (recall proof 
of Proposition \ref{convex}).
Hence, 
\[
\sum_{\hat \omega}\left[\Phi({P})(\hat \omega)\right]^\alpha\left[ \Phi(Q)(\hat \omega)\right]^{1-\alpha}
\geq 
\sum_{\hat \omega} \sum_{\omega}P(\omega)^\alpha Q(\omega)^{1-\alpha}\Phi(\omega,\hat \omega)=
\sum_{\omega}P(\omega)^\alpha Q(\omega)^{1-\alpha}, 
\]
and Part (7) follows. 

It remains to prove Part (8). For $\alpha \in \rr\setminus\{1\}$ set 
\[{\mathfrak S}_\alpha(P|Q)=\frac{1}{\alpha-1} \log \left(\sum_{\omega \in T}P(\omega)^\alpha Q(\omega)^{1-\alpha}\right).\]
Obviously,  $\alpha \mapsto {\mathfrak S}_\alpha(P|Q)$ is real-analytic on $\rr\setminus \{1\}$. Since 
\[\lim_{\alpha \uparrow 1} {\mathfrak S}_\alpha(P|Q)=\lim_{\alpha \downarrow 1} {\mathfrak S}_\alpha(P|Q)=S(P|Q),\]
$\alpha \mapsto {\mathfrak S}_\alpha(P|Q)$ extends to a real-analytic function on $\rr$ with  ${\mathfrak S}_1(P|Q)=S(P|Q)$.
Finally, Part (8) follows from the observation that  $S_\alpha(P|Q)={\mathfrak S}_\alpha(P|Q)$ for $\alpha \in ]0, 1[$.

\qed

Following on the discussion at  the end of Section \ref{sec-ren-nat}, we set
\[ {\widehat S}_{\alpha}(P|Q)=\log \left(\sum_{\omega\in T}P(\omega)^\alpha Q(\omega)^{1-\alpha}\right), \qquad \alpha \in \rr.
\]
If $P\ll Q$, then
\beq{\widehat S}_{\alpha}(P|Q)= \log {\mathbb E}_Q(\e^{\alpha S_{P|Q}}),
\label{cambridge-jar}
\eeq
and so $\widehat S_\alpha(P|Q)$ is the cumulant generating function for the relative entropy function 
$S_{P|Q}$ defined on the probability space $(T, P)$.  The discussion at the end of \ref{sec-ren-nat} can be now 
repeated verbatim (we will return to this point in Section \ref{sec-in-nat}). Whenever there is no danger 
of the confusion, we  shall also call $\widehat S_\alpha(P|Q)$  R\'enyi's relative entropy of the pair $(P, Q)$. Note that 
\beq
\widehat S_\alpha(P_{\rm ch}|P)= \widehat S_\alpha(P)-\alpha \log|\Omega|.
\label{hat-ch}
\eeq
Some care is needed in transposing the properties listed in Proposition \ref{renyi-relative} 
to ${\widehat S}_\alpha(P|Q)$. This point is discussed in the Exercise  \ref{f-cms}.

\begin{exo} 
\exop Describe the subset of ${\cal P}(\Omega)\times {\cal P}(\Omega)$ on which the function 
$(P, Q)\mapsto S_\alpha(P|Q)$ is strictly convex. 

\exop  Describe the subset of ${\cal P}(\Omega)\times {\cal P}(\Omega)$ on which 
$S_\alpha(\Phi({P})|\Phi(Q))<S_\alpha(P|Q)$. 

\exop Redo the Exercise \ref{f-cms-1} in Section \ref{sec-rel-entropy} and reprove Proposition \ref{data-sherry} following the proofs of Parts (7) and 
(8) of Proposition \ref{renyi-relative}. Describe the subset of ${\cal P}(\Omega)$ on which 
\[S(\Phi({P})|\Phi(Q))<S(P|Q).\] 

\end{exo}

\begin{exo}
\label{f-cms}
Prove the following properties of 
$\widehat S_\alpha(P|Q)$. 
\exop $\widehat S_\alpha(P|Q)=-\infty$ iff $P\perp Q$.

In the remaining statements we shall suppose that $P\not\perp Q$.

\exop The function $\rr\ni \alpha \mapsto \widehat S_\alpha(P|Q)$ is real-analytic and convex. This function is trivial 
({\sl i.e.}, identically equal to  zero) iff $P=Q$.  If $P/Q$ not constant on $T=\supp P \cap \supp Q$, then the function $\alpha \mapsto \widehat S_\alpha(P|Q)$ is strictly convex.
\exop If $Q\ll P$, then 
\[\frac{\d \widehat S_{\alpha}(P|Q)}{\d \alpha}\big|_{\alpha=0}=- S(Q|P).
\]
If $P\ll Q$, then 
\[\frac{\d \widehat S_{\alpha}(P|Q)}{\d \alpha}\big|_{\alpha=1}=S(P|Q).
\]
\exop If $P$ and $Q$ are mutually absolutely continuous, then $\widehat S_0(P|Q)=\widehat S_1(P|Q)=0$, $\widehat S_\alpha(P|Q)\leq 0$ for 
$\alpha \in [0,1]$, and $\widehat S_\alpha(P|Q)\geq 0$ for $\alpha \not\in [0,1]$.
Moreover, 
\[
\widehat S_\alpha(P|Q)\geq \max\{ -\alpha S(Q|P), (\alpha-1)S(P|Q)\}.
\]
\exop For $\alpha \in ]0,1[$ the function $(P, Q)\mapsto \widehat S_\alpha(P|Q)$ is continuous and jointly concave. Moreover, 
for any stochastic matrix $\Phi$, 
\[ \widehat S_\alpha(\Phi({P})|\Phi(Q))\geq \widehat S_\alpha(P|Q).
\] 
\end{exo}
\begin{exo}
 Prove that the fluctuation relation of Section \ref{sec-fluct-rel} is equivalent to the following statement: for all 
$\alpha \in \rr$, 
\[\widehat S_{\alpha}(P|P_{\Theta})=\widehat S_{1-\alpha}(P|P_\Theta).\]
\end{exo}


\section{Hypothesis testing}
\label{sec-hyp-test}
Let $\Omega$ be a finite set and $P, Q$ two distinct  probability measures on $\Omega$.  We shall assume that $P$ and $Q$ are faithful.

 Suppose that we know a  priori that  a probabilistic 
experiment 
is with probability $p$ described by  $P$ and with probability $1-p$ by  $Q$.   By performing 
an experiment  we wish to decide with minimal error probability what is the correct probability measure.  
For example, suppose that we are given  two coins, one fair ($P({\rm Head})=P({\rm Tail})=1/2$) and one 
unfair ($Q({\rm Head})=s, Q({\rm Tail})=1-s, s>1/2)$. We pick coin randomly   (hence $p=1/2$). The experiment is a coin toss. After  tossing a coin we wish to decide with minimal error probability whether we 
picked the fair or the unfair coin. The correct choice of obvious: if the outcome is  Head,  pick $Q$, if the outcome is  Tail, pick P.

The following procedure is known as {\sl hypothesis testing.} 
A {\em test} $T$  is a subset of $\Omega$. On the basis of the outcome of the experiment with respect to $T$ 
 one chooses between $P$ or $Q$. More
precisely, if the outcome of the experiment is  in  $T$, one chooses $Q$ 
(Hypothesis I: $Q$ is correct)  and if the outcome is not in $T$, one chooses $P$ 
(Hypothesis II: $P$ is correct). $P(T)$ is the conditional error probability of  accepting I if II is true and $Q(T^c)$ is the  conditional 
error probability  of accepting  II if I is true. The average error probability is 
\[
D_p(P, Q, T)=pP(T) + (1-p)Q(T^c),
\]
and we  are interested in minimizing $D_p(P, Q, T)$ w.r.t. $T$.  Let  
\[
D_p(P, Q)=\inf_T  D_p(P, Q, T).
\]
The  Bayesian distinguishability 
problem is to identify tests $T$ such that $D_p(P, Q, T)=D_p(P, Q)$. 
Let 
\[
T_{\rm opt}=\{\omega\,|\, pP(\omega)\leq (1-p)Q(\omega)\}.
\]
\bep\label{hypothesis-paris}\begin{enumerate}[{\rm (1)}]
\item $T_{\rm opt}$ is a minimizer of the function $T\mapsto D_p(P, Q, T)$. If $T$ is another minimizer, then 
$T\subset T_{\rm opt}$ and $pP(\omega)=(1-p)Q(\omega)$ for $\omega \in T_{\rm opt}\setminus T$.

\item 
\[ D_p(P, Q)=\int_\Omega \min \{1-p, p\Delta_{P|Q}(\omega)\}\d Q.\]
\item For $\alpha \in ]0,1[$, 
\[ D_p(P,Q)\leq p^{\alpha}(1-p)^{1-\alpha}\e^{\widehat S_\alpha(P|Q)}.\]
\item 
\[ D_p(P,Q)\geq \int_\Omega \frac{p\Delta_{P|Q}}{1 + \frac{p}{1-p}\Delta_{P|Q}}\d Q.
\]
\end{enumerate}
\eep
\begin{remark}
Part (1) of this proposition is called Neyman-Pearson lemma. Part (3) is called Chernoff bound. 
\end{remark}
\demo 
\[ D_p(P, Q, T)=1-p-\sum_{\omega \in T} \left((1-p) Q(\omega)-pP(\omega)\right)\geq 
1-p-\sum_{\omega \in T_{\rm opt}} \left((1-p) Q(\omega)-pP(\omega)\right),
\]
and Part (1) follows. Part (2) is a straightforward computation. Part (3) follows from (2) and the bound 
$\min\{x, y\}\leq x^\alpha y^{1-\alpha}$ that holds for $x, y\geq 0$ and $\alpha \in ]0, 1[$. Part (4) follows from (2) and the 
obvious estimate 
\[
\min \{1-p, p\Delta_{P|Q}(\omega)\}\geq \frac{p\Delta_{P|Q}}{1 + \frac{p}{1-p}\Delta_{P|Q}}.
\]
\qed

Obviously, the errors are  smaller  if the hypothesis testing is based on  repeated experiments. Let $P_N$ and $Q_N$ be the 
respective product probability measures on $\Omega^N$. 
\bet 
\[
\lim_{N\rightarrow \infty}\frac{1}{N}\log  D_p(P_N, Q_N)=\min_{\alpha \in [0,1]} \widehat S_\alpha(P|Q).
\]
\label{chernoff}
\eet
\demo 
By Part (2) of the last proposition, for any $\alpha \in ]0, 1[$, 
\[ D_p(P_N, Q_N)\leq p^\alpha (1-p)^{1-\alpha} \e^{\widehat S_\alpha(P_N|Q_N)}= p^\alpha (1-p)^{1-\alpha} \e^{N\widehat S_\alpha(P|Q)},
\]
and so 
\[
\frac{1}{N}\log D_p(P_N, Q_N)\leq \min_{\alpha \in [0,1]}\widehat S_\alpha(P|Q).
\]
This yields the upper bound:
\[
\limsup_{N\rightarrow \infty}\frac{1}{N}\log  D_p(P_N|Q_N)\leq \min_{\alpha \in [0,1]} \widehat S_\alpha(P|Q).
\]
To prove the lower bound we shall make use  of the lower bound in   Cram\'er's theorem (Corollary 
\ref{no-way-paris}). Note first that the function 
\[
x\mapsto \frac{px}{1 + \frac{p}{1-p}x}
\]
is increasing on $\rr_+$. Let $\theta >0$ be given. By Part (4) of the  last proposition, 
\[
D_p(P_N, Q_N)\geq \frac{p\e^{N\theta}}{1  + \frac{p}{1-p}\e^{N\theta}}
Q_N\left\{\omega \in \Omega^N\,|\, \Delta_{P_N|Q_N}(\omega)\geq \e^{N\theta}\right\}.
\]
Hence, 
\beq\label{paris-no-idea}
\liminf_{N\rightarrow \infty}\frac{1}{N}\log  D_p(P_N|Q_N)\geq 
\liminf_{N\rightarrow \infty}\frac{1}{N}\log Q_N\left\{\omega \in \Omega^N\,|\, \log \Delta_{P_N|Q_N}(\omega)\geq  N\theta\right\}.
\eeq
Let $X=\log \Delta_{P|Q}$ and ${\cal S}_N(\omega)=\sum_{k=1}^N X(\omega_k)$. Note that 
${\cal S}_N=\log \Delta_{P_N|Q_N}$.  The cummulant generating function of 
$X$ w.r.t. $Q$ is 
\[\log {\mathbb E}_Q(\e^{\alpha X})= \widehat S_\alpha(P|Q).\]
Since ${\mathbb E}_Q(X)=- S(Q|P)< 0$ and $\theta>0$, it follows from Corollary \ref{no-way-paris} that 
\beq\label{follows-paris}\lim_{N\rightarrow \infty}\frac{1}{N}\log Q_N\left\{\omega \in \Omega^N\,|\, \log \Delta_{P_N|Q_N}(\omega)\geq  N\theta\right\} 
\geq -I(\theta)
\eeq
Since 
\[\frac{\d \widehat S_\alpha}{\d \alpha}\big|_{\alpha=0}= -S(Q|P)<0, \qquad  \frac{\d \widehat S_\alpha}{\d \alpha}\big|_{\alpha=1}= S(P|Q)>0,
\]
the rate function $I(\theta)$ is continuous around zero, and it follows from (\ref{paris-no-idea}) and (\ref{follows-paris}) that
\[
\liminf_{N\rightarrow \infty}\frac{1}{N}\log  D_p(P_N|Q_N)\geq -I(0)=-\sup_{\alpha \in \rr} (-\widehat S_\alpha(P|Q)).
\]
Since $\widehat S_\alpha(P|Q)\leq 0$ for $\alpha \in [0,1]$ and $\widehat S_\alpha(P|Q)\geq 0$ for $\alpha \not \in [0,1]$, 
\[-\sup_{\alpha \in \rr} (-\widehat S_\alpha(P|Q))=\min_{\alpha \in [0,1]} \widehat S_\alpha(P|Q),
\]
and the lower bound follows: 
\[
\liminf_{N\rightarrow \infty}\frac{1}{N}\log  D_p(P_N|Q_N)\geq \min_{\alpha \in [0,1]}\widehat S_\alpha(P|Q).
\]
\qed
\section{Asymmetric hypothesis testing}
We continue with the framework and notation of the previous section. 
The asymmetric hypothesis testing concerns individual 
error probabilities $P_N(T_N)$ (type I-error) and $Q_N(T_N^c)$ (type II-error).  
For $\gamma \in ]0,1[$ the Stein error exponents are defined by 
\[ 
s_N(\gamma)=\min\left\{ P(T_N)\, \big|\, T_N\subset \Omega^N,\, Q(T_N^c)\leq \gamma\right\}.
\]
Theorem \ref{stein-lemma} gives 
\[
\lim_{N\rightarrow \infty}\frac{1}{N}\log s_N(\gamma)=-S(Q|P).
\]
The Hoeffding error exponents are similar to Stein's exponents, but with a tighter 
constraint on the family $(T_N)_{N\geq 1}$ of tests which are required to
ensure exponential decay of type-II errors with a minimal rate $s>0$. They are defined as
\[
\begin{split}
\bar h(s)&=\inf_{(T_N)}\left\{\limsup_{N\to\infty}\frac1N\log P_N(T_N)\,\bigg|\, 
\limsup_{N\to\infty}\frac1N\log Q_N(T_N^c)\leq -s\right\},\\[3mm]
\ubar h(s)&=\inf_{(T_N)}\left\{\liminf_{N\to\infty}\frac1N\log P_N(T_N)\,\bigg|\,
\limsup_{N\to\infty}\frac1T\log Q_N(T_N^c)\leq -s\right\},\\[3mm]
h(s)&=\inf_{(T_N)}\left\{\lim_{N\to\infty}\frac1N\log P_N(T_N)\,\bigg|\, 
\limsup_{N\to\infty}\frac1N\log  Q_N(T_N^c)\leq -s\right\},
\end{split}
\]
where in the last case the infimum is taken over all sequences  of tests $(T_N)_{N\geq 1}$ for which the limit
\[\lim_{N\to\infty}\frac1N\log P_N(T_N)\]
exists.
The analysis of these exponents is centred around the function 
\[\psi(s)=\inf_{\alpha \in [0,1[}\frac{s\alpha + \widehat S_\alpha(Q|P)}{1-\alpha}, \qquad s\geq 0.
\]
We first describe some basic properties of $\psi$. 
\bep
\begin{enumerate}[{\rm (1)}]
\item $\psi$ is continuous on $[0, \infty[$, $\psi(0)=-S(Q|P)$ and $\psi(s)=0$ for $s\geq S(P|Q)$.
\item $\psi$ is  strictly increasing and strictly concave on $[0, S(P|Q)]$, 
and real analytic on $]0, S(P|Q)[$. 
\item 
\[
\lim_{s\downarrow 0}\psi^{\prime}(s)=\infty, \qquad \lim_{s \uparrow S(P|Q)}\psi^\prime(s)= \left[{\widehat S}_\alpha^{\prime\prime}(Q|P)\big|_{\alpha=0}\right]^{-1}.
\]
\item For $\theta \in \rr$ set 
\[
\varphi(\theta)=\sup_{\alpha \in [0,1]}\left(\theta\alpha- \widehat S_\alpha(Q|P)\right), \qquad 
\hat \varphi(\theta)=\varphi(\theta)-\theta.
\]
Then for  all $s\geq 0$, 
\beq
\psi(s)=-\varphi(\hat \varphi^{-1}(s)).
\label{rome-holds}
\eeq
\end{enumerate}
\label{no-storm}\eep
\demo
Throughout the proof we shall often use Part 3  of the Exercise \ref{f-cms}.

We shall prove Parts (1)-(3) simultaneously.  Set 
\[F(\alpha)= \frac{s\alpha + \widehat S_\alpha(Q|P)}{1-\alpha}.\]
Then 
\[
F^\prime(\alpha)= \frac{G(\alpha)}{(1-\alpha)^2},
\]
where $G(\alpha)= s +\widehat S_\alpha(Q|P)+ (1-\alpha){\widehat S}_\alpha^{\prime}(Q|P)$. Futhermore, 
$G^\prime(\alpha)=(1-\alpha){\widehat S}_\alpha^{\prime \prime}(Q|P)$ and so $G^{\prime}(\alpha)>0$ for 
$\alpha \in [0,1[$. Note that  $G(0)=s- S(P|Q)$ and $G(1)= s$. It follows that if  $s=0$, then  $G(\alpha)<0$ for $\alpha \in[0,1[$
and  $F(\alpha)$ is decreasing on $[0,1[$. Hence, 
\[\psi(0)=\lim_{\alpha\rightarrow 1}\frac{\widehat S_\alpha(Q|P)}{1-\alpha}= - S(Q|P).
\]
On the other hand, if  $0 <s <S(P|Q)$, then $G(0)<0$, $G(1)>0$, and so there exists unique $\alpha_\ast(s) \in ]0, 1[$ such that 
\beq G(\alpha_\ast(s))=0.
\label{roma-wed}
\eeq
In this case, 
\beq \psi(s)=\frac{s\alpha_\ast(s) + \widehat S_{\alpha_\ast(s)}(Q|P)}{1-\alpha_\ast(s)}=-s - {\widehat S}^\prime_{\alpha_\ast(s)}(Q|P).
\label{roma-case}
\eeq
If $s\geq S(P|Q)$, then  $G(\alpha)\geq 0$ for $\alpha \in [0,1[$, and $\psi(s)=F(0)=0$. The analytic implicit 
function theorem yields that $s\mapsto \alpha_\ast(s)$ is analytic on $]0,S(P|Q)[$, and so  $\psi$ is real-analytic on $]0, S(P|Q)[$. The identity 
\beq
0=G(\alpha_\ast(s))= s +\widehat S_{\alpha_\ast(s)}(Q|P)+ (1-\alpha_\ast(s)){\widehat S}_{{\alpha_\ast(s)}}^{\prime}(Q|P),
\label{div-never}
\eeq
which holds for $s\in \,]0,S(P|Q)[$, gives that 
\beq \alpha_\ast^\prime(s)=-\frac{1}{(1-\alpha_\ast(s))G^\prime(\alpha_\ast(s))},
\label{mar-snow-1}
\eeq
and so $\alpha_\ast^\prime(s)<0$ for $s\in ]0,S(P|Q)[$. One computes 
\beq
\psi^\prime(s)=\frac{\alpha_\ast(s)- s\alpha_\ast^\prime(s)}{(1-\alpha_\ast(s))^2},
\label{mar-snow-2}
\eeq
and so $\psi$ is strictly increasing on $]0,S(P|Q)[$ and hence on $[0, S(P|Q)]$.  Since $\alpha_\ast(s)$ is strictly decreasing on $]0,S(P|Q)[$, 
the limits 
\[\lim_{s\downarrow 0}\alpha_\ast(s)=x, \qquad \lim_{s\uparrow S(P|Q)}\alpha_\ast(s)=y,\]
 exist. Obviously, 
$x, y\in [0,1]$, $x>y$, and the definition of $G$ and $\alpha_\ast$ give that 
\beq \widehat S_x(Q|P) +(1-x){\widehat S}^\prime_x(Q|P)=0, \qquad S(P|Q)+\widehat S_y(Q|P) +(1-y){\widehat S}^\prime_y(Q|P)=0.
\label{rome-snow-1}
\eeq
We proceed to show that $x=1$ and $y=0$. Suppose that $x<1$. The mean value theorem gives that for some $z\in\, ]x,1[$
\beq 
 -\widehat S_x(Q|P)=\widehat S_1(Q|P) -\widehat S_x(Q|P)= (1-x){\widehat S}^\prime_z(Q|P)>(1-x){\widehat S}^\prime_z(Q|P),
 \label{rome-snow-2}
 \eeq
where we used that $\alpha\mapsto {\widehat S}^\prime_\alpha(Q|P)$ is strictly increasing. Obviously, 
\eqref{rome-snow-2} contradicts the first equality in \eqref{rome-snow-1}, and so $x=1$. Similarly, if $y>0$, 
\[
\begin{split}
S(P|Q)+\widehat S_y(Q|P) +(1-y){\widehat S}^\prime_y(Q|P)&>S(P|Q)+\widehat S_y(Q|P) +(1-y){\widehat S}^\prime_0(Q|P)\\[3mm]
&=\widehat S_y(Q|P)-y{\widehat S}^\prime_0(Q|P)>0,
\end{split}
\]
contradicting the second equality in \eqref{rome-snow-1}. Since $x=1$ and $y=0$, \eqref{mar-snow-1} and \eqref{mar-snow-2} 
yield Part (3). Finally, to prove that $\psi$ is strictly concave on $[0, S(P| Q)]$ (in view of real analyticity of $\psi$ on 
$]0, S(P|Q)[$), it suffices to show that $\psi^\prime$ is not constant  on $]0, S(P|Q)[$. That follows from 
Part (3), and the proofs of Parts (1)-(3) are complete. 

We now turn to Part (4). The following basic properties of the "restricted Legendre transform" $\varphi$ are easily proven following 
the arguments in Section \ref{sec-rate} and we leave the details as an exercise for the reader: $\varphi$ is continuous, non-negative and convex on $\rr$, $\varphi(\theta)=0$ for $\theta \leq -S(P|Q)$, 
$\varphi$ is real analytic, strictly increasing and strictly convex on $]-S(P|Q), S(Q|P)[$, and $\varphi(\theta)=\theta$ for 
$\theta \geq S(Q|P)$. The properties of $\hat \varphi$ are now  deduced form those of $\varphi$ and we mention the following: 
$\hat \varphi$ is convex, continuous and decreasing, $\hat \varphi(\theta)=\theta$ for $\theta \leq -S(P|Q)$, and $\varphi(\theta)=0$ for $\theta \geq S(Q|P)$. Moreover, the map $\hat \varphi
:]-\infty, S(Q|P)]\rightarrow [0, \infty[$ is a bijection, and we denote by $\hat \varphi^{-1}$ its inverse. For $s\geq S(P|Q)$, 
$\hat \varphi^{-1}(s)=-s$ and $\varphi(-s)=0$, and so \eqref{rome-holds} holds for $s \geq S(P|Q)$. Since 
$\hat \varphi^{-1}(0)= S(Q|P)$ and $\varphi(S(Q|P))= S(Q|P)$, \eqref{rome-holds} also holds for $s=0$. 

It remains to consider the case $s\in\,]0, S(P|Q)[$. 
The map 
$\hat \varphi:\,]-S(P|Q), S(Q|P)[\rightarrow  ]0, S(P|Q)[$ is a strictly decreasing bijection. Since
\[
-\varphi(\hat\varphi^{-1}(s))= -s - \hat \varphi^{-1}(s),
\]
it follows from \eqref{roma-case} that it suffices to show that 
\[\hat \varphi^{-1}(s)=  {\widehat S}^\prime_{\alpha_\ast(s)}(Q|P),\]
or equivalently, that 
\beq  \varphi({\widehat S}^{\prime}_{\alpha_\ast(s)}(Q|P))=-s-{\widehat S}^{\prime}_{\alpha_\ast(s)}(Q|P)).
\label{roma-last}
\eeq
Since on  $ ]-S(P|Q), S(Q|P)[$ the function $\varphi$ coincides with the Legendre transform of ${\widehat S}_\alpha(P|Q)$, 
it follows from Part (1) of Proposition \ref{prop-rate} that 
\[
\varphi({\widehat S}^{\prime}_{\alpha_\ast(s)}(Q|P))=\alpha_\ast(s){\widehat S}^{\prime}_{\alpha_\ast(s)}(Q|P)- {\widehat S}_{\alpha_\ast(s)}(Q|P),
\]
and  \eqref{roma-last} follows from \eqref{div-never}.
\qed
\begin{exo} Prove the properties of $\varphi$ and $\hat \varphi$ that were stated and used  in the proof of Part (4) of 
Proposition \ref{no-storm}.
\end{exo}
The next result sheds additional light on the  function $\psi$. For $\alpha \in [0,1]$ we define $R_\alpha \in {\cal P}(\Omega)$ 
by 
\[
R_\alpha(\omega)=\frac{ Q(\omega)^\alpha P(\omega)^{1-\alpha}}{\sum_{\omega^\prime}Q(\omega^\prime)^\alpha 
P(\omega^\prime)^{1-\alpha}}.
\]
\bep
\begin{enumerate}[{\rm (1)}] \item For all $s\geq 0$, 
\beq
\psi(s)=-\inf\left\{ S(R|P)\,|\, R\in {\cal P}(\Omega), \, S(R|Q)\leq s\right\}.
\label{wait-w}
\eeq
\item For any  $s\in\, ]0, S(P|Q)[$, 
\[ S(R_{\alpha_\ast(s)}|Q)=s, \qquad S(R_{\alpha_\ast(s)}|P)=-\psi(s),
\]
where $\alpha_\ast(s)$ is given by \eqref{roma-wed}.
\end{enumerate}
\eep
\demo Denote by $\phi(s)$ the r.h.s. in \eqref{wait-w}. Obviously, $\phi(0)=-S(Q|P)$ and $\phi(s)=0$ for 
$s\geq S(P|Q)$. So we need to prove that $\psi(s)=\phi(s)$ for $s\in\, ]0, S(P|Q)[$. 

For any $R\in {\cal P}(\Omega)$ and $\alpha \in [0,1]$, 
\[S(R|R_\alpha)= \alpha S(R|Q) + (1-\alpha) S(R|P) + {\widehat S}_\alpha(Q|P).
\]
If $R$ is such that $S(R|Q)\leq  s$ and $\alpha \in [0,1[$, then 
\[\frac{S(R|R_\alpha)}{1-\alpha}\leq \frac{\alpha s + {\widehat S}_\alpha(Q|P)}{1-\alpha}  + S(R|P).\]
Since $S(R| R_{\alpha})\geq 0$, 
\[\inf_{\alpha \in [0,1[}\frac{\alpha s + {\widehat S}_\alpha(Q|P)}{1-\alpha}  + S(R|P)\geq 0.
\]
This gives that $\phi(s)\leq \psi(s)$.  If  Part (2) holds, then also $\phi(s)\geq\psi(s)$ for all $s\in\, ]0, S(P|Q)[$, and we have the equality 
$\phi=\psi$.  To prove  Part (2),  a simple computation gives
\[ S(R_\alpha|Q)=-(1-\alpha){\widehat S}_\alpha^\prime(Q|P)- {\widehat S}_\alpha(Q|P), \qquad 
S(R_\alpha|Q)=S(R_\alpha|P)+ {\widehat S}_\alpha^\prime(Q|P).
\]
After  setting $\alpha =\alpha_\ast(s)$ in these equalities, Part (2) follows from \eqref{div-never} and \eqref{roma-case}. \qed


The main result of this section is
\bet\label{thm-hoeffding}
 For all $s>0$, 
\beq
\bar h(s)=\ubar h(s)= h(s)=\psi(s).
\label{alp}
\eeq
\eet
\demo  Note that the functions $\bar h$, $\ubar h$, $h$ are non-negative and  increasing on $]0, \infty[$ and that 
\beq 
\ubar h(s)\leq \bar h(s)\leq h(s)
\label{no-storm-1}
\eeq
for all $s>0$. 

We shall prove that for all $s\in\, ]0, S(P|Q)[$, 
\beq
h(s)\leq \psi(s), \qquad \ubar h(s)\geq \psi(s).
\label{prove-no-storm}
\eeq
In view of \eqref{no-storm-1}, that proves \eqref{alp} for $s\in\,]0, S(P|Q)[$. Assuming that \eqref{prove-no-storm}
holds, the relations
$h(s)\leq h(S(P|Q))\leq 0$  for $s\in ]0, S(P|Q)[$  and 
\[\lim_{s\uparrow S(P|Q)}h(s)=\lim_{s\uparrow S(P|Q)}\psi(s)=0\]
give that $h(S(P|Q))=0$. Since $h$ is increasing, $h(s)=0$ for $s\geq S(P|Q)$ and so 
$h(s)=\psi(s)$ for $s\geq S(P|Q)$. In the same way one shows that $\bar h(s)=\ubar h(s)= \psi(s)$ for 
$s\geq S(P|Q)$.

We now prove the first inequality in \eqref{prove-no-storm}. Recall that the map  
$\hat \varphi:\,]-S(P|Q), S(Q|P)[\rightarrow  ]0, S(P|Q)[$ is a bijection. Fix $s\in \, ]0, S(P|Q)[$ and let 
$\theta \in ]-S(P|Q), S(Q|P[$ be such that $\hat\varphi(\theta)=s$. Let 
\beq 
T_N(\theta)=\left\{\omega\in \Omega^N\,|\, Q_N(\omega)\geq \e^{N\theta}P_N(\omega)\right\}.
\label{storm-given}
\eeq
Then 
\[
P_N(T_N(\theta))=P_N\left\{ \omega=(\omega_1, \cdots, \omega_N)\in \Omega^N\,|\, \frac{1}{N}\sum_{j=1}^N 
S_{Q|P}(\omega_j)\geq \theta\right\}.
\]
Since the cumulant generating function for $S_{Q|P}$ with respect to $P$ is $\widehat S_{\alpha}(Q|P)$, and the rate 
function $I$ for $S_{Q|P}$ with respect to $P$ coincides with $\varphi$ on  $\,]S(P|Q), S(Q|P)[$, it follows from Part (1) of  Corollary \ref{mm-2}
that 
\beq
\lim_{N\rightarrow \infty}\frac{1}{N}\log P_N(T_N(\theta))=-\varphi(\theta).
\label{str-f}
\eeq
Similarly, 
\[
Q_N([T_N(\theta)]^c)=Q_N\left\{ \omega=(\omega_1, \cdots, \omega_N)\in \Omega^N\,|\, \frac{1}{N}\sum_{j=1}^N 
S_{Q|P}(\omega_j)<\theta\right\}.
\]
The cumulant generating function for $S_{Q|P}$ with respect to $Q$ is ${\widehat S}_{\alpha +1}(Q|P)$, and 
the rate function for $S_{Q|P}$   with respect to $Q$ on   $\,]S(P|Q), S(Q|P)[$ is $\hat \varphi$. Part (2) of Corollary 
\ref{mm-2} yields 
\beq
\lim_{N\rightarrow \infty}\frac{1}{N}\log Q_N([T_N(\theta)]^c)=-\hat \varphi(\theta).
\label{str-f-1}
\eeq
The relations \eqref{str-f} and \eqref{str-f-1}  yield 
that $h(\hat \varphi(\theta))\leq -\varphi(-\theta)$. Since $\hat\varphi(\theta)=s$, the first inequality \eqref{prove-no-storm} 
follows from Part (4) of Proposition \ref{no-storm}.

We now turn to the second inequality in \eqref{prove-no-storm}.
For $\theta \in \, ]-S(P|Q), S(Q|P)[$ and $T_N\subset \Omega^N$ we set 
\[ D_N(T_N, \theta)= Q_N([T_N]^c)+ \e^{\theta N}P_N(T_N).\]
Arguing in the same way as in the proof of Parts (1)-(3) of  Proposition \ref{hypothesis-paris}, one shows 
that for any $T_N$, 
\[D_N(T_N, \theta)\geq D_N(T_N(\theta), \theta).\]
The relations \eqref{str-f} and \eqref{str-f-1} yield 
\[
\lim_{N\rightarrow \infty}\frac{1}{N}\log D_N(T_N(\theta), \theta))=-\hat \varphi(\theta).
\]
Fix now $s\in\,]0, S(P|Q)[$ and let $\theta \in\, ]-S(P|Q), S(Q|P)[$ be such that $\hat \varphi(\theta)=s$. 
Let $(T_N)_{N\geq 1}$ be a sequence of tests such that 
\[
\limsup_{N\rightarrow \infty}\frac{1}{N}\log Q_N(T_N^c)\leq -s.
\]
Then, for any $\theta^\prime$ satisfying $\theta <\theta^\prime < S(Q|P)$ we have 
\beq \begin{split}
-\hat \varphi(\theta^\prime)&=\lim_{N\rightarrow \infty}\frac{1}{N}
\log\left( Q_N([T_N(\theta^\prime)]^c) + \e^{\theta^\prime N}P_N(T_N(\theta^\prime)\right)\\[3mm]
&\leq \liminf_{N\rightarrow \infty}\frac{1}{N}
\log\left( Q_N(T_N^c) + \e^{\theta^\prime N}P_N(T_N)\right)\\[3mm]
&\leq \max\left(\liminf_{N\rightarrow \infty}\frac{1}{N}\log Q_N(T_N^c), \theta^\prime + \liminf_{N\rightarrow \infty}\frac{1}{N}\log P_N(T_N)\right)\\[3mm]
&\leq \max\left(-\hat\varphi(\theta), \theta^\prime + \liminf_{N\rightarrow \infty}\frac{1}{N}\log P_N(T_N)\right).
\end{split}
\label{storm-yields}\eeq
Since $\hat \varphi$ is strictly decreasing on $]-S(P|Q), S(Q|P)[$ we have that  $-\hat \varphi(\theta^\prime)>-\varphi(\theta)$, and \eqref{storm-yields} 
gives 
\[
\liminf_{N\rightarrow \infty}\frac{1}{N}\log P_N(T_N)\geq -\theta^\prime-\hat\varphi(\theta^\prime)=-\varphi(\theta^\prime).
\]
Taking $\theta^\prime \downarrow \theta$, we derive
\[
\liminf_{N\rightarrow \infty}\frac{1}{N}\log P_N(T_N)\geq -\varphi(\theta)=-\varphi(\hat\varphi^{-1}(s))=\psi(s), 
\]
and so   $\ubar h(s)\geq \psi(s)$. \qed

\begin{remark} Theorem \ref{thm-hoeffding} and its proof give the following. 
For any sequence of tests $(T_N)_{N\geq 1}$ such that 
\beq
\limsup_{N\to\infty}\frac1N\log Q_N(T_N^c)\leq -s
\label{hoeff-desp}
\eeq
one has 
\[
\liminf_{N\to\infty}\frac1N\log P_N(T_N)\geq \psi(s).
\]
On the other hand, if  $s\in ]0, S(P|Q)[$, $\hat \varphi(\theta)=s$, and $T_N(\theta)$ is defined by \eqref{storm-given}, 
then 
\[
\limsup_{N\to\infty}\frac1N\log Q_N([T_N(\theta)]^c)= -s\qquad\hbox{and}\qquad \lim_{N\to\infty}\frac1N\log P_N(T_N(\theta))=\psi(s).
\]
\end{remark}

\begin{exo}
Set
\[
\begin{split}
\bar h(0)&=\inf_{(T_N)}\left\{\limsup_{N\to\infty}\frac1N\log P_N(T_N)\,\bigg|\, 
\limsup_{N\to\infty}\frac1N\log Q_N(T_N^c)< 0\right\},\\[3mm]
\ubar h(0)&=\inf_{(T_N)}\left\{\liminf_{N\to\infty}\frac1N\log P_N(T_N)\,\bigg|\,
\limsup_{N\to\infty}\frac1N\log Q_N(T_N^c)< 0\right\},\\[3mm]
h(0)&=\inf_{(T_N)}\left\{\lim_{N\to\infty}\frac1N\log P_N(T_N)\,\bigg|\, 
\limsup_{N\to\infty}\frac1N\log  Q_N(T_N^c)<0 \right\},
\end{split}
\]
where in the last case the infimum is taken over all sequences  of tests $(T_N)_{N\geq 1}$ for which the limit
\[\lim_{N\to\infty}\frac1N\log P_N(T_N)\]
exists. Prove that 
\[\bar h(0)=\ubar h(0)=h(0)=-S(Q|P).
\]
Compare with Exercise \ref{Stein-0}.

\end{exo}
\section{Notes and references}
The relative entropy $S(P|P_{\rm ch})$ already appeared in Shannon's work \cite{Sha}. The definition \eqref{def-rel-ent}
is commonly attributed to Kullback and Leibler \cite{KullLe}, and the relative entropy is sometimes called 
the Kullback-Leibler divergence. From a historical perspective, it is interesting to note that 
  the symmetrized relative entropy $S(P|Q)+ S(Q|P)$ was introduced 
by Jeffreys in \cite{Jeff} (see Equation (1)) in 1946.

The basic properties of the relative entropy described in Section \ref{sec-rel-entropy} are so well-known 
that it is difficult to trace the original sources. The statement of Proposition \ref{pos} is sometimes called Gibbs's 
inequality and sometimes Shannon's inequality. For the references regarding Theorem  \ref{pos-fine} and Exercise 
\ref{csi-ex} see Exercise 17 in Chapter 3 of \cite{CsiK\"o} (note the typo regarding the value of  the constant $c$).

The variational principles discussed in Section \ref{sec-var-pri} are of fundamental importance in statistical 
mechanics and we postpone their discussion to  Part II of the lecture notes. 

The attribution of Theorem \ref{stein-lemma} to statistician Charles Stein appears to be historically 
inaccurate; for a hilarious account of the events that has led to this see the footnote on the page 85 of \cite{John}.
Theorem \ref{stein-lemma} was proven by Hermann Chernoff in  \cite{Che}. To avoid further confusion, 
we have used the usual terminology. 
To the best of my knowledge, the Large Deviations  arguments behind the proof of  Stein's Lemma, which were implicit 
in the original work \cite{Che}, were brought to the surface for the first time in \cite{Ana, Sow}, allowing for a substantial generalization of the original results.\footnote{By this I mean that essentially the same argument yields the proof of Stein's Lemma in a very general 
probabilistic setting.} Our 
proof follows  \cite{Sow}.  

The Fluctuation Relation described in Section \ref{sec-fluct-rel} is behind the    spectacular developments in non-equilibrium 
statistical mechanics mentioned in the Introduction. We will return to this topic in Part II of the lecture notes. 

The choice of the name for Jensen-Shannon entropy (or diveregence) and metric  is unclear; see \cite{Lin}.   To the best of my knowledge, Theorem \ref{sj-metric}
 was first  proven in \cite{EndSc,\"OstVa}. Our proof follows closely \cite{EndSc}. For additional information 
see \cite{FugTo}.

The definition of the R\'enyi relative entropy is usually attributed to \cite{R\'en}, although the "un-normalized"  $\widehat S_\alpha(P|Q)$ already appeared in the work of Chernoff  \cite{Che} in 1952.

The hypothesis testing is an essential procedure in statistics. Its relevance to modern developments in non-equilibrium 
statistical mechanics will be discussed in Part II of the lecture notes. Theorem \ref{chernoff}  is due to 
Chernoff \cite{Che}.  As in the case of Stein's Lemma,  the   LDP based  proof allows to considerably generalize the original result. The Hoeffding error exponents were  first introduced and studied in \cite{Hoe} and the previous remarks regarding the proof applies to them as well. For additional information about hypothesis testing see \cite{LeRo}.

\chapter{Why is the relative entropy natural?}
\label{sec-re-nat}
\section{Introduction}
\label{sec-in-nat}
This chapter is a continuation of Section \ref{sec-e-natural} and concerns naturalness of the relative 
entropy. 

{\bf 1. Operational interpretation.} Following on Shannon's quote in Section \ref{sec-notre-entropy}, 
Stein's Lemma gives an operational interpretation of the relative entropy $S(P|Q)$. Chernoff and Hoeffding error 
exponents, Theorems \ref{chernoff} and \ref{thm-hoeffding}, give an operational interpretation of  R\'enyi's relative entropy $\widehat S_\alpha(P|Q)$ and, 
via formula \eqref{hat-ch}, of R\'enyi's entropy  $\widehat S_\alpha(P)$ as well. Note that this operational interpretation 
of R\'enyi's entropies is rooted in the LDP's for respective entropy functions which are behind the proofs of Theorems \ref{chernoff} 
and \ref{thm-hoeffding}.

{\bf 2. Axiomatic characterizations.} 
Recall that ${\cal A}(\Omega)=\{(P, Q)\in {\cal P}(\Omega)\,|\, P\ll Q\}$. Set ${\cal A}=\cup _\Omega {\cal A}(\Omega)$. 
The axiomatic characterizations of relative entropy concern choice of  a function 
${\mathfrak S}: {\cal A}\rightarrow \rr$ that should qualify as a measure of {\em entropic distinguishability} 
of a pair $(P, Q)\in {\cal A}$. 
The goal is to show that  intuitive natural demands uniquely specify ${\mathfrak S}$ up to a choice of units, 
namely  that for some $c>0$ and all $(P, Q)\in {\cal A}$, ${\mathfrak S}(P, Q)= c S(P|Q)$.

We list    basic  properties that any candidate ${\mathfrak S}$ for  relative entropy should satisfy. 
The obvious ones are 
\beq {\mathfrak S}(P, P)=0, \qquad {\mathfrak S}(P, Q)\geq 0, \qquad \exists\, (P, Q)\,\,\hbox{such that}\,\, {\mathfrak S}(P, Q)>0.
\label{non-train}
\eeq
Another obvious requirement  is that if $|\Omega_1|= |\Omega_2|$ and $\theta:\Omega_1\rightarrow \Omega_2$ is a bijection, 
then for any $(P, Q)\in {\cal A}$, 
\[{\mathfrak S}(P, Q)={\mathfrak S}(P\circ \theta, Q\circ \theta).\]
In other words. 
the distinguishability of a pair  $(P, Q)$ should not depend on the labeling  of the elementary events. 
This requirement gives  that ${\mathfrak S}$ is completely specified by its restriction 
${\mathfrak  S}: \cup_{L\geq 1} {\cal A}_L\rightarrow [0, \infty[$, 
where 
\[{\cal A}_L=\{((p_1,\cdots, p_L), (q_1, \cdots, q_L))\in {\cal P}_L\times {\cal P}_L\,|\, q_k=0\, \Rightarrow p_k=0\},
\]
and that this restriction satisfies 
\beq  {\mathfrak S}((p_1, \cdots, p_L), (q_1, \cdots, q_L)))= {\mathfrak S}((p_{\pi(1)}, \cdots, p_{\pi(L)}), 
(q_{\pi(1)}, \cdots q_{\pi(L)}))
\label{perm-inv-rel}
\eeq
for any $L\geq 1$ and any  permutation $\pi$ of $\{1, \cdots, L\}$. 
In the proofs of Theorems \ref{khin-rel} and \ref{ax-rel-1} we shall assume that \eqref{non-train} and 
\eqref{perm-inv-rel} 
are satisfied.

{\em  Split additivity characterization.}
This axiomatic characterization is the relative entropy analog of Theorem \ref{khin}, and has its  
 roots in the identity (recall Proposition \ref{convex-rel})
\[
\begin{split}
S(p_1P_1 + \cdots +p_nP_n|&q_1Q_1+\cdots+ q_nQ_n )\\[2mm]
& = p_1 S(P_1|Q_1) + \cdots +p_nS(P_n|Q_n) + S((p_1, \cdots, p_n)|(q_1, \cdots q_n))
\end{split}
\]
which holds if $(\supp P_j \cup \supp Q_j)\cap(\supp P_k \cup \supp Q_k)=\emptyset $ for all $j\not=k$. 
\bet
\label{khin-rel}
Let ${\mathfrak S}: {\cal A}\rightarrow [0,\infty[$ be a function such that:
\begin{enumerate}[{\rm (a)}]

\item ${\mathfrak S}$ is continuous on ${\cal A}_2$.  

\item For any finite collection of disjoint sets $\Omega_j$, $j=1, \cdots, n$,  any $(P_j, Q_j )\in {\cal A}(\Omega_j)$, and any 
$p=(p_1, \cdots, p_n), q=(q_1, \cdots, q_n)\in {\cal P}_n$, 
\beq  {\mathfrak  S}\left(\bigoplus_{k=1}^n p_kP_k, \bigoplus_{k=1}^n q_kQ_k\right)=\sum_{k=1}^n p_k {\mathfrak S}(P_k, Q_k) + {\mathfrak  S}(p|q).
\label{rain-sun-rel}
\eeq
\end{enumerate}
Then there exists $c>0$ such that for all $(P, Q)\in {\cal A}$, 
\beq  {\mathfrak S}(P, Q)=c S(P|Q).
\label{long-dm-rel}
\eeq
\eet
\begin{remark} 
If the  positivity and non-triviality assumptions are dropped, then  the proof gives that  \eqref{long-dm-rel} holds for some $c\in \rr$. 
\end{remark} 
 \begin{exo} Following on Remark \ref{split-ver}, can you verbalize the split-additivity property \eqref{rain-sun-rel}?
 \end{exo}
We shall prove Theorem \ref{khin-rel} in Section \ref{sec-khin-rel}. The vanishing assumption 
${\mathfrak S}(P, P)=0$ for all $P$  plays a very important role in the argument. Note that 
\[{\mathfrak S}(P, Q)=-\sum_{\omega}P(\omega)\log Q(\omega)\]
satisfies  (a) and (b) of Theorem \ref{khin-rel} and  assumptions \eqref{non-train} apart from 
${\mathfrak S}(P, P)=0$.

{\em Stochastic monotonicity + super additivity characterization.} This characterization is related to Theorem \ref{ax-entropy}, 
although its proof is both conceptually different and  technically simpler.
The characterization asserts that two  intuitive requirements, the stochastic monotonicity (Proposition \ref{data-sherry}) and 
super-additivity (Proposition \ref{sup-add}) uniquely specify relative entropy.

\bet
\label{ax-rel-1}
Let ${\mathfrak S}: {\cal A}\rightarrow [0,\infty[$ be a function such that:
\begin{enumerate}[{\rm (a)}]

\item ${\mathfrak S}$ is continuous on ${\cal A}_L$ for all $L\geq 1$.

\item For any $P, Q\in {\cal A}(\Omega)$ and any stochastic map $\Phi: {\cal P}(\Omega)\rightarrow 
{\cal P}(\hat \Omega)$ (note that $(\Phi(P), \Phi(Q))\in {\cal A}(\hat \Omega)$), 
\beq {\mathfrak S}(\Phi(P), \Phi(Q))\leq {\mathfrak S}(P, Q).
\label{rain-sun-rel-1}
\eeq
\item For any $P$ and $Q=Q_l\otimes Q_r$ in ${\cal A}(\Omega_l\times \Omega_r)$, 
\beq  {\mathfrak S}(P_l, Q_l) + {\mathfrak S}(P_r, Q_r)\leq {\mathfrak S}(P, Q),
\label{super-late}\eeq
with the equality  iff $P=P_l\otimes P_r$.
\end{enumerate}
Then there exists $c>0$ such that for all $(P, Q)\in {\cal A}$, 
\beq  {\mathfrak S}(P, Q)=c S(P|Q).
\label{long-dm-rel-1}
\eeq
\eet

We shall prove Theorem \ref{ax-rel-1} in Section \ref{sec-ax-rel-1}.  Note that neither assumptions (a) $\wedge$ (b) nor 
(a) $\wedge$  (c) are 
sufficient to deduce
\eqref{long-dm-rel-1}: (a) and (b) hold for the R\'enyi relative entropy  $(P, Q)\mapsto { S}_\alpha(P, Q)$ if 
$\alpha \in ]0, 1[$ ((c) fails here), while (a) and (c) hold for the entropy $(P, Q)\mapsto S(P)$ ((b) fails here, recall 
Exercise \ref{sto-e-f}).

{\bf  4. Sanov's theorem.}  This result  is a deep refinement of Cr\'amer's theorem and the basic indicator 
of the central role the relative entropy plays in the theory of Large Deviations. 
 We continue with our framework: $\Omega$ is a finite set and 
$P$ a given probability measure on $\Omega$. We shall assume that $P$ is faithful. 

To avoid confusion, we shall occasionally denote the generic element of $\Omega$ with  a  letter $a$ (and list the elements of 
$\Omega$ as  
$\Omega=\{a_1, \cdots, a_L\}$). For $\omega \in \Omega$ we denote by $\delta_{\omega}\in {\cal P}(\Omega)$ the pure probability 
measure concentrated at $\omega$: $\delta_\omega(a)=1$ if $a=\omega$ and zero otherwise. For 
$\omega=(\omega_1, \cdots, \omega_N)$ we set 
\[
\delta_{\omega}=\frac{1}{N}\sum_{k=1}^N \delta_{\omega_k}.
\]

Obviously, $\delta_\omega\in {\cal P}(\Omega)$ and 
\[
\delta_{\omega}(a)=\frac{\hbox{the number of times $a$ appears in the sequence $\omega=(\omega_1, \cdots, \omega_N)$}}{N}.
\]
Sanov's theorem concerns the statistics of the map $\Omega^N\ni \omega \mapsto \delta_\omega \in {\cal P}(\Omega)$ w.r.t.  the product 
probability measure $P_N$. The starting point is  the corresponding  law of large numbers.

\bep For any $\epsilon >0$, 
\[
\lim_{N\rightarrow \infty}{ P}_N\left\{\omega \in \Omega^N\,|\, d_V(\delta_\omega, P)\geq \epsilon \right\}=0.
\]
\label{sanov-lln}
\eep
Sanov's theorem concerns  fluctuations in the above LLN, or more precisely, for a given  $\Gamma \subset {\cal P}(\Omega)$,
it estimates the probabilities 
\[
P_N\left\{\omega \in \Omega^N\,|\, \delta_\omega \in \Gamma\right\}
\]
in the limit of large $N$.

\bet For any closed   set $\Gamma\subset {\cal P}(\Omega)$, 
\[
\limsup_{N\rightarrow \infty}\frac{1}{N}\log P_N\left\{\omega \in \Omega^N\,|\, \delta_\omega \in \Gamma\right\}
\leq  -\inf_{Q\in \Gamma}S(Q|P),
\]
and for any open  set $\Gamma \subset {\cal P}(\Omega)$,
\[
\liminf_{N\rightarrow \infty}\frac{1}{N}\log P_N\left\{\omega \in \Omega^N\,|\, \delta_\omega \in \Gamma\right\}
\geq  -\inf_{Q\in \Gamma}S(Q|P).
\]
\label{sanov-ldp}
\eet
We shall prove Proposition \ref{sanov-lln} and Theorem \ref{sanov-ldp} in Section \ref{sec-sanov} where the reader 
can also find  additional information about Sanov's theorem.

\section{Proof of Theorem 5.1}
\label{sec-khin-rel}
The function 
\[F(t)={\mathfrak S}((1,0), (t, 1-t)), \qquad t\in ]0,1],\]
will play an important role in the proof. Obviously,  $F$ is continuous on $]0,1]$ and $F(1)=0$. 

We split the proof into five  steps.

{\bf Step 1.} Let $(P, Q)\in {\cal A}(\Omega)$, where $\Omega=\{\omega_1, \cdots, \omega_n\}$, 
and suppose that $P(\omega_j)=0$ for $j>k$. Set  $\Omega_1=\{\omega_1,\cdots, \omega_k\}$,  
$P_1(\omega_j)=P(\omega_j)$,  and 
\[ Q_1(\omega_j) =\frac{Q(\omega_j)}{Q(\omega_1) +\cdots+ Q(\omega_k)}.
\]
It is obvious that  $(P_1, Q_1)\in {\cal A}(\Omega_1)$. We then have 
\beq {\mathfrak S}(P, Q)= F(q_1+\cdots+ q_k)+ {\mathfrak S}( P_1 , Q_1).
\label{obvious-ottawa}\eeq

Note that if $k=n$,  then \eqref{obvious-ottawa} follows from $F(1)=1$. Otherwise, 
 write $\Omega =\Omega_1\oplus \Omega_2$, with 
$\Omega_2=\{\omega_{k+1}, \cdots, \omega_n\}$. Take any $P_2\in {\cal P}(\Omega_2)$, write 
\[
(P, Q)= (1\cdot P_1\oplus 0 \cdot P_2, t Q_1\oplus  (1-t) Q_2),
\]
where $t= q_1+\cdots +q_k$,  $Q_2$ is arbitrary if $t=1$, and $Q_2(\omega_j)= Q(\omega_j)/(1-t)$ if $t<1$, 
and observe that the statement follows from \eqref{rain-sun-rel-1}.

{\bf Step 2.} $F(ts)= F(t) + F(s)$ for all $s, t\in ]0,1]$. 

Consider ${\mathfrak S}((1,0,0), (ts, t(1-s), 1-t))$. Applying Step 1 with $k=1$ we get 
\[
{\mathfrak S}((1,0,0), (ts, t(1-s), 1-t))=F(ts) + {\mathfrak S}((1), (1))=F(ts).
\]
Applying Step 1 with $k=2$ gives 
\[
{\mathfrak S}((1,0,0), (ts, t(1-s), 1-t))= F(t) + {\mathfrak S}((1,0), (s, 1-s))= F(t) + F(s),
\]
and the  statement follows. 

{\bf Step 3.} For some $c\in \rr$, $F(t)=-c\log t$ for all $t\in ]0, 1]$.

Set $H(s)=F(\e^{-s})$. Then $H$ is continuous on $[0, \infty[$ and satisfies $H(s_1+s_2)= H(s_1) + H(s_2)$. 
It is now a standard exercise to show that $H(s)=c s$ where $c=H(1)$. Setting $t=\e^{-s}$ gives $F(t)=-c\log t$.

This is the only point where the regularity assumption (a) has been used (implying the continuity of $F$), and so obviously (a) can be relaxed.\footnote{It suffices that $F$ is Borel measurable.} Note that \eqref{non-train} implies  $c\geq 0$.

{\bf Step 4.} We now prove  that for any $n\geq 2$ and any pair  $(p, q)\in {\cal A}_n$ of faithful probability measures, 
\beq
{\mathfrak S}(p, q)= c S(p|q),
\label{train-faithful}
\eeq
where  $c$ is the constant from Step 3.

Let $p=(p_1,\cdots, p_n)$, $q=(q_1,\cdots, q_n)$, and choose  $t\in ]0,1]$ such that $q_k- t p_k \geq 0$ for all $k$. Set 
\[
K={\mathfrak S}((p_1, \cdots, p_n, 0,\cdots, 0), (tp_1, \cdots, tp_n, q_1- tp_1,\cdots, q_n-t p_n)).
\]
It follows from Steps 1 and  3 that 
\beq K= F(t) + {\mathfrak S}(p,p)=-c\log t.
\label{train-early}
\eeq
On the other hand, \eqref{perm-inv-rel} and \eqref{rain-sun-rel} yield 
\[
\begin{split}
K&={\mathfrak S}((p_1,0, \cdots, p_n, 0), (tp_1, q_1-tp_1, \cdots, tp_n, q_n-tp_n))\\[2mm]
&={\mathfrak S}\left((p_1(1,0), \cdots, p_n(1,0)), \left(q_1\left(\frac{tp_1}{q_1}, 1-\frac{tp_1}{q_1}\right)\right), \cdots, 
q_n\left(\frac{tp_n}{q_n}, 1-\frac{tp_n}{q_n}\right)\right)\\[2mm]
&=\sum_{k=1}^n p_k F\left(\frac{tp_k}{q_k}\right) + {\mathfrak S}(p, q),
\end{split}
\]
and it follows from Step 3 that 
\beq
\label{train-early-1}
K= -c \log t - c S(p|q) + {\mathfrak S}(p,q).
\eeq
Comparing \eqref{train-early} and \eqref{train-early-1} we derive \eqref{train-faithful}.

{\bf Step 5.} We now show that \eqref{train-faithful} also holds for non-faithful $p$'s and complete the proof of 
Theorem \ref{khin-rel}. By \eqref{perm-inv-rel} we may assume that $p_j>0$ for $j\leq k$ and $p_j=0$ for $j>k$, where 
$k<n$. 
Then, setting $s=q_1+\cdots q_k$, Steps 1 and  3 yield 
\[
{\mathfrak S}(p, q)= -c \log s +{\mathfrak S}((p_1, \cdots, p_k), (q_1/s, \cdots, q_k/s)),
\]
and it follows from  Step 4 that
\[
{\mathfrak S}(p, q)= -c \log s +c S((p_1, \cdots, p_k) |(q_1/s, \cdots, q_k/s)).
\]
On the other hand, a direct computation gives 
\[
S(p|q)=- \log s + S((p_1, \cdots, p_k) |(q_1/s, \cdots, q_k/s)),
\]
and so ${\mathfrak S}(p, q)=cS(p|q)$. 

The non-triviality assumption that ${\mathfrak S}$ is not vanishing on ${\cal A}$ gives that $c>0$. 

\section{Proof of Theorem 5.2}
\label{sec-ax-rel-1}
We shall need the following preliminary result which is of independent interest and which we will prove at the end 
of this section. Recall that 
if $P$ is a probability measure on $\Omega$, then $P_N=P\otimes \cdots \otimes P$ is  the 
product probability measure on $\Omega^N=\Omega\times \cdots\times \Omega$.

\bep Suppose that $(P, Q)\in {\cal A}(\Omega)$  and  $(\widehat P, \widehat Q)\in {\cal A}(\widehat \Omega)$  are such that
$S(P|Q)> S(\widehat P| \widehat Q)$. Then there exists  a sequence of stochastic maps $(\Phi_N)_{N\geq 1}$, 
$\Phi_N : {\cal P}(\Omega^N)\rightarrow {\cal P}(\widehat \Omega^N)$ such that   
$\Phi_N(Q_N)=\widehat Q_N$ for all $N\geq 1$ and 
\[
\lim_{N\rightarrow \infty}d_V(\Phi_N(P_N), \widehat P_N)=0.
\]
\label{sleep-train}
\eep

We now turn to the proof of  Theorem \ref{ax-rel-1}.
Recall our standing assumptions \eqref{non-train}. 
Let $(P^{(0)}, Q^{(0)})\in {\cal A}$ be such that ${\mathfrak S}(P^{(0)}, Q^{(0)})>0$, and let  $c>0$ 
be such that 
\[ {\mathfrak S}(P^{(0)}, Q^{(0)})= cS(P^{(0)}|Q^{(0)}).\]
Let  $(P, Q)\in {\cal A}$, $P\not=Q$, be given and  let $L, M, L^\prime, M^\prime$ be positive integers 
such that 
\beq
\frac{L^\prime}{M^\prime}S(P^{(0)}|Q^{(0)})< S(P|Q)<\frac{L}{M}S(P^{(0)}|Q^{(0)}).
\label{train-recal}
\eeq
We work first with the r.h.s. of this inequality which can be rewritten as 
\[
S(P_M|Q_M)<S(P_L^{(0)}|Q_L^{(0)}).
\]
It follows from Proposition \ref{sleep-train} that there exists a sequence of stochastic maps 
$(\Phi_N)_{N\geq 1}$ such that $\Phi_N(Q_{LN}^{(0)})= Q_{MN}$ and 
\beq
\lim_{N\rightarrow \infty}d_V(\Phi_N(P_L^{(0)}), P_{MN})=0.
\label{train-never}
\eeq
We now turn to ${\mathfrak S}(P, Q)$ and note that 
\beq
\label{train-voila}
\begin{split}
M{\mathfrak S}(P, Q)&= {\mathfrak S}(P_M, Q_M)=\frac{1}{N}{\mathfrak S}(P_{MN}, Q_{MN})\\[3mm]
&=\frac{1}{N}\left[ {\mathfrak S}(P_{MN}, Q_{MN})- {\mathfrak S}(\Phi_N(P_L^{(0)}), Q_{MN})\right] 
+ \frac{1}{N}{\mathfrak S}(\Phi_N(P_L^{(0)}), \Phi_N (Q^{(0)}_{LN}))\\[3mm]
&\leq \frac{1}{N}\left[ {\mathfrak S}(P_{MN}, Q_{MN})- {\mathfrak S}(\Phi_N(P_L^{(0)}), Q_{MN})\right]  +
\frac{1}{N}{\mathfrak S}(P_{LN}^{(0)}, Q^{(0)}_{LN})\\[3mm]
&=\frac{1}{N}\left[ {\mathfrak S}(P_{MN}, Q_{MN})- {\mathfrak S}(\Phi_N(P_L^{(0)}), Q_{MN})\right] +
L{\mathfrak S}(P_{L}^{(0)}, Q^{(0)}).
\end{split}
\eeq
Write $Q_{MN}= Q_M\otimes \cdots \otimes Q_M$ and denote by 
$R_{k, N}$ the marginal of $\Phi_N(P_L^{(0)})$ with the respect to the $k$-th component of 
this decomposition. Assumption (c) gives
\beq 
\frac{1}{N}\left[ {\mathfrak S}(P_{MN}, Q_{MN})- {\mathfrak S}(\Phi_N(P_L^{(0)}), Q_{MN})\right]
\leq \frac{1}{N}\sum_{k=1}^N \left[ {\mathfrak S}(P_M, Q_M)-{\mathfrak S}(R_{k, N}, Q_M)
\right].
\label{train-really}
\eeq
One easily shows that \eqref{train-never} implies that for any $k$, 
\beq \lim_{N\rightarrow\infty}d_V(R_{k,N}, P_M)=0.
\label{train-to}
\eeq
 It then follows from \eqref{train-really} 
that 
\beq
\limsup_{N\rightarrow \infty}
\frac{1}{N}\left[ {\mathfrak S}(P_{MN}, Q_{MN})- {\mathfrak S}(\Phi_N(P_L^{(0)}), Q_{MN})\right]\leq 0.
\label{train-gives}
\eeq
Returning to \eqref{train-voila}, \eqref{train-gives} yields 
\beq {\mathfrak S}(P, Q)\leq \frac{L}{M}{\mathfrak  S}(P^{(0)}, Q^{(0)})=\frac{L}{M}c S(P^{(0)}| Q^{(0)}).
\label{train-derive}
\eeq
Since the only constraint regarding the choice of  $L$ and $M$ is that \eqref{train-recal} holds, 
we derive from \eqref{train-derive} that 
\[{\mathfrak S}(P, Q)\leq c S(P|Q).\]
 Starting  with the l.h.s. of the inequality \eqref{train-recal} and repeating the above argument 
 one derives that ${\mathfrak S}(P, Q)\geq c S(P|Q)$. Hence,  ${\mathfrak S}(P, Q)= c S(P|Q)$ for all $(P, Q)\in{\cal A}$ with 
 $P\not=Q$. Since this relation holds trivially for $P=Q$,  the proof is complete.
 \qed
 
 \begin{exo} Prove that \eqref{train-never} implies \eqref{train-to}.
 \end{exo}
 
 {\bf Proof of Proposition \ref{sleep-train}.} The statement is trivial if $\widehat P=\widehat Q$, so we assume  that 
 $\widehat P\not=\widehat Q$ (hence  $S(\widehat P|\widehat Q)>0$). Let  $t, \hat t $ be such  that 
 \[ S(\widehat P|\widehat Q) <\hat t < t < S(P|Q).\]
It follows from Stein's Lemma  that one can find a sequence of sets $(T_N)_{N\geq 1}$,  $ T_N\subset  \Omega_N$, such that 
\[
\lim_{N\rightarrow \infty}P_N( T_N)=1, \qquad  Q_N( T_N) \leq C_1\e^{-Nt},
\]
for some constant $C_1>0$. Let $\Psi_N: {\cal P}( \Omega)\rightarrow {\cal P}(\{0,1\})$ be a stochastic 
map induced by the  matrix 
\[\Psi_N(\omega, 0)=\chi_{ T_N}(\omega), \qquad \Psi_N(\omega, 1)=\chi_{ T_N^c}(\omega),
\]
where $\chi_{T_N}$ and $\chi_{ T_N^c}$ are the characteristic functions of $T_N$ and its complement $T_N^c$. 
It follows that 
\[\Psi_N( P_N)=(p_N, \bar p_N), \qquad \Psi( Q_N)=(q_N, \bar q_N),\]
where 
\[p_N= P_N(T_N), \qquad  q_N= Q(T_N).\]
Obviously $\bar p_N=1-p_N$, $\bar q_N=1-q_N$.

It follows again from Stein's Lemma that one can find a sequence of sets $(\widehat T_N)_{N\geq 1}$, $\widehat T_N \subset \widehat \Omega_N$,
 such that 
\[
\lim_{N\rightarrow \infty} \widehat P_N(\widehat T_N)=1, \qquad Q_N(\widehat T_N^c)>C_2\e^{-N \hat t},
\]
for some constant $C_2>0$. 
We now construct a stochastic map $\widehat \Psi_N: {\cal P}(\{0,1\})\rightarrow {\cal P}(\widehat \Omega)$ as follows. 
Let  $\delta_0=(1,0)$,  $\delta_1=(0,1)$. We set first 
\[
\widehat \Psi_N(\delta_0)(\omega)=\frac{\widehat P_N(\omega)}{\sum_{\omega^\prime \in \widehat T_N}\widehat P_N(\omega^\prime)}\qquad \hbox{if}\,\, 
\omega \in \hat T_N,
\]
$\hat \Psi_N(\delta_0)(\omega)=0$ otherwise, 
and observe that 
\[
d_V(\widehat \Psi_N(\delta_0), \widehat P_N)\leq \widehat P_N(\widehat T_N^c)+ \frac{1- \widehat P_N(\widehat T_N)}{\widehat P_N(\widehat T_N)}.
\]
Hence, 
\[\lim_{N\rightarrow \infty}d_V(\widehat \Psi_N(\delta_0), \widehat P_N)=0.\]
Let 
\[
D_N(\omega)= \widehat Q_N(\omega)- q_N \Phi_N(\delta_0)(\omega).
\]
If $\omega \not \in \widehat T_N$, then obviously $D_N(\omega)= \widehat Q_N(\omega)\geq 0$, and if 
$\omega \in \widehat T_N$, 
\[
D_N(\omega)\geq C_2^{-\hat t N}- c_1\e^{-t N}.
\]
Since $0 <\hat t <t$, there is $N_0$ such that for $N\geq N_0$ and all $\omega \in \widehat \Omega$, $D_N(\omega)\geq 0$.
From now on we assume that $N\geq N_0$, set 
\[ \widehat \Psi_N(\delta_1)=\frac{1}{\bar q_N} (Q_N- q_N \Phi_N(\delta_0)),
\]
and define  $\widehat \Psi_N:{\cal P}(\{0,1\}) \rightarrow {\cal P}(\widehat \Omega)$ by 
\[\widehat \Psi_N (p,q)= p\Psi(\delta_0) + q \Psi(\delta_1).\]
The map $\widehat \Psi_N$ is obviously stochastic and 
\[\widehat \Psi_N(q_N, \bar q_N)=\widehat Q_N.
\]
Moreover, 
\[
\begin{split}
d_V(\widehat \Psi_N(p_N, \bar  p_N), \widehat P_N)&\leq d_V( \widehat \Psi_N(p_N, \bar p_N), \widehat \Psi_N(\delta_0)) + 
d_V(\widehat \Psi_N(\delta_0), \widehat P_N)\\[2mm]
&\leq 2(1-p_N) + d_V(\widehat \Psi_N(\delta_0), \widehat P_N),
\end{split}
\]
and so 
\[
\lim_{N\rightarrow \infty}d_V(\widehat \Psi_N(p_N, \bar  p_N), \widehat P_N)=0.
\]
For $N<N_0$ we take for $\Phi_N$ an arbitrary stochastic map satisfying $\Phi_N(Q_N)=\widehat Q_N$ 
and  for $N\geq N_0$ we set $\Phi_N =\widehat \Psi_N \circ \Psi_N$. Then $\Phi_N(Q_N)= \widehat Q_N$ for all 
$N\geq 1$ and 
\[\lim_{N\rightarrow \infty}d_V(\Phi_N(P_N), \widehat P_N)=0,
\]
proving the proposition. \qed

\begin{exo} Write down the stochastic matrix that induces $\widehat \Psi_N$.
\end{exo}
\section{Sanov's theorem}
\label{sec-sanov}
We start with

{\bf Proof of Proposition \ref{sanov-lln}. } Recall that $L=|\Omega|$. We have 
\[
d_V(\delta_\omega, P)=\sum_{a\in \Omega}
\left|\frac{\sum_{k=1}^N\delta_{\omega_k}(a)}{N} - P(a)\right|, 
\]
and 
\[
\left\{\omega \in \Omega^N\,|\, d_V(\delta_\omega, P)\geq \epsilon \right\}\subset 
\bigcup_{a\in \Omega}\left\{\omega \in \Omega^N\,\big|\,\left|\frac{\sum_{k=1}^N\delta_{\omega_k}(a)}{N} - P(a)\right|\geq 
\frac{\epsilon}{L}\right\}.
\]
Hence, 
\beq\label{beau-d}{P}_N\left\{\omega \in \Omega^N\,|\, d_V(\delta_\omega, P)\geq \epsilon \right\}
\leq \sum_{a\in \Omega}P_N\left\{\omega \in \Omega^N\,\big|\,\left|\frac{\sum_{k=1}^N\delta_{\omega_k}(a)}{N} - P(a)\right|\geq 
\frac{\epsilon}{L}\right\}.
\eeq
For given $a\in \Omega$, consider a random variable $X:\Omega \rightarrow \rr$ defined by $X(\omega)=\delta_\omega(a)$. 
Obviously, ${\mathbb E}(X)=P(a)$ and the LLN yields that 
\[
\lim_{N\rightarrow \infty}P_N\left\{\omega \in \Omega^N\,\big|\,\left|\frac{\sum_{k=1}^N\delta_{\omega_k}(a)}{N} - P(a)\right|\geq 
\frac{\epsilon}{L}\right\}=0.
\]
The proposition follows by combining this observation with inequality (\ref{beau-d}). \qed

We now turn to the proof of Sanov's theorem. Recall the assumption that $P$ is faithful.  We start with the upper bound. 
\bep\label{sanov-upper}Suppose that $\Gamma \subset {\cal P}(\Omega)$ is a closed set. Then 
\[
\limsup_{N\rightarrow \infty}\frac{1}{N}\log P_N\left\{\omega \in \Omega^N\,|\, \delta_\omega \in \Gamma\right\}
\leq -\inf_{Q\in \Gamma}S(Q|P).
\]
\eep
\begin{remark} Recall  that the map ${\cal P}(\Omega)\ni Q\mapsto S(Q|P)\in [0, \infty[$ is continuous 
($P$ is faithful). Since $\Gamma$ is compact,  there exists $Q_m\in {\cal P}(\Omega)$ such that 
\[
\inf_{Q\in \Gamma}S(Q|P)=S(Q_m|P).
\]
\end{remark}
\demo 
Let $\epsilon >0$ be given.
Let $Q\in \Gamma$. By Exercise \ref{exe-var}, 
\[S(Q|P)=\sup_{X:\Omega \rightarrow \rr}\left(\int_\Omega X\d Q- \log \int_\Omega \e^X\d P\right).
\]
Hence, we can find $X$ such that 
\[
S(Q|P)-\epsilon <\int_\Omega X \d Q-\log \int_\Omega \e^X \d P.
\]
Let 
\[ U_{\epsilon}(Q)=\left\{ Q^\prime \in {\cal P}(\Omega)\,\big|\, \left|\int_\Omega X\d Q - \int_{\Omega}X \d Q^\prime\right|<\epsilon 
\right\}.
\] 
Since the map ${\cal P}(\Omega)\ni Q^\prime \mapsto \int_\Omega X \d Q^\prime$ is continuous, $U_\epsilon(Q)$ is an open 
subset of ${\cal P}(\Omega)$. We now estimate 
\[
\begin{split}
P_N\left\{ \delta_{\omega}\in U_\epsilon(Q)\right\}&= 
P_N\left\{\left|\int_\Omega X\d Q - \int_\Omega X\d \delta_\omega\right|<\epsilon\right\}\\[2mm]
&\leq P_N\left\{\int_\Omega X\delta_\omega > \int_{\Omega}X \d Q-\epsilon\right\}\\[2mm]
&= P_N\left\{ \sum_{k=1}^N X(\omega_k)> N\int_{\Omega}X \d Q-N\epsilon\right\}\\[2mm]
&=P_N\left\{\e^{\sum_{k=1}^N X(\omega_k)}> \e^{N\int_{\Omega}X \d Q-N\epsilon}\right\}\\[2mm]
&\leq \e^{-N\int_\Omega X \d Q + N\epsilon}{\mathbb E}(\e^X)^N\\[2mm]
&=\e^{-N \int_\Omega X \d Q + N\log \int_\Omega \e^X \d P + N\epsilon}\\[2mm]
&\leq \e^{-N S(Q|P) + 2N\epsilon}
\end{split}
\]
Since $\Gamma$ is compact, we can find $Q_1, \cdots, Q_M\in \Gamma$  such that 
\[ \Gamma \subset \bigcup_{j=1}^M U_\epsilon(Q_j).
\]
Then 
\[
\begin{split}
P_{N}\left\{\delta_\omega \in \Gamma\right\}&\leq \sum_{j=1}^M P_N\left\{\delta_\omega \in U_\epsilon (Q_j)\right\}\\[2mm]
&\leq \e^{2N\epsilon}\sum_{j=1}^M\e^{-NS(Q_j|P)}\\[2mm]
&\leq \e^{2N\epsilon}M\e^{-N \inf_{Q\in \Gamma}S(Q|P)}.
\end{split}
\]
Hence
\[
\limsup_{N\rightarrow \infty}\frac{1}{N}\log P_N\left\{\omega \in \Omega^N\,|\, \delta_\omega \in \Gamma\right\}
\leq -\inf_{Q\in \Gamma}S(Q|P) + 2 \epsilon.
\]
Since $\epsilon >0$ is arbitrary, the statement follows. \qed

We now turn to the lower bound. 

\bep\label{sanov-lower} For any open set $\Gamma\subset {\cal P}(\Omega)$, 
\[
\liminf_{N\rightarrow \infty}\frac{1}{N}\log P_N\left\{\omega \in \Omega^N\,|\, \delta_\omega \in \Gamma\right\}
\geq  -\inf_{Q\in \Gamma}S(Q|P).
\]
\eep
\demo Let $Q\in \Gamma$ be faithful.  Recall that $S_{Q|P}=\log \Delta_{Q|P}$ and 
\[
\int_\Omega S_{P|Q}\d \delta_\omega = \frac{S_{Q|P}(\omega_1)+ \cdots + S_{Q|P}(\omega_N)}{N}.
\]
Let $\epsilon >0$ and 
\[ 
R_{N, \epsilon}=\left\{\delta_\omega \in \Gamma\,\big|\, \left|\int_{\Omega} S_{Q|P}\d \delta_\omega - S(Q|P)\right| <\epsilon\right\}.
\]
Then 
\[
\begin{split}
P_N\left\{\delta_\omega \in \Gamma\right\}&\geq P_{N}(R_{N, \epsilon})=
\int_{R_{N, \epsilon}}\Delta_{P_N|Q_N}\d Q_N=\int_{R_{N, \epsilon}}\Delta_{Q_N|P_N}^{-1}\d Q_N\\[2mm]
&=\int_{R_{N, \epsilon}}\e^{-\sum_{k=1}^N S_{Q|P}(\omega_k)}\d Q_N\\[3mm]
&\geq 
\e^{- N S(Q|P) - N\epsilon}Q_N(R_{N, \epsilon}).
\end{split}
\]
Note that for $\epsilon$ small enough ($\Gamma$ is open!) 
\[
R_{N, \epsilon}\supset \left\{\omega \in \Omega^N\,|\, d_{V}(Q, \delta_\omega)<\epsilon\right\}\cap 
\left\{\omega \in \Omega^N\,\big|\, \left| \frac{S_{Q|P}(\omega_1)+ \cdots + S_{Q|P}(\omega_N)}{N} - S(Q|P)\right|<\epsilon\right\}.
\]
By the LLN, 
\[\lim_{N\rightarrow \infty} Q_{N}(R_{N, \epsilon})=1.
\]
Hence,  for any faithful $Q\in \Gamma$, 
\beq\label{home-home}
\liminf_{N\rightarrow \infty}\frac{1}{N}\log P_N\left\{\omega \in \Omega^N\,|\, \delta_\omega \in \Gamma\right\}
\geq  -S(Q|P).
\eeq
Since $\Gamma$ is open and the map ${\cal P}(\Omega)\ni Q \rightarrow S(Q|P)$ is continuous, 
\beq\label{exer}
\inf_{Q\in \Gamma\cap {\cal P}_{\rm f}(\Omega)}S(Q|P)=\inf_{Q\in \Gamma}S(Q|P).
\eeq
The relations  (\ref{home-home}) and (\ref{exer}) imply 
\[
\liminf_{N\rightarrow \infty}\frac{1}{N}\log P_N\left\{\omega \in \Omega^N\,|\, \delta_\omega \in \Gamma\right\}
\geq  -\inf_{Q\in \Gamma}S(Q|P).
\]
\qed

\begin{exo}  Prove the identity (\ref{exer}).
\end{exo}

A set $\Gamma\in {\cal P}(\Omega)$ is called  {\em Sanov-nice} if 
\[
\inf_{Q\in {\rm int}\, \Gamma}S(Q|P)=\inf_{Q\in {\rm cl}\, \Gamma}S(Q|P), 
\]
where int/cl stand for the interior/closure. 
If $\Gamma$ is Sanov-nice, then 
\[
\lim _{N\rightarrow \infty}\frac{1}{N}\log P_N\left\{\omega \in \Omega^N\,|\, \delta_\omega \in \Gamma\right\}=-\inf_{Q\in \Gamma}
S(Q|P).
\]
\begin{exo}
\exop Prove that any open set $\Gamma\subset {\cal P}(\Omega)$ is Sanov-nice.
\exop Suppose that $\Gamma\subset {\cal P}(\Omega)$ is convex and has non-empty interior. 
Prove that $\Gamma$ is Sanov-nice.
\end{exo}

We now show that  Sanov's theorem implies  Cram\'er's theorem. The argument we shall use  is an example  of the  powerful 
{\em contraction principle} in theory of Large Deviations. 

Suppose that in addition to $\Omega$ and $P$ we are given a random variable $X:\Omega \rightarrow \rr$.
$C$ and $I$ denote the cumulant generating function and the rate function of $X$. Note that 
\[
\frac{{\cal S}_N(\omega)}{N}=
\frac{X(\omega_1) + \cdots+ X(\omega_N)}{N}=\int_{\Omega}X\d\delta_\omega.
\]
Hence, for any $S\subset  \rr$, 
\[
\frac{{\cal S}_N(\omega)}{N}\in S \,\, \Leftrightarrow\,\, \delta_\omega \in \Gamma_S,
\]
where 
\[ \Gamma_S=\left\{ Q\in {\cal P}(\Omega)\,\big|\, \int_\Omega X\d Q \in S\right\}.
\]
\begin{exo} 
Prove that 
\[{\rm int}\, \Gamma_S=\Gamma_{{\rm int} S}, \qquad {\rm cl}\, \Gamma_S=\Gamma_{{\rm cl}S}.
\]
\end{exo}
Sanov's theorem and the last exercise yield 
\bep\label{san-cra} For any $S\subset \rr$, 
\[
\begin{split}
-\inf_{Q\in  \Gamma_{{\rm int}S}}
S(Q|P)&\leq \liminf _{N\rightarrow \infty}\frac{1}{N}\log P_N\left\{\omega \in \Omega^N\,\big|\, \frac{{\cal S}_N(\omega)}{N}\in S\right\}\\[2mm]
&\leq \limsup_{N\rightarrow \infty}\frac{1}{N}\log P_N\left\{\omega \in \Omega^N\,\big|\, \frac{{\cal S}_N(\omega)}{N}\in S  \right\}\leq 
-\inf_{Q\in  \Gamma_{{\rm cl}S}}S(Q|P),
\end{split}
\]
\eep
To relate this result to  Cram\'er's theorem we need:
\bep For any $S\subset \rr$,  
\beq\label{flash}\inf_{\theta\in S}I(\theta)=\inf_{Q\in \Gamma_S}S(Q|P).
\eeq
\label{contraction}\eep
\demo Let $Q\in {\cal P}(\Omega)$. An application of Jensen's inequality gives that for all $\alpha \in \rr$, 
\[
\begin{split}
C(\alpha)&=\log \left(\sum_{\omega \in \Omega} \e^{\alpha X(\omega)}P(\omega)\right)\\[2mm]
&\geq \log \left(\sum_{\omega \in {\rm supp}Q} \e^{\alpha X(\omega)}\frac{P(\omega)}{Q(\omega)}Q(\omega)\right)\\[2mm]
&\geq \sum_{\omega\in \supp Q}Q(\omega)\log \left[\e^{\alpha X(\omega)}\frac{P(\omega)}{Q(\omega)}\right].
\end{split}
\]
Hence, 
\beq C(\alpha)\geq \alpha \int_\Omega X\d Q - S(Q|P).
\label{l-snow}
\eeq
If $Q$ is such that  $\theta_0 =\int_\Omega X \d Q\in S$,  then \eqref{l-snow} gives
\[ S(Q|P)\geq \sup_{\alpha \in \rr}\left(\alpha \theta_0   - C(\alpha)\right)= I(\theta_0)\geq \inf_{\theta \in S}I(\theta),
\]
and so 
\beq \inf_{Q\in \Gamma_S}S(Q|P)\geq \inf_{\theta \in S}I(\theta).
\label{well-st}
\eeq
One the other hand,  if $\theta \in ]m, M[$, where  $m=\min _{\omega \in \Omega}X(\omega)$ and $M=\max_{\omega \in 
\Omega}X(\omega)$, and $\alpha=\alpha(\theta)$ is such that $C^\prime(\alpha(\theta))=\theta$, then, with $Q_\alpha$ 
defined by \eqref{def-Q-alpha} (recall also the proof of Cramer's theorem), 
$\theta=\int_\Omega X\d Q_\alpha$ and $S(Q_\alpha|P)=\alpha\theta - C(\alpha)= I(\theta)$. Hence, if $S\subset\, ]m, M[$, then for 
any $\theta_0\in S$, $\inf_{Q\in \Gamma_S}S(Q|P)\leq I(\theta_0)$, and so 
\beq  \inf_{Q\in \Gamma_S}S(Q|P)\leq \inf_{\theta \in S}I(\theta).
\label{well-st-1}
\eeq
It follows from \eqref{well-st} and \eqref{well-st-1} that \eqref{flash} holds for $S\subset\,]m, M[$. 
One checks directly  that 
\beq\label{exercise-new} I(m)=\inf_{Q: \int_\Omega X \d Q=m} S(Q|P), \qquad  I(M)=\inf_{Q: \int_\Omega X \d Q=M} S(Q|P).
\eeq
If $S\cap [m, M]=\emptyset$, then both sides in (\ref{flash}) are $\infty$ (by definition, $\inf \emptyset =\infty$). Hence, 
\[
\inf_{\theta\in S}I(\theta)=\inf_{\theta\in S\cap [m, M]}I(\theta)=\inf_{Q\in \Gamma_{S\cap [m, M]}}S(Q|P)=
\inf_{Q\in \Gamma_{S}}S(Q|P),
\]
and the statement follows. \qed

\begin{exo}Prove the identities (\ref{exercise-new}).
\end{exo}

Propositions \ref{san-cra} and \ref{contraction}  yield the following generalization of Cram\'er's theorem:
\bet\label{cra-gen} For any $S\subset \rr$, 
\[
\begin{split}
-\inf_{\theta \in {\rm int} S} I(\theta)&\leq \liminf _{N\rightarrow \infty}\frac{1}{N}\log P_N\left\{\omega \in \Omega^N\,\big|\, \frac{{\cal S}_N(\omega)}{N}\in S\right\}\\[2mm]
&\leq \limsup_{N\rightarrow \infty}\frac{1}{N}\log P_N\left\{\omega \in \Omega^N\,\big|\, \frac{{\cal S}_N(\omega)}{N}\in S  \right\}\leq 
-\inf_{\theta \in {\rm cl} S}I(\theta).
\end{split}
\]
\eet
A set $S$ is called {\em Cramer-nice} if 
\[
\inf_{\theta \in {\rm int} S}I(\theta)=\inf_{\theta \in {\rm cl } S}I(\theta).
\]
Obviously, if $S$ is Cramer-nice, then 
\[
\lim _{N\rightarrow \infty}\frac{1}{N}\log P_N\left\{\omega \in \Omega^N\,\big|\, \frac{{\cal S}_N(\omega)}{N}\in S\right\}=-\inf_{\theta \in S}
I(\theta).
\]
\begin{exo} 

\exop Is it true that any open/closed interval is Cram\'er-nice?

\exop Prove that any open set $S\subset ]m, M[$ is Cram\'er-nice.

\exop Describe all open sets that are Cram\'er-nice.
\end{exo}

\section{Notes and references}
Theorem \ref{khin-rel} goes back to the work of  Hobson \cite{Hob} in 1969. Following in Shannon's step, 
Hobson has proved Theorem \ref{khin-rel} under the additional assumptions that ${\mathfrak  S}$ is continuous 
on ${\cal A}_L$ for all $L\geq 1$, and that the function 
\[(n, n_0) \mapsto {\mathfrak S}\left(\left(\frac{1}{n}, \cdots, \frac{1}{n}, 0, \cdots, 0\right), \left(\frac{1}{n_0}, 
\cdots, \frac{1}{n_0}\right)\right),
\]
defined for $n\leq  n_0$, is an increasing function of $n_0$ and a decreasing function of $n$.  
Our proof of Theorem \ref{khin-rel} follows closely \cite{Lei} where the reader can find additional information 
about the history of this result. 

The formulation and the proof of Theorem \ref{ax-rel-1} are based on the recent works \cite{Mat, WiGaEi}. 

For additional information about axiomatizations of relative entropy we refer the reader to Section 7.2 in 
\cite{AczDa}.

Regarding Sanov's theorem, for the original references and additional information we refer the reader 
to \cite{DeZe, CovTh}. In these monographs one can also find a purely combinatorial proof 
of Sanov's theorem and we urge the reader to study this alternative proof. As in the case of 
Cram\'er's theorem, the proof presented 
here has the advantage that it extends to a  much more general setting that will be discussed in the Part II 
of the lecture notes. 

\chapter{Fisher entropy}
\label{sec-fisher-entropy}
\section{Definition and basic properties}
\label{sec-fisher-def}
Let $\Omega$ be a finite set and $[a, b]$ a bounded closed interval  in $\rr$.  To avoid trivialities, 
we shall always assume that $|\Omega|=L>1$. Let 
$\{P_{\theta}\}_{\theta \in [a,b]}$, $P_\theta \in {\cal P}_{\rm f}(\Omega)$, be a family of faithful probability measures 
on $\Omega$ indexed by points $\theta\in [a,b]$. We shall assume that the functions $[a, b]\ni \theta \mapsto P_\theta(\omega) $ are 
$C^2$ (twice  continuously differentiable) for all $\omega\in \Omega$. The expectation and variance with respect to $P_\theta$ are 
denoted by ${\mathbb E}_\theta$ and ${\rm Var}_\theta$. The entropy function is denoted by 
$S_\theta=-\log P_\theta$. The derivatives w.r.t. $\theta$ are denoted as $\dot f(\theta)=\partial_\theta f(\theta)$, 
$\ddot f(\theta)= \partial_\theta^2f(\theta)$, etc. Note that 
\[
\dot S_\theta=-\frac{\dot P_\theta}{P_\theta}, \qquad \ddot S_\theta=-\frac{\ddot P_\theta}{P_\theta} + \frac{\dot P_\theta^2}{P_\theta^2},\qquad {\mathbb E}_\theta(\dot S_\theta)=0.
\]

The Fisher entropy  of $P_\theta$  is defined by 
\[ {\cal I}(\theta)={\mathbb E}_\theta ([\dot S_\theta]^2)=\sum_{\omega \in \Omega}\frac{[\dot P_\theta(\omega)]^2}{P_\theta(\omega)}.\]
Obviously, 
\[
 {\cal I}(\theta)= {\rm Var}_\theta(\dot S_\theta)= {\mathbb E}_\theta (\ddot S_\theta).
 \]
\begin{example} Let $X: \Omega \rightarrow \rr$ be a random variable and  
\[P_\theta(\omega)=\frac{\e^{\theta X(\omega)}}{\sum_{\omega^\prime} \e^{\theta X(\omega^\prime)}}. \]
Then 
\[
{\cal I}(\theta)={\rm Var}_\theta(X). 
\]
\end{example}
The Fisher entropy arises by considering  local relative entropy distortion of $P_\theta$. Fix 
$\theta \in I$ and set 
\[  L(\epsilon)= S(P_{\theta +\epsilon}|P_\theta), \qquad R(\epsilon)= S(P_\theta|P_{\theta +\epsilon}).
\]
The functions $\epsilon\mapsto L(\epsilon)$ and $\epsilon \mapsto R(\epsilon)$ are well-defined in a neighbourhood 
of $\theta$ (relative to the interval $[a,b]$). An elementary computation yields:
\bep 
\[
\lim_{\epsilon \rightarrow 0}\frac{1}{\epsilon^2}L(\epsilon)= \lim_{\epsilon \rightarrow 0}\frac{1}{\epsilon^2}R(\epsilon)=
\frac{1}{2}{\cal I}(\theta). 
\]
\label{fisher-1}
\eep
In terms of the Jensen-Shannon  entropy and metric we have 
\bep
\[ 
\begin{split}
 \lim_{\epsilon \rightarrow 0}\frac{1}{\epsilon^2} S_{\rm JS}(P_{\theta+\epsilon}, P_\theta)&=\frac{1}{4} {\cal I}(\theta),\\[3mm]
\lim_{\epsilon \rightarrow 0}\frac{1}{|\epsilon|}d_{\rm JS}(P_{\theta +\epsilon}, P_\theta)&=\frac{1}{2}\sqrt{{\cal I}(\theta)}.
\end{split}
\]
\label{fisher-2}
\eep
\begin{exo} Prove Propositions \ref{fisher-1} and \ref{fisher-2}.
\end{exo}
Since the relative entropy is stochastically monotone, Proposition \ref{fisher-1} implies  that the Fisher entropy  is 
also stochastically monotone. More precisely, let $[\Phi(\omega, \hat \omega)]_{(\omega, \hat \omega)\in \Omega \times \hat \Omega}$ be a stochastic matrix and $\Phi: {\cal P}(\Omega) 
\rightarrow {\cal P}(\hat \Omega)$ the induced stochastic map. Set 
\[\widehat P_\theta=\Phi(P_\theta), 
\]
and note that $\widehat P_\theta$ is faithful. Let $\widehat{\cal I}(\theta)$ be the Fisher entropy of $\widehat P_\theta$. 
Then 
\[
\widehat {\cal I}(\theta)=\lim_{\epsilon \rightarrow 0}\frac{1}{\epsilon^2}
S(\widehat P_{\theta +\epsilon}|\widehat P_\theta)\leq \lim_{\epsilon \rightarrow 0}\frac{1}{\epsilon^2}
S( P_{\theta +\epsilon}|P_\theta)={\cal I}(\theta).
\]
The inequality $\widehat{\cal I}(\theta)\leq {\cal I}(\theta)$ can be directly proven as follows. Since the function $x\mapsto x^2$ is convex, 
the Jensen inequality yields 
\[
\begin{split}
\left(\sum_{\omega} \Phi(\omega, \hat \omega) \dot P_\theta(\omega)\right)^2&=
\left(\sum_{\omega} \Phi(\omega, \hat \omega)P_\theta(\omega) \frac{\dot P_\theta(\omega)}{P_\theta(\omega)}\right)^2\\[3mm]
&\leq \left(\sum_{\omega} \Phi(\omega, \hat \omega)\frac{[\dot P_\theta(\omega)]^2}{P_\theta(\omega)}\right)
\left(\sum_{\omega}\Phi(\omega, \hat \omega)P_\theta(\omega)\right).
\end{split}
\]
Hence, 
\[
\begin{split}
\widehat {\cal I}(\theta)&=\sum_{\hat \omega} \left(\sum_{\omega}\Phi(\omega, \hat \omega)P_\theta(\omega)\right)^{-1}\left(\sum_{\omega} \Phi(\omega, \hat \omega) \dot P_\theta(\omega)\right)^2\\[3mm]
&\leq 
\sum_{\hat \omega}\sum_{\omega}\Phi(\omega, \hat \omega)P_\theta(\omega) \frac{[\dot P_\theta(\omega)]^2}{P_\theta(\omega)}\\[3mm]
&={\cal I}(\theta).
\end{split}
\]
\section{Entropic geometry}
We continue with the framework of the previous section. In this section we again identify ${\cal P}_{\rm f}(\Omega)$ with 
\[
{\cal P}_{L, {\rm f}}=\left\{(p_1, \cdots, p_L)\in \rr^L\,|\, p_k>0, \sum_k p_k=1\right\}.
\]
We view ${\cal P }_{L, {\rm f}}$ as a  surface in $\rr^L$ and write   $p=(p_1, \cdots, p_L)$. 
The family $\{P_\theta\}_{\theta \in [a,b]}$  is viewed as a map (we will also call it a path)
\[ [a,b] \ni \theta \mapsto p_\theta=(p_{\theta 1}, \cdots, p_{\theta L}) \in {\cal P}_{L, {\rm f}},
\]
where $p_{\theta k}=P_\theta(\omega_k)$. For the purpose of this section it suffices to assume that  all such  path are $C^1$ 
(that is, continuously differentiable).  
The tangent vector $\dot p_\theta=(\dot p_{\theta 1}, \cdots, \dot p_{\theta L})$ satisfies  $\sum_k\dot p_{\theta k}=0$ and hence belongs to the hyperplane 
\[{\cal T}_L=\left\{\zeta=(\zeta_1, \cdots, \zeta_L)\,|\, \sum_k \zeta_k=0\right\}.
\]
The tangent space of the surface ${\cal P}_{L, {\rm f}}$ is $T_L ={\cal P}_{L, {\rm f}}\times {\cal T}_L$.

A Riemannian structure (abbreviated RS) on ${\cal P}_{L, {\rm f}}$ is a family $g_L=\{g_{L,p}(\cdot, \cdot)\}_{p\in {\cal P}_L}$ of real inner products 
on ${\cal T}_L$ such that for all  $\zeta, \eta\in {\cal T}_L$ the map 
\beq  {\cal P}_L \ni p\mapsto g_{L,p}(\zeta, \eta)
\label{rs-req}
\eeq

is continuous. The geometric notions (angles, length of curves, curvature...) on ${\cal P}_L$ are defined with respect to the RS 
(to define some of them one needs additional regularity of the maps \eqref{rs-req}). For example, the energy 
of the path $\theta\mapsto p_\theta$ is 
\[{\cal E}([p_\theta])=\int_a^b g_{L, p_\theta}(\dot p_\theta, \dot p_\theta)\d \theta,\]
and its length is 
\[{\cal L}([p_\theta])=\int_a^b \sqrt{g_{L, p_\theta}(\dot p_\theta, \dot p_\theta)}\d \theta.\]
Jensen's inequality for integrals (which is proven by applying Jensen's inequality to Riemann sums) gives that 
\beq
{\cal L}([p_\theta])\geq \left[(b-a){\cal E}([p_\theta])\right]^{1/2}.
\label{jens-fish}
\eeq

The Fisher Riemannian structure (abbreviated FRS) is defined by 
\[g_p^F(\zeta, \eta)=\sum_k \frac{1}{p_k}\zeta_k\eta_k.\] In this case, 
\[
g_{p(\theta)}^F(\dot p_\theta, \dot p_\theta)= {\cal I}(\theta), 
\]
where ${\cal I}(\theta)$ is the Fisher entropy of $P_\theta$.
Hence. 
\[{\cal E}([p_\theta])=\int_a^b {\cal I}(\theta)\d \theta, \qquad {\cal L}([p_\theta])=\int_a^b \sqrt{{\cal I}(\theta)}\d \theta.\]
We have the following general bounds:
\bep
\beq\label{low-fis}
\int_a^b {\cal I}(\theta)\d \theta\geq \frac{1}{b-a}d_V(p_a, p_b)^2, \qquad\int_a^b\sqrt{ {\cal I}(\theta)}\d \theta\geq d_V(p_a, p_b),
\eeq
where $d_V$ is the variational distance defined by \eqref{var-dist}.
\eep
\begin{remark}
The  first inequality in \eqref{low-fis} yields the "symetrized"  version of  Theorem \ref{pos-fine}.  
Let $p, q\in {\cal P}_{L, {\rm f}}$ and 
consider the path $p_\theta =\theta p+ (1-\theta)q$, $\theta \in [0,1]$. Then 
\[
\int_0^1 {\cal I}(\theta)\d \theta= S(p|q) + S(q|p),
\]
and  the  first inequality in \eqref{low-fis}  gives
\[
S(p|q) + S(q|p)\geq d_V(p, q)^2.
\label{mar-mor}
\]
\end{remark}
\demo  To prove the first inequality, 
note that Jensen's inequality gives 
\beq
\label{fish-low}
{\cal I}(\theta)=\sum_{k=1}^L \frac{\dot p_{\theta k}^2}{p_{\theta_k}}=\sum_{k=1}^L \left[ 
\frac{\dot p_{\theta k}}{p_{\theta k}}\right]^2p_{\theta k}\geq \left(\sum_{k=1}^L |\dot p_{\theta k}|\right)^2.
\eeq
Hence, 
\[ 
\int_a^b {\cal I}(\theta)\d \theta \geq \int_a^b \left(\sum_{k=1}^L |\dot p_{\theta k}|\right)^2\d \theta\geq 
\frac{1}{b-a}\left(\sum_{k=1}^L \int_a^b|\dot p_{\theta k}|\d\theta \right)^2,
\label{there-fi}
\]
where the second inequality follows from Jensen's integral inequality. The last inequality and 
\beq\label{fis-end}
\int_a^b|\dot p_{\theta k}|\d \theta \geq \left|\int_a^b \dot p_{\theta k}\d\theta\right|=|p_{b k}- p_{ak}|
\eeq
yield the statement.

Note that the first inequality in \eqref{low-fis} and \eqref{jens-fish} imply the second. Alternatively, the second 
inequality follows immediately from \eqref{fish-low} and \eqref{fis-end}. \qed

The geometry induced by the  FRS can be easily understood in terms of the  surface 
\[{\mathfrak S_L}=\{s=(s_1, \cdots, s_L)\in \rr^L\,|\, s_k>0, \sum_k s_k^2=1\}. \]
The respective tangent space is ${\mathfrak S}_L\times \rr^{L-1}$ which we equip with the Euclidian  RS
\[e_s(\zeta, \eta)=\sum_k \zeta_k\eta_k.
\]
Note that $e_s(\zeta, \eta)$ does not depend on $s \in {\mathfrak S}_L$ and we will drop the  subscript $s$. Let now 
$\theta \mapsto p_\theta=(p_{\theta_1}, \cdots, p_{\theta L})$ be a path connecting $p=(p_1, 
\cdots, p_L)$ and $q=(q_1, \cdots, q_L)$ in ${\cal P}_{L, {\rm f}}$. Then, 
\[\theta \mapsto s_\theta=(\sqrt{p_{\theta_1}}, \cdots, \sqrt{p_{\theta L}})\]
 is a path in 
${\mathfrak S}_L$ connecting $s=(\sqrt{p_1}, 
\cdots, \sqrt{p_L})$ and $u=(\sqrt{q_1}, \cdots,\sqrt{ q_L})$. The  
map $[p_\theta]\mapsto [s_\theta]$ is  a bijective correspondences between all  $C^1$-paths in ${\cal P}_{L, {\rm f}}$ 
connecting $p$ and $q$ and all $C^1$-paths in ${\mathfrak S}_L$ connecting $s$ and $u$. Since 
\[
e(\dot s_\theta, \dot s_\theta)=\frac{1}{4} g_{p(\theta)}^F(\dot p_\theta, \dot p_\theta)=\frac{1}{4} {\cal I}(\theta),
\]
the geometry on ${\cal P}_{L, {\rm f}}$ induced by the FRS is identified with the Euclidian geometry of ${\mathfrak  S}_L$ via the map 
$[p_\theta]\mapsto [s_\theta]$. 
\begin{exo} The geodesic distance between  $p, q\in {\cal P}_{L, {\rm f}}$ w.r.t. the FRS is defined by 
\beq \gamma(p,q)=\inf \int_a^b \sqrt{g_{p(\theta)}^F(\dot p_\theta, \dot p_\theta)}\d \theta,
\label{geod}
\eeq
where $\inf$ is taken over all $C^1$-paths $[a, b]\ni \theta \mapsto p_\theta\in {\cal P}_{L, {\rm f}}$ such that 
$p_a=p$ and $p_b=q$. Prove that 
\[\gamma(p, q)=\arccos\left(\sum_{k=1}^L \sqrt{p_kq_k}\right).\]
Show that the r.h.s. in \eqref{geod} has a unique minimizer and identify this minimizer.

\end{exo}

The obvious  hint for a solution of this exercise is  to use the correspondence between the Euclidian geometry of the 
sphere and the FRS geometry of  ${\cal P}_{L, {\rm f}}$. We leave it to the interested reader familiar with basic notions of 
differential geometry to explore this connection further. For example, can you compute the sectional curvature of ${\cal P}_{L, 
{\rm f}}$ w.r.t. the FRS?

\section{Chentsov's theorem}

Let $(g_L)_{L\geq 2}$ be a sequence of RS, where $g_L$ is a RS  on ${\cal P}_{L, {\rm f}}$. The sequence 
$(g_L)_{L\geq 2}$ is called stochastically monotone if for any $L, \widehat L\geq 2$ and any stochastic map $\Phi: {\cal P}_{L, 
{\rm f}}\rightarrow{\cal P}_{{\widehat L}, {\rm f}}$,
\[
 g_{\widehat L, \Phi(p)}(\Phi(\zeta), \Phi(\zeta))\leq g_{L, p}(\zeta,\zeta)
\]
for all  $p\in {\cal P}_{L, {\rm f}}$ and $\zeta \in {\cal T}_L$. 
Here we used that, in the obvious way, $\Phi$ defines a linear map $\Phi: \rr^L \mapsto \rr^{\widehat L}$ which maps ${\cal T}_L$ to 
${\cal T}_{\widehat L}$. 
\bep The sequence $(g_L^F)_{L\geq 1}$ of the FRS is stochastically monotone. 
\eep
\demo The argument is a repetition of the direct proof of the inequality ${\cal I}(\theta)\leq \widehat {\cal I}(\theta)$ 
given in Section \ref{sec-fisher-def}. The details are as follows. 

Let $[\Phi(i,j)]_{1\leq i\leq L, 1\leq j\leq \widehat L}$ be a stochastic matrix defining $\Phi: {\cal P}_{L, {\rm f}} \rightarrow 
{\cal P}_{{\widehat L}, {\rm f}}$, i.e., for any $v=(v_1, \cdots, v_L)\in \rr^L$, 
$\Phi(v)\in \rr^{\widehat L}$ is given by 
\[
(\Phi(v))_j=\sum_{i=1}^L \Phi(i, j)v_i.
\]
For $p\in {\cal P}_L$ and $\zeta\in {\cal T}_L$ the convexity  gives
\[
\begin{split}
\left(\sum_i \Phi(i, j) \zeta_i\right)^2&=
\left(\sum_i \Phi(i,j) p_i \frac{\zeta_i}{p_i}\right)^2\leq \left(\sum_i \Phi(i, j)\frac{\zeta_i^2}{p_i}\right)
\left(\sum_i\Phi(i,j)p_i\right)\\[3mm]
&=\left(\sum_i \Phi(i, j)\frac{\zeta_i^2}{p_i}\right)(\Phi(p))_j.
\end{split}
\]
Hence, 
\[
\begin{split}
g_{\widehat L}^F(\Phi(\zeta), \Phi(\zeta))&=\sum_j \frac{1}{(\Phi(p))_j}\left(\sum_i \Phi(i, j) \zeta_i\right)^2\\[3mm]
&\leq 
\sum_{j}\sum_i\Phi(i, j)\frac{\zeta_i^2}{p_i}=\sum_i\frac{\zeta_i^2}{p_i}=g_{L, p}^F(\zeta, \zeta).
\end{split}
\]
\qed

The main result of this section is:
\bet Suppose that a sequence $(g_L)_{L\geq 2}$ is stochastically monotone. Then there exists a constant $c>0$ such that 
$g_L=cg_L^F$ for all $L\geq 2$.
\eet
\demo We start the proof by  extending each $g_{L, p}$ to a bilinear map $G_{L, p}$ on $\rr^L\times \rr^L$ as follows. Set 
$\nu_L=(1, \cdots, 1)\in \rr^L$ and note that any $v\in \rr^L$ can be uniquely written as 
$v=a \nu_L + \zeta$, where $a\in \rr$ and $\zeta \in {\cal T}_L$. If $v=a\nu_L +\zeta$ and 
$w=a^\prime\nu_L +\zeta^\prime$, we set 
\[
G_{L, p}(v, w)=g_{L, p}(\zeta, \zeta^\prime).
\]
The map $G_{L, p}$ is obviously bilinear, symmetric ($G_{L,p}(v, w)=G_{L, p}(w,v)$),  and non-negative ($G_{L, p}(v, v)\geq 0$). In particular, 
the polarization identity holds:
\beq
G_{L, p}(v, w)=\frac{1}{4}\left( G_{L,p}(v+w, v+w)- G_{L, p}(v-w, v-w)\right).
\label{polarization}
\eeq
Note however 
that $G_{L, p}$ is not an inner product since $G_{L, p}(\nu_L, \nu_L)=0$. 

In what follows $p_{L, {\rm ch}}$ denotes the chaotic probability distribution in ${\cal P}_L$, i.e., $p_{L, {\rm ch}}=(1/L, 
\cdots, 1/L)$. A basic observation is that if the stochastic map $\Phi: {\cal P}_{L, {\rm f}}\rightarrow {\cal P}_{{\widehat L}, {\rm f}}$ is stochastically 
invertible (that is, there exists a stochastic map $\Psi: {\cal P}_{{\widehat L}, {\rm f}}\rightarrow {\cal P}_{L, {\rm f}}$ such that 
$\Phi \circ \Psi(p)=p$ for all $p\in {\cal P}_{L, {\rm f}}$) and $\Phi(p_{L, {\rm ch}})= p_{\widehat L, {\rm ch}}$, then 
for all $v, w\in \rr^L$, 
\beq 
\label{cold-1}
G_{\widehat  L, p_{\widehat L, {\rm ch}}}(\Phi(v), \Phi(w))=G_{  L, p_{ L, {\rm ch}}}(v, w).
\eeq 
To prove this, note that since $\Phi$ preserves the chaotic probability distribution, we have that $\Phi(\nu_{L})=L{\hat L}^{-1}\nu_{\widehat L}$. 
Then, writing $v=a\nu_L +\zeta$, we have 
\beq\label{cold-2}
\begin{split}
G_{  L, p_{ L, {\rm ch}}}(v, v)&=g_{  L, p_{ L, {\rm ch}}}(\zeta, \zeta)\geq g_{  \widehat L, p_{ \widehat L, {\rm ch}}}(\Phi(\zeta), \Phi(\zeta))
\\[3mm]
&=G_{  \widehat L, p_{ \widehat L, {\rm ch}}}\left(aL{\widehat L}^{-1}\nu_{\widehat L} +\Phi(\zeta), aL{\widehat L}^{-1}\nu_{\widehat L} +\Phi(\zeta)
\right)\\[3mm]
&=G_{  \widehat L, p_{ \widehat L, {\rm ch}}}\left(a\Phi(\nu_L) +\Phi(\zeta), a\Phi(\nu_L) +\Phi(\zeta)
\right)\\[3mm]
&=G_{  \widehat L, p_{ \widehat L, {\rm ch}}}(\Phi(v), \Phi(v)).
\end{split}
\eeq
If $\Psi: {\cal P}_{{\widehat L}, {\rm f}}\rightarrow {\cal P}_{L, {\rm f}}$ is the stochastic inverse of $\Phi$, then 
$\Psi(p_{\widehat L, {\rm ch}})= p_{ L, {\rm ch}}$ and so by repeating the above argument we get 
\beq\label{cold-3}
G_{  \widehat L, p_{ \widehat L, {\rm ch}}}(\Phi(v), \Phi(v))\geq G_{ L, p_{ L, {\rm ch}}}(\Psi(\Phi(v)), \Psi(\Phi(v))=
G_{ L, p_{ L, {\rm ch}}}(v, v).
\eeq
The inequalities \eqref{cold-2} and \eqref{cold-3} yield \eqref{cold-1} in the case  $v=w$. The polarization identity \eqref{polarization} 
then yields  the statement for all vectors $v$ and $w$.

We proceed to identify $G_{  \widehat L, p_{ \widehat L, {\rm ch}}}$ and  $g_{  \widehat L, p_{ \widehat L, {\rm ch}}}$. The identity  \eqref{cold-1} will play a central role in this part of the argument. Let  $e_{L, k}$, $k=1, \cdots, L$, be the standard basis of  
$\rr^L$. Let $\pi$ be a permutation of $\{1, \cdots, L\}$.  Then  for all  $1\leq j,k\leq L$, 
\beq
\label{cold-4}
G_{p_{L, {\rm ch}}}(e_{L, j}, e_{L, k})= G_{p_{L, {\rm ch}}}(e_{L, \pi(j)}, e_{L, \pi(k)}).
\eeq
To establish \eqref{cold-4},  we use \eqref{cold-1} with $\Phi: {\cal P}_{L, {\rm f}}\rightarrow  {\cal P}_{L, {\rm f}}$ defined by 
\[\Phi((p_1, \cdots p_L))=(p_{\pi(1)},\cdots,  p_{\pi(L)}).\]
Note that $\Phi$ is stochastically invertible with the inverse 
\[\Psi((p_1, \cdots p_L))=(p_{\pi^{-1}(1)},\cdots,  p_{\pi^{-1}(L)}),\]
and that $\Phi(p_{L, {\rm ch}})=p_{L, {\rm ch}}$.
An immediate consequence of the \eqref{cold-4} is that for all $k, j$, 
\beq
G_{p_{L, {\rm ch}}}(e_{L, j}, e_{L, j})= G_{p_{L, {\rm ch}}}(e_{L, k}, e_{L, k}),
\label{insu-1}
\eeq
and that for all pairs $(j, k)$, $(j^\prime, k^\prime)$ with $j\not=j^\prime$ and $k\not=k^\prime$, 
\beq
G_{p_{L, {\rm ch}}}(e_{L, j}, e_{L, k})= G_{p_{L, {\rm ch}}}(e_{L, j^\prime}, e_{L, k^\prime}).
\label{insu-2}
\eeq
We introduce the constants 
\[
c_L= G_{p_{L, {\rm ch}}}(e_{L, j}, e_{L, j}), \qquad 
b_L=G_{p_{L, {\rm ch}}}(e_{L, j}, e_{L, k}),
\]
where $j\not=k$. By \eqref{insu-1} and \eqref{insu-2}, these constants  do not depend on the choice of $j,k$. 
We now show that there exist constants $c, b\in \rr$ such that for all $L\geq 2 $, $c_L= cL +b$ and $b_L=b$. 
To prove this, let $L, L^\prime\geq 2$ and consider the  stochastic map 
$\Phi: {\cal P}_{L, {\rm f}} \rightarrow {\cal P}_{LL^\prime, {\rm f}}$ defined by 
\[
\Phi((p_1, \cdots, p_L))=\left(\frac{p_1}{L^\prime}, \cdots, \frac{p_1}{L^\prime}, \cdots, 
\frac{p_L}{L^\prime}, \cdots, \frac{p_L}{L^\prime}\right),
\]
where each term $p_k/L^\prime$ is repeated $L^\prime$ times.
This map is stochastically invertible with the inverse 
\[
\Psi\left((p_1^{(1)}, \cdots, p_{L^\prime}^{(1)}, \cdots, p_{1}^{(L)}, \cdots, p_{L^\prime}^{(L)})\right)=\left(
\sum_{k=1}^{L^\prime}p_k^{(1)}, \cdots, \sum_{k=1}^{L^\prime}p_k^{(L)}\right).
\]
Since  $\Phi(p_{L, {\rm ch}})= p_{LL^\prime, {\rm ch}}$,  \eqref{cold-1} holds. Combining 
\eqref{cold-1} with the definition $b_L$, we derive  that   
\[
b_L= b_{LL^\prime}=b_{L^\prime}.
\]
Set $b=b_L$. Then, for $L, L^\prime\geq 2$, \eqref{cold-1} 
and  the definition  of $c_L$ give
\[
c_L= \frac{1}{L^\prime} c_{LL^\prime}+  \frac{L^\prime(L^\prime-1)}{(L^\prime)^2} b_{LL^\prime}=
\frac{1}{L^\prime} c_{LL^\prime} + \frac{L^\prime(L^\prime-1)}{(L^\prime)^2},
\]
and so 
\[c_L-b = \frac{1}{L^\prime}(c_{LL^\prime}-b).\]
Hence, 
\[
\frac{1}{L}(c_L-b)=\frac{1}{LL^\prime}(c_{LL^\prime}-b)=\frac{1}{L^\prime}(c_{L^\prime}-b),
\]
and we conclude that 
\[ c_L= cL + b\]
for some $c\in \rr$.  It follows that for $v,w\in \rr^L$, 
\[
G_{p_{L, {\rm ch}}}(v, w)= cL\sum_{k=1}^L v_k w_k + b \left(\sum_{k=1}^Lv_k\right)\left(\sum_{k=1}^L w_k\right).
\]
and  that for $\zeta, \eta\in {\cal T}_L$, 
\beq
g_{p_{L, {\rm ch}}}(\zeta, \eta)= cL\sum_{k=1}^L \zeta_k\eta_k.
\label{call}
\eeq
The last relation implies in particular that $c>0$. Note that \eqref{call} can be written as $g_{L, p_{L, {\rm ch}}}=c g^F_{L, p_{\rm ch}}$, 
proving the statement of the theorem for the special values $p=p_{L, {\rm ch}}$. 

The rest of  the argument is based on the relation \eqref{call}. By essentially  repeating the proof of the identity 
\eqref{cold-1} one easily shows that if $\Phi: {\cal P}_{L, {\rm f}} \rightarrow {\cal P}_{{\widehat L}, {\rm f}}$ is stochastically 
invertible, then for all  $p\in {\cal P}_{L, {\rm f}}$ and $\zeta, \eta \in {\cal T}_L$, 
\beq 
\label{cold-1-again}
g_{L, \Phi(p)}(\Phi(\zeta), \Phi(\eta))=g_{L, p}(\zeta, \eta).
\eeq

Let now $\bar p=(\bar p_1, \cdots, \bar p_L) \in {\cal P}_{L, {\rm f}}$ be such that all $\bar p_k$'s are rational numbers. We can write 
\[ \bar p= \left(\frac{\ell_1}{L^\prime}, \cdots, \frac{\ell_L}{L^\prime}\right).
\]
where all $\ell_k$'s are integers $\geq 1$ and $\sum_k \ell_k=L^\prime$. Let 
$\Phi: {\cal P}_{L, {\rm f}} \rightarrow {\cal P}_{L^\prime, {\rm f}}$ be a stochastic map defined by 
\[\Phi((p_1, \cdots, p_L))=\left(\frac{p_1}{\ell_1}, \cdots, \frac{p_1}{\ell_1}, \cdots, \frac{p_L}{\ell_L}, \cdots, 
\frac{p_L}{\ell_L}\right),\]
where each term $p_k/\ell_k$ is repeated $\ell_k$ times. The map $\Phi$ is stochastically invertible and its inverse is 
\[
\Psi((p_{1}^{(1)}, \cdots, p_{\ell_1}^{(1)}, \cdots, p_1^{(\ell_L)}, \cdots, p_{\ell_L}^{(\ell_L)}))=
\left(\sum_{k=1}^{\ell_1}p_k^{\ell_1}, \cdots, \sum_{k=1}^{\ell_L}p_k^{(\ell_L)}\right).
\]
Note that $\Phi(\bar p)=p_{L^\prime, {\rm ch}}$, and so 
\beq
g_{L, \bar p}(\zeta, \eta)= g_{L^\prime, p_{L^\prime, {\rm ch}}}(\Phi(\zeta), \Phi(\eta))= c\sum_{k=1}^L\frac{L^\prime}{\ell_k}\zeta_k\eta_k
=c g_{L, \bar p}^F(\zeta, \eta).
\label{cold-eve}
\eeq
Since the set of all $\bar p$'s in ${\cal P}_{L, {\rm f}}$ whose all components are rational is dense in ${\cal P}_{L, {\rm f}}$ and since the map 
$p\mapsto  g_{L, p}(\zeta, \eta)$ is continuous, it follows from \eqref{cold-eve}  that for all $L\geq 2$ and all $p\in {\cal P}_{L, {\rm f}}$, 
\[g_{L, p}=c g_{L, p}^F.\]
This completes the proof of  Chentsov's theorem. \qed

\section{Notes and references}
The Fisher entropy (also often called Fisher information) was introduced by Fisher in  \cite{Fis1} and plays a fundamental 
role in statistics (this is the topic of the next chapter). Although Fisher's work precedes Shannon's by twenty three years, it apparently 
played no role in the genesis  of  the information theory. The first mentioning of the Fisher entropy in context of information theory 
goes back to \cite{KullLe} where Proposition \ref{fisher-1} was stated. 

The geometric interpretation of the Fisher entropy is basically built in its definition. We shall return to this 
point in the Part II of the lecture notes where the reader can find references to the vast literature on this topic.  

Chentsov's theorem goes back to \cite{Cen}. Our proof is based on the elegant arguments of Campbel \cite{Cam}.

\chapter{Parameter estimation}
\label{sec-estimation}
\section{Introduction}
Let ${\cal A}$ be a set and $\{P_\theta\}_{\theta \in {\cal A}}$ a family of probability measures  on a finite set 
$\Omega$. We shall refer to the elements of ${\cal A}$ as {\em parameters}. Suppose that a probabilistic experiment is described by one unknown member of this family. By performing a
trial we wish to choose the unknown parameter $\theta$ such that   $P_\theta$ is the most likely 
description of the experiment. To predict $\theta$  one choses a function   $\hat \theta:\Omega\rightarrow {\cal A}$ which, 
in the present context, is called an {\em estimator}. If the outcome of a trial is $\omega \in \Omega$, then the value 
$\theta=\hat \theta (\omega)$ is the prediction of the unknown parameter and the probability. Obviously, a  reasonable estimator should satisfy a reasonable 
requirements, and we will return to this point shortly.   

The hypothesis testing, described in Section \ref{sec-hyp-test}, is the simplest non-trivial example of the above setting with 
${\cal A}=\{0, 1\}$, $P_0=P$ and  $P_1= Q$ (we also  assume that the priors are $p=q=1/2$.)  The estimators are  identified with 
characteristic  functions $\hat \theta =\chi_T$, $T\subset  \Omega$. With an obvious change of vocabulary, the mathematical theory 
described 
in Section \ref{sec-hyp-test} can be viewed as a theory of  parameter estimation  in the case where ${\cal A}$ has two elements.

Here we shall assume that ${\cal A}$ is a bounded closed interval $[a, b]$ and we shall 
explore the conceptual and mathematical aspects the continuous set of parameters brings to the problem of estimation. 
The Fisher entropy will play an important  role in this development.  We continue with the notation and assumptions introduced  in the 
beginning of Section \ref{sec-fisher-def}, and start with some preliminaries. 

A {\em loss function} is a map $L: \rr \times [a, b]\rightarrow \rr_+$ such that 
$L(x, \theta)\geq 0$ and $L(x, \theta)=0$ iff $x=\theta$.  To a given loss function and the estimator $\hat \theta$, one associates the 
{\em risk function} by 
\[
R(\hat \theta, \theta)= E_\theta(L(\hat \theta, \theta))=\sum_{\omega \in \Omega}L(\hat \theta (\omega), \theta)P_\theta(\omega).
\]
Once a choice of the 
loss function is made, the goal is to find an estimator that will minimize the risk function subject to  appropriate 
consistency requirements. 

We shall  work  only 
with the quadratic loss function $L(x, \theta)=(x-\theta)^2$. In this case, the risk function is 
\[E_\theta((\hat \theta -\theta)^2)={\rm Var}_\theta(\hat \theta).\] 

\section{Basic facts}
The following  general 
estimate is known as the Cram\'er-Rao bound. 

\bep For  any estimator $\hat \theta$ and all $\theta \in [a, b]$,
\[
\frac{[\dot E_\theta(\hat \theta)]^2}{{\cal I}(\theta)}\leq E_\theta((\hat \theta-\theta)^2).
\]
\eep
\demo 
\[\dot E_\theta(\hat \theta)=\sum_{\omega \in \Omega}\hat \theta(\omega)\dot P_\theta(\omega)=
\sum_{\omega \in \Omega} (\hat \theta(\omega)-\theta)\dot P_\theta(\omega).\]
Writing $\dot P_\theta(\omega)=\dot P_\theta(\omega)\sqrt{P_\theta(\omega)}/\sqrt{P_\theta(\omega)}$ and applying 
the Cauchy-Schwartz inequality one gets
\[
\begin{split}
|\dot E_\theta(\hat \theta)|&\leq \left(\sum_{\omega \in \Omega}(\hat \theta(\omega)-\theta)^2P_\theta(\omega)\right)^{1/2}
\left(\sum_{\omega\in  \Omega}\frac{[\dot P_\theta(\omega)]^2}{P_\theta(\omega)}\right)^{1/2}\\[2mm]
&=\left(E_\theta((\hat \theta-\theta)^2)\right)^{1/2}\sqrt {{\cal I}(\theta)}.
\end{split}
\]
 \qed

As  in the case of hypothesis testing, multiple trials  improve the errors in the parameter estimation.
Passing to the product space $\Omega^N$ and the product probability measure $P_{\theta N}$, and denoting by $E_{\theta N}$ the expectation 
w.r.t. $P_{\theta N}$, the Cram\'er-Rao  bound takes the following form. 
\bep For  any estimator $\hat \theta_N: \Omega^N \rightarrow [a, b]$ and all $\theta \in [a, b]$,
\[
\frac{1}{N}\frac{[\dot E_{N\theta}(\hat \theta_N)]^2}{{\cal I}(\theta)}\leq E_{\theta N}((\hat \theta_N-\theta)^2).
\]
\eep
\demo 
\[
\begin{split}
\dot E_{\theta N}(\hat \theta_N)&=\sum_{\omega=(\omega_1, \cdots, \omega_N) \in \Omega^N}\sum_{k=1}^N(\hat \theta_N(\omega)-\theta) P_\theta(\omega_1)\cdots \dot P_\theta(\omega_k)\cdots P_\theta(\omega_N)\\[3mm]
&=\sum_{\omega=(\omega_1, \cdots, \omega_N) \in \Omega^N}\left(\sum_{k=1}^N\frac{\dot P_\theta(\omega_k)}
{P_\theta(\omega_k)}\right)(\hat \theta_N(\omega)-\theta) P_{\theta N}(\omega).
\end{split}
\]
Applying the Cauchy-Schwarz inequality 
\[\int_{\Omega^N}fg \d P_{\theta N}\leq \left(\int_{\Omega^N}f^2\d P_{\theta N}\right)^{1/2}\left(
\int_{\Omega^N}g^2 \d P_{\theta N}\right)^{1/2}
\]
with 
\[
f(\omega)=\sum_{k=1}^N\frac{\dot P_\theta(\omega_k)}
{P_\theta(\omega_k)},\qquad g(\omega)= \hat \theta_N(\omega)-\theta,
\]
one gets
\[
\begin{split}
|\dot E_{\theta N}(\hat \theta)| &\leq  \left(\sum_{\omega \in \Omega^N}(\hat \theta_N(\omega)-\theta)^2P_{\theta N}(\omega)\right)^{1/2}
\left(\sum_{\omega=(\omega_1, \cdots, \omega_N)}\sum_{k=1}^N \frac{[\dot P_\theta(\omega_k)]^2}{[P_\theta(\omega_k)]^2} P_{\theta N}(\omega)\right)^{1/2}\\[2mm]
&=\left(E_{\theta N}((\hat \theta_N-\theta)^2)\right)^{1/2}\sqrt {N{\cal I}(\theta)}.
\end{split}
\]
 \qed

We now describe  the {\em consistency} requirement.  In a nutshell, the consistency states 
that if the experiment is described by $P_\theta$, then the estimator should statistically return the value $\theta$. 
An ideal consistency would  
be  $E_{\theta N}(\hat \theta_N)=\theta$ for all $\theta \in [a, b]$. However, it is clear that in  our setting 
such estimator cannot exists. 
Indeed, using that $\hat \theta$ takes values in 
$[a,b]$, the relations  $E_{a N}(\hat \theta_N)=a$  and $E_{b N}(\hat \theta_N)=b$ give  that $\hat \theta_N(\omega)=a$  and 
$\hat \theta_N(\omega)=b$ for all $\omega \in \Omega^N$.  Requiring $E_{\theta N}(\hat \theta_N)=\theta$ only for 
 $\theta \in ]a, b[$  does not help, and  the remaining possibility is to formulate the 
consistency in an asymptotic setting. 

\begin{definition} A sequence of estimators $\hat \theta_N :\Omega^N \rightarrow [a,b]$, $N=1,2, \cdots$,  is called consistent if 
\[
\lim_{N\rightarrow \infty}E_{\theta N}(\hat \theta_N)=\theta
\]
for  all $\theta \in [a, b]$, and uniformly consistent if 
\[
\lim_{N\rightarrow \infty}\sup_{\theta \in [a, b]}E_{\theta N}(|\hat \theta -\theta|)=0.
\]
\label{consistent}
\end{definition}

Finally, we introduce the notion of  {\em efficiency}.  
\begin{definition} Let  $\hat \theta_N :\Omega^N \rightarrow [a,b]$, $N=1,2, \cdots$ be a sequence of estimators. A continuous function 
${\cal E}:\, ]a, b[\rightarrow \rr_+$ is called the efficiency of $(\hat \theta_N)_{N\geq 1}$ if 
\beq \lim_{N\rightarrow \infty}N E_{\theta N}\left((\hat \theta-\theta)^2\right)={\cal E}(\theta)
\label{in-cr}
\eeq
for all $\theta \in\, ]a,b[$. The sequence $(\hat \theta_N)_{N\geq 1}$ is called uniformly efficient  if in addition for any $[a^\prime, b^\prime]
\subset\,]a, b[$, 
\beq 
\limsup_{N\rightarrow \infty}\sup_{\theta \in [a^\prime,b^\prime]}\left| N E_{\theta N}\left(\hat \theta -\theta)^2\right)- {\cal E}(\theta)\right|=0.
\label{ny-rain}
\eeq
\end{definition}
To remain on a  technically elementary level, we will work only with  uniformly efficient  estimators. 
The reason for staying away from the boundary points $a$ and $b$ in the definition of efficiency is somewhat subtle 
and we will elucidate it in Remark \ref{last-re}.

\bep  Let  $(\hat \theta_N)_{N\geq 1}$ be a uniformly  efficient consistent  sequence of estimators. Then its efficiency 
${\cal E}$ satisfies 
\[
{\cal E}(\theta)\geq \frac{1}{{\cal I}(\theta)}
\]
for all $\theta \in\, ]a, b[$. 
\label{ny-never}
\eep
\demo Fix $\theta_1, \theta_2\in \,]a,b[$, $\theta_1 <\theta_2$. 
The consistency gives 
\beq\theta_2-\theta_1= \lim_{N\rightarrow\infty}\left[E_{\theta_{2}N}(\hat \theta_N)-E_{{\theta_1}N}(\hat \theta_N)\right].
\label{ny-1}
\eeq
The Cram\'er-Rao bound yields the  estimate
\beq
\begin{split}
E_{\theta_2 N}(\hat \theta_N)-E_{\theta_1 N}(\hat \theta_N)&=\int_{\theta_1}^{\theta_2} \dot E_{\theta N}(\hat \theta_N)
\d \theta \leq \int_{\theta_1}^{\theta_2} |\dot E_{\theta N}(\hat \theta_N)|
\d \theta \\[3mm]
&\leq \int_{\theta_1}^{\theta_2}\left[ N {\cal I}(\theta) E_{\theta N}\left((\hat \theta_N-\theta)^2)\right)\right]^{1/2}\d \theta.
\end{split}
\label{ny-2}
\eeq
Finally, the uniform efficiency  gives 
\beq
\begin{split}
\lim_{N\rightarrow \infty}
\int_{\theta_1}^{\theta_2}\left[ N {\cal I}(\theta) E_{\theta N}\left((\hat \theta_N-\theta)^2)\right)\right]^{1/2}\d \theta
&=\int_{\theta_1}^{\theta_2}\lim_{N\rightarrow \infty}
\left[ N {\cal I}(\theta) E_{\theta N}\left((\hat \theta_N-\theta)^2)\right)\right]^{1/2}\d \theta\\[3mm]
&=\int_{\theta_1}^{\theta_2} \sqrt{{\cal I}(\theta){\cal E}(\theta)}\d \theta.
\end{split}
\label{ny-3}
\eeq
Combining \eqref{ny-1}, \eqref{ny-2}, and \eqref{ny-3}, we derive that 
\[\theta_2-\theta_1\leq \int_{\theta_{1}}^{\theta_2}\sqrt{{\cal I}(\theta){\cal E}(\theta)}\d \theta
\]
for all $a\leq \theta_1<\theta_2 \leq b$. Hence, $\sqrt{{\cal I}(\theta){\cal E}(\theta)}\geq 1$ for 
all $\theta \in \,]a, b[$, 
and the statement follows. \qed

In  Section \ref{sec-MLE} we shall construct a uniformly consistent and uniformly efficient  sequence of estimators whose efficiency is equal to 
$1/{\cal I}(\theta)$ for all $\theta \in  \,]a, b[$. This sequence of estimators saturates the bound of Proposition \ref{ny-never} 
and in that sense is the best possible one. In Remark \ref{last-re} we shall also exhibit 
a concrete example of such estimator sequence for which the limit \eqref{in-cr} also  exists for $\theta=a$ and satisfies  
${\cal E}(a)< 1/ {\cal I}(a)$. This shows that Proposition \ref{ny-never} is an optimal result.

\section{Two remarks}

The first remark  is that the existence of a consistent estimator sequence 
obviously implies that
\beq 
\theta_1\not=\theta_2\,\,\Rightarrow P_{\theta_1}\not=P_{\theta_2}.
\label{ny-iden}
\eeq
In Section \ref{sec-MLE} we shall  assume that \eqref{ny-iden} holds and refer to it  as the {\em identifiability} property of our starting family  of probability measures $\{P_\theta\}_{\theta \in [a, b]}$. 

The second remark  concerns the  LLN adapted to the parameter setting, which will play a  central role in the proofs 
of the next section. This variant of  the LLN is of independent interest, and for this reason we state it and prove it separately.
\bep Let $X_\theta: \Omega \rightarrow \rr$, $\theta \in [a, b]$, be random variables such that the map 
$[a, b]\ni \theta \mapsto X_\theta(\omega)$ is continuous for all $\omega\in \Omega$. Set 
\[{\cal S}_{\theta N}(\omega=(\omega_1, \cdots, \omega_N))=\sum_{k=1}^N X_{\theta}(\omega_k).\]
Then for any $\epsilon >0$,
\beq 
\lim_{N\rightarrow \infty} \sup_{\theta \in [a, b]}P_{\theta N}\left\{\omega \in \Omega^N\,|\, 
\sup_{\theta^\prime \in [a,b]}\left| \frac{{\cal S}_{\theta^\prime N}(\omega)}{N}-E_{\theta} (X_{\theta^\prime})\right|\geq \epsilon
\right\}=0.
\label{ny-verona-first}
\eeq
Moreover, \eqref{ny-verona-first} can be refined as follows. For any $\epsilon >0$ there 
are constants $C_\epsilon >0$ and $\gamma_\epsilon >0$ such that for all $N\geq 1$,
\beq
\sup_{\theta \in [a, b]}P_{\theta N}\left\{\omega \in \Omega^N\,|\,\sup_{\theta^\prime\in [a,b]} \left| \frac{{\cal S}_{\theta^\prime N}(\omega)}{N}-E_\theta (X_{\theta^\prime})\right|\geq \epsilon
\right\}\leq C_\epsilon \e^{-\gamma_\epsilon N}.
\label{ny-second}
\eeq
\label{ny-par-LLN}
\eep
\begin{remark}The point of this result is uniformity in $\theta$ and $\theta^\prime$. Note that 
\[
\lim_{N\rightarrow \infty} P_{\theta N}\left\{\omega \in \Omega^N\,|\, 
\left| \frac{{\cal S}_{\theta^\prime N}(\omega)}{N}-E_{\theta} (X_{\theta^\prime})\right|\geq \epsilon
\right\}=0
\]
is  the statement of the LLN, while 
\[
P_{\theta N}\left\{\omega \in \Omega^N\,|\, \left| \frac{{\cal S}_{\theta^\prime N}(\omega)}{N}-E_\theta (X_{\theta^\prime})\right|\geq \epsilon
\right\}\leq C_\epsilon \e^{-\gamma_\epsilon N},
\]
with $C_\epsilon$ and $\gamma_\epsilon$ depending on $\theta, \theta^\prime$, is the statement of the strong LLN formulated in Exercise \ref{ex-strong-LLN}.
\end{remark}
\demo  By uniform continuity, there exists $\delta >0$ such that for all 
$u, v \in [a,b]$ satisfying 
$|u-v|<\delta$ one has 
\[ \sup_{u^\prime \in [a, b]}|E_{u^\prime}(X_u)- E_{u^\prime}(X_v)| <\frac{\epsilon}{4} \qquad 
\hbox{and}\qquad  \sup_{\omega \in \Omega}|X_u(\omega)- X_{v}(\omega)|<\frac{\epsilon}{4}.
\]
 Let  $a=\theta_0^\prime <\theta_1^\prime<\cdots <\theta_n^\prime=b$ be such that $\theta_k^\prime-\theta_{k-1}^\prime<\delta$. 
Then, for all $\theta \in [a,b]$, 
\beq \left\{\omega\in \Omega^N\,|\, 
\sup_{\theta^\prime \in [a,b]}\left| \frac{{\cal S}_{\theta^\prime N}(\omega)}{N}- E_{\theta}(X_{\theta^\prime})\right|
\geq \epsilon\right\}\subset \bigcup_{k=1}^n \left\{\omega\in \Omega^N\,|\, 
\left| \frac{{\cal S}_{\theta_k^\prime N}(\omega)}{N}- E_{\theta}(X_{\theta_k^\prime})\right|
\geq \frac{\epsilon}{2}\right\}.
\label{verona-1}
\eeq
 It follows that (recall the proof of the LLN, Proposition \ref{LLN})
\beq
\begin{split}
P_{\theta N} \left\{\omega\in \Omega^N\,|\, 
\sup_{\theta^\prime \in [a,b]}\left| \frac{{\cal S}_{\theta\prime N}}{N}- E_\theta (X_{\theta^\prime})\right|
\geq \epsilon\right\} &\leq 
\sum_{k=1}^n P_{\theta N}\left\{\omega\in \Omega^N\,|\, 
\left| \frac{{\cal S}_{\theta_k^\prime N}(\omega)}{N}- E_\theta(X_{\theta_k^\prime})\right|
\geq \frac{\epsilon}{2}\right\}\\[3mm]
&\leq \frac{4}{\epsilon^2}\sum_{k=1}^n
 E_{\theta N}\left(\left|\frac{{\cal S}_{\theta_k^\prime N}}{N}- E_\theta(X_{\theta_k^\prime})\right|^2\right)\\[3mm]
 &\leq \frac{4}{\epsilon^2}\frac{1}{N}\sum_{k=1}^n E_{\theta}
 \left(|X_{\theta_k^\prime}- E_\theta(X_{\theta_k^\prime})|^2\right).
\end{split}
\label{verona-ny}
\eeq
Setting 
\[ C=\max_{1\leq k\leq n}\max_{\theta, \theta^\prime \in [a,b]}E_{\theta}
  \left(|X_{\theta^\prime}- E_\theta(X_{\theta^\prime})|^2\right),
 \]
 we derive that 
 \[
\sup_{\theta\in [a,b]} P_{\theta N} \left\{\omega\in \Omega^N\,|\, 
\sup_{\theta^\prime \in [a,b]}\left| \frac{S_{\theta^\prime N}(\omega)}{N}- E_{\theta}(X_{\theta^\prime})\right|
\geq \epsilon\right\} \leq \frac{4}{\epsilon^2}\frac{Cn}{N},
 \] 
 and \eqref{ny-verona-first} follows. 
 
 The proof of \eqref{ny-second} also starts with \eqref{verona-1} and follows the argument of Proposition \ref{bound-upper}
 (recall the Exercise \ref{ex-strong-LLN}).  The details are as follows. Let $\alpha >0$. Then for  any $\theta$ and $k$, 
 \beq
 \begin{split}
P_{\theta N}\left\{\omega\in \Omega^N\,|\, 
 \frac{{\cal S}_{\theta_k^\prime N}(\omega)}{N}- E_{\theta}(X_{\theta_k^\prime})
\geq \frac{\epsilon}{2}\right\}&=P_{\theta N}\left\{\omega\in \Omega^N\,|\, 
 {\cal S}_{\theta_k^\prime N}(\omega)\geq  N\frac{\epsilon}{2}+ NE_{\theta}(X_{\theta_k^\prime})\right\}\\[3mm]
 &=P_{\theta N}\left\{\omega\in \Omega^N\,|\, 
 \e^{\alpha {\cal S}_{\theta_k^\prime N}(\omega)}\geq \e^{\alpha N\epsilon/2}\e^{\alpha NE_{\theta}(X_{\theta_k^\prime})}\right\}\\[3mm]
 &\leq \e^{-\alpha N\epsilon/2} \e^{-\alpha NE_{\theta}(X_{\theta_k^\prime})}E_{\theta N}
 \left( \e^{\alpha {\cal S}_{\theta_k^\prime N}}\right)\\[3mm]
 &\leq \e^{-\alpha N\epsilon/2} \e^{-\alpha NE_{\theta}(X_{\theta_k^\prime})}\e^{N C_{\theta}^{(k)}(\alpha)},
 \end{split}
\label{verona-last}
 \eeq
 where 
 \[
 C_{\theta}^{(k)}(\alpha)= \log E_\theta\left(\e^{\alpha X_{\theta_k^\prime}}\right).
 \]
 We write 
 \[ 
 C_{\theta}^{(k)}(\alpha)- \alpha E_\theta(X_{\theta^\prime_k})=
 \int_{0}^\alpha \left[\left(C_{\theta}^{(k)}\right)^\prime(u)- E_\theta(X_{\theta^\prime_k})\right]\d u,
 \]
 and estimate
 \[
 |C_{\theta}^{(k)}(\alpha)- \alpha E_\theta(X_{\theta^\prime_k})|\leq \alpha \sup_{u\in [0, \alpha]}
\left| \left(C_{\theta}^{(k)}\right)^\prime(u)- E_\theta(X_{\theta^\prime_k})\right|.
 \]
 Since  $\left(C_{\theta}^{(k)}\right)^\prime(0)=E_\theta(X_{\theta^\prime_k})$, the uniform continuity gives
 \[
 \lim_{\alpha \rightarrow 0}\sup_{\theta \in [a,b]}\sup_{u\in [0, \alpha]}
\left| \left(C_{\theta}^{(k)}\right)^\prime(u)- E_\theta(X_{\theta^\prime_k})\right|=0.
 \]
 It follows that there exists $\alpha_\epsilon^+ >0$ such that for all $k=1, \cdots, n$, 
 \[
 \sup_{\theta \in [a, b]} \left|C_{\theta}^{(k)}(\alpha_\epsilon^+)- \alpha_\epsilon^+ E_\theta(X_{\theta^\prime_k})\right|\leq \frac{\epsilon}{4},
 \]
 and \eqref{verona-last} gives that for all $k$,
 \[
 \sup_{\theta \in [a, b]}P_{\theta N}\left\{\omega\in \Omega^N\,|\, 
 \frac{{\cal S}_{\theta_k^\prime N}(\omega)}{N}- E_{\theta}(X_{\theta_k^\prime})
\geq \frac{\epsilon}{2}\right\}\leq \e^{-\alpha_\epsilon^+ N\epsilon/4}.
 \]
 Going back to first inequality in  \eqref{verona-ny}, we conclude that 
 \beq
\sup_{\theta\in [a,b]} P_{\theta N} \left\{\omega\in \Omega^N\,|\, 
\sup_{\theta^\prime \in [a,b]} \left(\frac{S_{\theta^\prime N}(\omega)}{N}- E_{\theta}(X_{\theta^\prime})\right)
\geq \epsilon\right\}\leq n \e^{-\alpha_\epsilon^+ N\epsilon/4}. 
\label{verona-final}
\eeq
By  repeating the above argument (or by simply  applying the final estimate \eqref{verona-final} to the random variables $-X_\theta$), one 
derives  
\beq
\sup_{\theta\in [a,b]} P_{\theta N} \left\{\omega\in \Omega^N\,|\, 
\inf_{\theta^\prime \in [a,b]} \left(\frac{S_{\theta^\prime N}(\omega)}{N}- E_{\theta}(X_{\theta^\prime})\right)
\leq - \epsilon\right\}\leq n \e^{-\alpha_\epsilon^- N\epsilon/4}
\label{verona-final-1}
\eeq
for a suitable $\alpha_\epsilon^->0$. 
Finally, since 
\[
\begin{split}
\left\{\omega\in \Omega^N\,|\, 
\sup_{\theta^\prime \in [a,b]} \left|\frac{S_{\theta^\prime N}(\omega)}{N}- E_{\theta}(X_{\theta^\prime})\right|
\geq   \epsilon\right\}&\subset  \left\{\omega\in \Omega^N\,|\, 
\sup_{\theta^\prime \in [a,b]} \left(\frac{S_{\theta^\prime N}(\omega)}{N}- E_{\theta}(X_{\theta^\prime})\right)
\geq \epsilon\right\}\\[3mm]
&\, \cup\, \left\{\omega\in \Omega^N\,|\, 
\inf_{\theta^\prime \in [a,b]} \left(\frac{S_{\theta^\prime N}(\omega)}{N}- E_{\theta}(X_{\theta^\prime})\right)
\leq - \epsilon\right\},
\end{split}
\]
 \eqref{ny-second} follows from \eqref{verona-final} and \eqref{verona-final-1}. \qed
 
 \begin{exo} Prove the relation \eqref{verona-1}.
 \end{exo}

\section{The maximum likelihood estimator}
\label{sec-MLE}
For each $N$ and $\omega=(\omega_1, \cdots, \omega_N)\in \Omega^N$, 
consider the function 
\beq
[a, b]\ni \theta \mapsto P_{\theta N}(\omega_1, \cdots, \omega_N)\in ]0,1[. 
\label{mle}
\eeq
By continuity, this function achieves its global maximum on the interval $[a,b]$. We denote by 
$\hat \theta_{ML, N}(\omega)$ a point where this  maximum is achieved (in the case where there are several such points, we select one 
arbitrarily but  always choosing $\hat \theta_{ML, N}(\omega)\in \,]a, b[$ whenever such  possibility exists).  This defines a  random variable 
\[\hat \theta_{ML, N}: \Omega^N\rightarrow [a,b]\]
that is  called the {\em maximum likelihood estimator} (abbreviated MLE)  of order $N$. We shall also refer to the sequence $(\hat \theta_{ML, N})_{N\geq 1}$ 
as the MLE.  

Note that maximizing (\ref{mle}) is equivalent to minimizing the entropy function 
\[
[a, b]\ni \theta \mapsto S_{\theta N}(\omega)=\sum_{k=1}^N-\log P_{\theta}(\omega_k).
\]
Much of our analysis of the MLE will make use of this elementary observation and will be centred  around the entropy function 
$S_{\theta  N}$. We set 
\[
S(\theta, \theta^\prime)=E_{\theta}(S_{\theta^\prime})=-\sum_{\omega\in \Omega} P_{\theta}(\omega)\log P_{\theta^\prime}(\omega).
\]
Obviously, $S(\theta, \theta)=S(P_\theta)$ and 
\beq
S(\theta, \theta^\prime)- S(\theta, \theta)= S(P_{\theta}|P_{\theta^\prime}).
\label{ident-1}
\eeq
The last relation and the identifiability \eqref{ny-iden}, which we assume throughout, give that 
\beq
S(\theta, \theta^\prime) > S(\theta, \theta)\qquad\hbox{for}\qquad \theta\not=\theta^\prime.
\label{verona-fire}
\eeq
Applying  Proposition \ref{ny-par-LLN} to $X_\theta=-\log P_\theta$, we derive
\bep For any $\epsilon>0$, 
\[
\lim_{N\rightarrow \infty}\sup_{\theta \in [a,b]}P_{\theta N}\left\{\omega\in \Omega^N\,|\, 
\sup_{\theta^\prime \in [a,b]}\left| \frac{S_{\theta^\prime N}(\omega)}{N}- S(\theta, \theta^\prime)\right|
\geq \epsilon\right\}=0.
\]
Moreover,  for  any $\epsilon>0$ there is $C_\epsilon>0$ and $\gamma_\epsilon>0$ such that for all $N\geq 1$, 
\[
\sup_{\theta \in [a,b]}P_{\theta^\prime N}\left\{\omega\in \Omega^N\,|\, 
\sup_{\theta^\prime \in [a,b]}\left| \frac{S_{\theta^\prime N}(\omega)}{N}- S(\theta, \theta^\prime)\right|
\geq \epsilon\right\}\leq C_\epsilon\e^{-\gamma_\epsilon N}.
\]
 \label{verona-night}
 \eep
 
 The first result of this section is:
 \bet For any $\epsilon >0$, 
 \[
 \lim_{N\rightarrow \infty}\sup_{\theta\in [a,b]}P_{\theta N}\left\{\omega \in \Omega^N\,|\, 
 |\hat \theta_{ML, N}(\omega)-\theta|\geq \epsilon\right\}=0.
 \]
 Moreover, for any $\epsilon>0$ there exists $C_\epsilon>0$ and $\gamma_\epsilon>0$ such that for 
 all $N\geq 1$, 
 \[
 \sup_{\theta\in [a,b]}P_{\theta N}\left\{\omega \in \Omega^N\,|\, 
 |\hat \theta_{ML, N}(\omega)-\theta|\geq \epsilon\right\}\leq C_\epsilon\e^{-\gamma_\epsilon N}.
 \]
 \label{majolino}
 \eet
 \demo Let 
 \[I_{\epsilon}=\left\{(u, v)\in [a, b]\times [a, b]\,|\,|u-v|\geq \epsilon \right\}.
 \]
It follows from \eqref{verona-fire} and continuity that 
\beq
\delta= \sup_{(u, v)\in I_\epsilon}\left[S(u, v)- S(u, u)\right]>0.
\label{verona-0}
\eeq
Fix $\theta\in [a,b]$ and set $I_\epsilon(\theta)=\{\theta^\prime\in [a,b]\,|\, |\theta-\theta^\prime|\geq \epsilon\}$. 
Let 
\[
A=\left\{\omega\in \Omega^N\,|\, \sup_{\theta^\prime \in I_\epsilon(\theta)}\left| \frac{S_{\theta^\prime N}(\omega)}{N}- S(\theta, \theta^\prime)\right|<\frac{\delta}{2}\right\},
\]
\[
B=\left\{\omega\in \Omega^N\,|\, \sup_{\theta^\prime \in [a, b]\setminus I_\epsilon(\theta)}\left| \frac{S_{\theta^\prime N}(\omega)}{N}- S(\theta, \theta^\prime)\right|<\frac{\delta}{2}\right\}.
\]
For $\omega\in A$ and  $\theta^\prime \in I_\epsilon(\theta)$, 
\beq
\frac{S_{\theta^\prime  N}(\omega)}{N}<S(\theta, \theta^\prime)+\frac{\delta}{2}\leq   S(\theta, \theta)-\frac{\delta}{2}.
\label{verona-2}\eeq
On the other hand, for $\omega\in B$ and $\theta \in[a, b]\setminus  I_\epsilon(\theta)$, 
\beq
\frac{S_{\theta^\prime N}(\omega)}{N}>S(\theta, \theta^\prime)-\frac{\delta}{2}\geq S(\theta, \theta)-\frac{\delta}{2}.
\label{verona-3}\eeq
Since $\hat \theta_{ML, N}(\omega)$ minimizes the map $[a, b]\ni \theta^\prime \mapsto S_{\theta^\prime N}(\omega)$, 
\[\omega\in A\cap B\qquad \Rightarrow\qquad  |\hat \theta_{ML, N}(\omega)-\theta|<\epsilon.\]
 It follows that 
\[
\left\{\omega\in \Omega^N\,|\, |\hat \theta_{ML, N}(\omega)-\theta|\geq \epsilon\right\}\subset A^c\cup B^c=\left\{\omega\in \Omega^N\,|\, \sup_{\theta^\prime\in [a, b]}\left| \frac{S_{\theta^\prime N}(\omega)}{N}- S(\theta, \theta^\prime)\right|\geq \frac{\delta}{2}\right\},
\label{belgrade-1}
\]
and so 
\[
\sup_{\theta \in [a,b]} P_{\theta N}\left\{\omega\in \Omega^N\,|\, |\hat \theta_{ML, N}(\omega)-\theta|\geq \epsilon\right\}\leq \sup_{\theta \in [a,b]}P_{\theta N}\left\{\omega\in \Omega^N\,|\, \sup_{\theta^\prime  \in [a, b]}\left| \frac{S_{\theta^\prime N}(\omega)}{N}- S(\theta, \theta^\prime)\right|\geq \frac{\delta}{2}\right\}.
\]
Since $\delta$ depends only on the choice of $\epsilon$ (recall \eqref{verona-0}), the last 
inequality and Proposition \ref{verona-night} yield the statement. \qed

Theorem \ref{majolino} gives that the MLE is consistent in a very strong sense, and in particular  that is uniformly consistent.

\begin{corollary} 
\[
\lim_{N\rightarrow\infty}\sup_{\theta \in [a,b]}E_{\theta N}(|\hat\theta_{ML,N}-\theta|)=0.
\]
\label{verona-thu}
\end{corollary}
\demo Let $\epsilon >0$. Then 
\[
\begin{split}
E_{\theta N}(|\hat\theta_{ML,N}-\theta|)&=\int_{\Omega^N} |\hat\theta_{ML,N}-\theta|\d P_{\theta N}\\[3mm]
&=\int_{|\hat\theta_{ML,N}-\theta|<\epsilon}|\hat\theta_{ML,N}-\theta|\d P_{\theta N} + 
\int_{|\hat\theta_{ML,N}-\theta|\geq \epsilon}|\hat\theta_{ML,N}-\theta|\d P_{\theta N}\\[3mm]
&
\leq \epsilon +   (b-a) P_{\theta N}\left\{\omega \in \Omega^N\,|\, |\hat\theta_{ML,N}(\omega)-\theta|\geq\epsilon\right\}.
\end{split}
\]
Hence, 
\[
\sup_{\theta \in [a,b]} E_{\theta N}(|\hat\theta_{ML,N}-\theta|)\leq 
 \epsilon +   (b-a) \sup_{\theta \in [a,b]} P_{\theta N}\left\{\omega \in \Omega^N\,|\, |\hat\theta_{ML,N}(\omega)-\theta|\geq\epsilon\right\},
\]
and the result follows from Proposition \ref{majolino}. \qed

We note that so far all results of this section hold under the sole assumptions that the maps $[a, b]\ni \theta \mapsto P_\theta(\omega)$ are 
continuous for all $\omega \in \Omega$ and that the identifiability condition  \eqref{ny-iden} is satisfied. 

We now turn to study of the efficiency of the MLN and prove the second main result  of this section. We strengthen our standing assumptions and assume that the maps $[a, b]\ni \theta \mapsto P_\theta(\omega)$ are $C^3$ for all 
$\omega\in \Omega$. 
\bet Suppose that $[a^\prime, b^\prime]\subset\, ]a, b[$. 
Then 
\[
\lim_{N\rightarrow \infty}\sup_{\theta \in [a^\prime, b^\prime]}\left| NE_{\theta N}(|\hat \theta_{ML, N}-\theta|^2)- \frac{1}{{\cal I}
(\theta)}\right|=0.
\]
\label{mle-eff}
\eet
\demo
Recall that 
\[
[a, b]\ni \theta \mapsto S_{\theta N}(\omega=(\omega_1, \cdots, \omega_N))=-\sum_{k=1}^N \log P_\theta(\omega_k)
\]
achieves its minimum at  $\hat \theta_{ML, N}(\omega)$ and that $\hat \theta_{ML, N}(\omega)\in \,]a, b[$ unless 
a strict minimum is achieved at either $a$ or $b$. Let 
\[B_N(a)=\left\{\omega\in \Omega^N\, |\, \hat \theta_{ML, N}(\omega)=a \right\}, \qquad B_N(b)=\left\{\omega\in \Omega^N\, |\, \hat \theta_{ML, N}(\omega)=b \right\},\]
and
\[\zeta=\min \left( \inf_{\theta \in [a^\prime, b^\prime]}S(P_\theta|P_a),  \inf_{\theta \in [a^\prime, b^\prime]}S(P_\theta|P_b)\right).
\]
Since the maps $\theta \mapsto S(P_\theta|P_a)$, $\theta \mapsto S(P_\theta|P_b)$ are continuous, 
 the identifiability \eqref{ny-iden} yields that $\zeta>0$. Then, for $\theta \in [a^\prime, b^\prime]$,
 \[
 \begin{split}
 P_{\theta N}(B_N(a))&\leq P_{\theta N}\left\{\omega\in \Omega^N\,|\, \frac{1}{N}\sum_{k=1}^N \log \frac{P_\theta(\omega_k)}
 {P_a(\omega_k)} <0\right\}\\[3mm]
 & \leq P_{\theta N}\left\{\omega\in \Omega^N\,|\, \frac{1}{N}\sum_{k=1}^N \log \frac{P_\theta(\omega_k)}
 {P_a(\omega_k)}- S(P_{\theta}|P_a) \leq -\zeta \right\},\\[3mm]
\end{split}
 \]
and similarly, 
\[
P_{\theta N}(B_N(b))\leq P_{\theta N}\left\{\omega\in \Omega^N\,|\, \frac{1}{N}\sum_{k=1}^N \log \frac{P_\theta(\omega_k)}
 {P_b(\omega_k)}- S(P_{\theta}|P_b) \leq -\zeta\right\}.
\]
Proposition \ref{ny-verona-first} now  yields that for some constants $K_\zeta >0$ and $k_\zeta>0$, 
\[
\sup_{\theta \in [a^\prime, b^\prime]} P_{\theta N}(B_N(a)\cup B_N(b))\leq K_\zeta \e^{-k_\zeta N}
\]
for all $N\geq 1$. A simple but important observation is that 
if $\omega \not\in B_N(a)\cup B_N(b)$, then    $\hat \theta_{ML, N}(\omega)\in ]a, b[$ and so 
\beq\dot S_{\hat \theta_{ML, N}(\omega)N}(\omega)=0.
\label{verona-lunch}
\eeq
The  Taylor expansion gives that for any $\omega \in \Omega^N$ and $\theta \in [a, 
b]$  there is $\theta^\prime(\omega)$ between $\hat \theta_{ML, N}(\omega)$ and $\theta$ such that
\beq
\dot S_{\hat \theta_{ML, N}(\omega)N}(\omega)- \dot S_{\theta N}(\omega)= (\hat \theta_{ML, N}(\omega) - 
\theta)\left[\ddot S_{\theta N} + \frac{1}{2}(\hat \theta_{ML, N}(\omega) - 
\theta)\dddot S_{\theta^\prime(\omega) N}\right].
\label{verona-lunch-1}
\eeq
Write 
\[
E_{\theta N}\left(\left(\dot S_{\hat \theta_{ML, N}(\omega)N}(\omega)- \dot S_{\theta N}(\omega)\right)^2\right)=
L_N(\theta)+ E_{\theta N}\left(\left[\dot S_{ \theta N}\right]^2\right),
\]
where 
\beq 
L_N(\theta)=E_{\theta N}\left(\left[\dot S_{ \hat \theta_{ML, N}(\omega)N}\right]^2\right) +
2 E_{\theta N}\left(\dot S_{ \hat \theta_{ML, N}(\omega)N}\dot S_{\theta N}\right).
\label{titi}\eeq
It follows from \eqref{verona-lunch} that in \eqref{titi} $E_{\theta N}$ reduces to integration over $B_N(a)\cup B_N(b)$, 
and we arrive at the estimate 
\beq
\sup_{\theta \in [a^\prime, b^\prime]}|L_N(\theta)|\leq KN^2\sup_{\theta \in [a^\prime, b^\prime]}P_{\theta N}(B_N(a)\cup B_N(b))
\leq KN^2K_\zeta \e^{-k_\zeta N}
\label{insudis}
\eeq
for some uniform constant $K>0$, where by uniform we mean that $K$ does not depend 
on $N$. It is easy to see that one can take 
\[K= 3 \sup_{\theta \in [a, b], \omega \in \Omega}\left( \frac{\dot P_\theta(\omega)}{P_\theta(\omega)}\right)^2.
\]
In Exercise \ref{theend} the reader is asked to   estimate    other uniforms constant that will appear in the proof. 

Squaring both sides in \eqref{verona-lunch-1}, taking the expectation, and dividing 
both sides with $N^2$, we derive the identity
\beq
\frac{1}{N^2}L_N(\theta)+\frac{1}{N^2}E_{\theta N}\left(\left[\dot S_{ \theta N}\right]^2\right)=
E_{\theta N}\left((\hat \theta_{ML, N} - 
\theta)^2\left[\frac{\ddot S_{\theta N}}{N} + \frac{1}{2N}(\hat \theta_{ML, N} - 
\theta)\dddot S_{\theta^\prime N}\right]^2\right).
\label{verona-thur}
\eeq
An easy computation gives 
\[
\frac{1}{N^2}E_{\theta N}\left(\left[\dot S_{ \theta N}\right]^2\right)=\frac{1}{N}{\cal I}(\theta).
\]
Regarding the right hand side in \eqref{verona-thur}, we write 
it as 
\[
E_{\theta N}\left((\hat \theta_{ML, N} - 
\theta)^2\left[\frac{\ddot S_{\theta N}}{N}\right]^2\right) + R_N(\theta),
\]
where the remainder $R_N(\theta)$ can be estimated as 
\beq|R_N(\theta)|\leq C_1E_{\theta N}\left(|\hat \theta_{ML, N}-\theta|^3\right)
\label{verona-mass}
\eeq
for some uniform constant $C_1>0$.

With these simplifications, an algebraic 
manipulation of the identity \eqref{verona-thur} gives
\beq
\label{verona-mass-1}
NE_{\theta N}\left((\hat \theta_{ML, N}-\theta)^2\right)- \frac{1}{{\cal I}(\theta)}= - 
D_N(\theta)- \frac{N R_N(\theta)}{{\cal I}(\theta)^2} + \frac{1}{N}\frac{L_N(\theta)}{{\cal I}(\theta)^2},
\eeq
where
\beq
D_{N}(\theta)= N E_{\theta N}\left( (\hat \theta_{ML}-\theta)^2 \left( \left[
\frac{\ddot S_{\theta N}}{N}\right]^2\frac{1}{{\cal I}(\theta)^2}- 1 \right)\right).
\label{verona-mass-2}
\eeq
Writing 
\[
\left[
\frac{\ddot S_{\theta N}}{N}\right]^2\frac{1}{{\cal I}(\theta)^2}- 1=\frac{1}{{\cal I}(\theta)^2}\left( 
\frac{\ddot S_{\theta N}}{N} + {\cal I}(\theta)\right)\left( 
\frac{\ddot S_{\theta N}}{N} - {\cal I}(\theta)\right)
\]
and using that ${\cal I}(\theta)$ is continuous and strictly positive on $[a,b]$, we derive the estimate 
\beq
|D_N(\theta)|\leq C_2 N E_{\theta N}\left( (\hat \theta_{ML}-\theta)^2 \left|\frac{\ddot S_{\theta N}}{N} - {\cal I}(\theta)\right|
\right)
\label{bonny}
\eeq
for some uniform constant $C_2>0$. 

Fix $\epsilon >0$, and choose $C_\epsilon >0$ and $\gamma_\epsilon>0$ such that 
\beq
\sup_{\theta \in [a,b]}P_{\theta N}\left\{\omega \in \Omega^N\,|\, 
|\hat \theta_{ML, N}(\omega)-\theta|\geq \epsilon\right\}\leq C_\epsilon\e^{-\gamma_\epsilon N}, 
\label{bonny-bed-1}
\eeq
\beq
\sup_{\theta \in [a,b]}P_{\theta N}\left\{\omega \in \Omega^N\,|\, 
\left|\frac{\ddot S_{\theta N}}{N}- {\cal I}(\theta)\right|\geq \epsilon\right\}\leq C_\epsilon\e^{-\gamma_\epsilon N}. 
\label{bonny-bed-2}
\eeq
Here, \eqref{bonny-bed-1} follows from Theorem \ref{majolino}, while \eqref{bonny-bed-2} follows from 
Proposition \ref{ny-verona-first} applied to $X_\theta=-\frac{d^2}{\d \theta^2}\log P_\theta$ (recall that $E_\theta(X_\theta)={\cal I}(\theta)$).

Let $\delta=\inf_{u\in [a, b]}{\cal I}(u)$. Then, for all $\theta \in [a, b]$, 
\beq
\begin{split}
\frac{N |{\cal R}_N(\theta)|}{{\cal I}(\theta)^2}&\leq \frac{C_1N}{\delta^2}\int_{\Omega^N}|\hat \theta_{ML, N}-\theta|^3\d P_{\theta N}
\\[3mm]
&=\frac{C_1N}{\delta^2}\int_{|\hat\theta_{ML, N}-\theta|<\epsilon}|\hat \theta_{ML, N}-\theta|^3\d P_{\theta N} +
\frac{C_1N}{\delta^2}\int_{|\hat\theta_{ML, N}-\theta|\geq \epsilon}|\hat \theta_{ML, N}-\theta|^3\d P_{\theta N}\\[2mm]
&\leq \epsilon \frac{C_1}{\delta^2} NE_{\theta N}\left((\hat \theta_{ML, N}-\theta)^2\right)+ \frac{C_1(b-a)^3 N}{\delta^2}C_\epsilon 
\e^{-\gamma_{\epsilon N}}.
\end{split}
\label{bonny-sleep}
\eeq
Similarly, splitting the integral  (that is, $E_{\theta N}$) on the r.h.s. of  \eqref{bonny} into the sum of  integrals  over the sets 
\[
\left|\frac{\ddot S_{\theta N}}{N} - {\cal I}(\theta)\right|<\epsilon, \qquad \left|\frac{\ddot S_{\theta N}}{N} - {\cal I}(\theta)\right|\geq \epsilon,
\]
we derive that for all $\theta \in [a, b]$, 
\beq
|D_N(\theta)|\leq \epsilon C_2N E_{\theta N}\left((\hat \theta_{ML, N}-\theta)^2\right) + C_2C_3 NC_\epsilon 
\e^{-\gamma_{\epsilon N}},
\label{bonny-sleep-1}
\eeq
where $C_3>0$ is a uniform constant. Returning to \eqref{verona-mass-1} and taking $\epsilon=\epsilon_0$ such that 
\[\epsilon_0\frac{C_1}{\delta^2}<\frac{1}{4}, \qquad \epsilon_0C_2 <\frac{1}{4},
\]
the estimates \eqref{insudis},  \eqref{bonny-sleep}, and \eqref{bonny-sleep-1} give that for all $\theta \in [a^\prime,b^\prime]$, 
\[N E_{\theta N}\left(
(\hat\theta_{ML, N}-\theta)^2\right)\leq  \frac{2}{{\cal I}(\theta)} + C_{\epsilon_0}^\prime N\e^{-\gamma_{\epsilon_0}N} + 
\frac{2K}{\delta^2} K_\zeta \e^{-k_\zeta N},
\]
where $C_{\epsilon_0}^\prime>0$ is a uniform constant (that of course depends on $\epsilon_0$). It follows that 
\beq 
C^\prime=\sup_{N\geq 1}\sup_{\theta \in [a^\prime, b^\prime]}N E_{\theta N}\left((\hat \theta_{ML, N}-\theta)^2\right)<\infty.
\label{belgrade-night}
\eeq
Returning to  \eqref{bonny-sleep}, \eqref{bonny-sleep-1}, we then have that for any $\epsilon>0$, 
\beq
\sup_{\theta \in [a^\prime, b^\prime]}\frac{N |{\cal R}_N(\theta)|}{{\cal I}(\theta)^2}\leq \epsilon \frac{C_1}{\delta^2} C^\prime+ \frac{C_1(b-a)^3 N}{\delta^2}C_\epsilon 
\e^{-\gamma_{\epsilon N}},
\label{bonny-sleep-2}
\eeq
\beq
\sup_{\theta \in [a^\prime,b^\prime]}|D_N(\theta)|\leq \epsilon C_2C^\prime + C_2C_3 NC_\epsilon. 
\e^{-\gamma_{\epsilon N}}.
\label{bonny-sleep-3}
\eeq
Finally, returning once again to \eqref{verona-mass-1}, we derive that for any $\epsilon >0$, 
\[
\begin{split}
\sup_{\theta \in [a^\prime,b^\prime]} \left|NE_{\theta N}\left((\hat \theta_{ML, N}-\theta)^2\right)- \frac{1}{{\cal I}(\theta)}\right|
&\leq  \sup_{\theta \in [a^\prime,b^\prime]}|D_N(\theta)| + 
\sup_{\theta \in [a^\prime,b^\prime]} \frac{N |R_N(\theta)|}{{\cal I}(\theta)^2}+ \sup_{\theta \in [a^\prime, b^\prime]}
\frac{|L_N(\theta)|}{N{\cal I}(\theta)^2}\\[3mm]
&\leq \epsilon C^{\prime\prime} + C_{\epsilon}^{\prime\prime}N\e^{-\gamma_\epsilon N} + K K_\zeta\e^{-k_\zeta N},
\end{split}
\]
where $C^{\prime\prime}>0$  is a uniform constant and $C_\epsilon^{\prime\prime}>0$ depends only 
on $\epsilon$. 
Hence, 
\[
\limsup_{N\rightarrow \infty}
\sup_{\theta \in [a^\prime,b^\prime]} \left|NE_{\theta N}\left((\hat \theta_{ML, N}-\theta)^2\right)- \frac{1}{{\cal I}(\theta)}\right|
\leq  \epsilon C^{\prime\prime}.
\]
Since $\epsilon >0$ is arbitrary, the result follows. \qed

\begin{exo}  Write an explicit estimate for all uniform constants that have  appeared in the 
above proof. \label{theend}
\end{exo}
\begin{remark}The proof of Theorem \ref{mle-eff} hints at the special role the boundary points $a$ and $b$ of the chosen 
parameter interval may play in study of the  efficiency. The MLE is selected with respect to  the $[a,b]$ and 
$\hat \theta_{ML, N}(\omega)$ may take value $a$ or $b$ without the derivative $\dot S_{\hat \theta_{ML, N}(\omega)N}(\omega)$ 
vanishing. That forces the estimation of the probability of the set $B_N(a)\cup B_N(b)$ and the argument requires 
that $\theta$ stays away from the boundary points. If the parameter interval is replaced by a circle, there would be no boundary 
points and the above proof  then gives  that the uniform efficiency of the MLE  holds with respect to the entire parameter set. One may wonder 
whether a different type of argument may yield the same result in the case of $[a,b]$. The following example shows that this is 
not the case.

Let $\Omega=\{0,1\}$ and let $P_\theta(0)=1-\theta$, $P_\theta(1)=\theta$, where $\theta \in\,]0,1[$. One computes 
${\cal I}(\theta)=(\theta-\theta^2)^{-1}$. If $[a, b]\subset\, ]0,1[$   is selected as the estimation interval, the MLE $\theta_{ML, N}$ takes the following form:
\[
\hat \theta_{ML, N}(\omega_1, \cdots, \omega_N)=\frac{\omega_1+ \cdots + \omega_N}{N}\qquad \hbox{if}\qquad 
\frac{\omega_1+\cdots+\omega_N}{N}\in [a, b], 
\]
\[\hat \theta_{ML, N}(\omega_1, \cdots, \omega_N)=a \qquad \hbox{if}\qquad 
\frac{\omega_1+\cdots+\omega_N}{N}< a,\]
\[\hat \theta_{ML, N}(\omega_1, \cdots, \omega_N)=b \qquad \hbox{if}\qquad 
\frac{\omega_1+\cdots+\omega_N}{N}>b.\]
We shall indicate the dependence of $\hat \theta_{ML, N}$ on $[a, b]$ by  $\hat \theta_{ML, N}^{[a, b]}$. 
It follows from Theorem \ref{mle-eff} that 
\[
\lim_{N\rightarrow \infty}N E_{(\theta=1/2)N}\left(\left(\hat \theta_{ML, N}^{[\frac{1}{3}, \frac{2}{3}]}- \frac{1}{2}\right)^2\right)=
\left[{\cal I}\left(\frac{1}{2}\right)\right]^{-1}=\frac{1}{4}.
\]
On the other hand, a moment's reflection shows that 
\[
\frac{1}{2}E_{(\theta=1/2)N}\left(\left(\hat \theta_{ML, N}^{[\frac{1}{3}, \frac{2}{3}]}- \frac{1}{2}\right)^2\right)=
E_{(\theta=1/2)N}\left(\left(\hat \theta_{ML, N}^{[\frac{1}{2}, \frac{2}{3}]}- \frac{1}{2}\right)^2\right),
\]
and so 
\[
\lim_{N\rightarrow \infty}N E_{(\theta=1/2)N}\left(\left(\hat \theta_{ML, N}^{[\frac{1}{2}, \frac{2}{3}]}- \frac{1}{2}\right)^2\right)= 
\frac{1}{8}.
\]
Thus, in this case even the  bound of Proposition \ref{ny-never} fails at the boundary point $1/2$ at which 
the MLE becomes "superefficient". In general, such artificial boundary effects are difficult to quantify and we feel 
it is best that they are  excluded from the theory. These observations hopefully elucidate our  definition of efficiency which 
excludes the boundary points of the interval of parameters.
\label{last-re}
\end{remark}

\section{Notes and references}
For additional information and references  about parameter estimation  the reader may consult  \cite{LeCa, Vaa}. 
For additional information  about the Cram\'er-Rao bound and its history we refer the reader to the respective 
Wikipedia and Scholarpedia  articles. 

The modern theory of the MLE started  with the seminal work of Fisher \cite{Fis1}; for the fascinating history of 
the subject see \cite{Sti}.  Our analysis of the MLE follows the standard route, but I have followed no 
particular reference. In particular, I am not aware whether Theorem \ref{mle-eff} as formulated have appeared 
previously in the literature. 



\end{document}